\documentclass[journal]{IEEEtran}
\usepackage{graphicx}
\usepackage{amssymb}
\usepackage{epstopdf}
\usepackage{color}
\usepackage{subfigure}
\usepackage[cmex10]{amsmath}
\usepackage{hyperref,empheq}
\usepackage{url}
\usepackage[normalem]{ulem} 
\usepackage{graphicx}
\usepackage{amssymb}
\usepackage{xcolor}
\usepackage{subfigure}
\usepackage[cmex10]{amsmath}
\usepackage{cite}
\usepackage{url}
\usepackage[normalem]{ulem} 
\usepackage{etoolbox} 
\usepackage{empheq}
\usepackage{psfrag}

\usepackage{enumitem} 
\setlist[enumerate]{leftmargin=1.25em, itemsep=1ex} 
\setlist[itemize]{leftmargin=1.0em, itemsep=1ex} 

\newtoggle{long}
\toggletrue{long} 
\iftoggle{long}{
  \newcommand{\mytextstyle}{} 
}{
  \newcommand{\mytextstyle}{\textstyle} 
}
\newcommand{\longcolor}{black} 
\newcommand{\textl}[1]{\textcolor{\longcolor}{#1}}

{\bf}{\it}
{\em}

\DeclareMathAlphabet{\mathsfbf}{OT1}{cmss}{bx}{n}
\DeclareSymbolFont{bbold}{U}{bbold}{m}{n}
\DeclareSymbolFontAlphabet{\mathbbold}{bbold}
\renewcommand{\hat}{\widehat}

\renewcommand{\bar}{\overline}
\newcommand{\Dirac}{\mathbbold{1}}

\newcommand{\non}{\nonumber}

\makeatletter
\newcommand{\vast}{\bBigg@{3}} 
\newcommand{\Vast}{\bBigg@{4}}
\makeatother

\newcommand{\ba}[1]{\begin{array}{#1}}
\newcommand{\ea}{\end{array}}
\newcommand{\beq}{\begin{equation}}
\newcommand{\eeq}{\end{equation}}
\newcommand{\beqar}{\begin{eqnarray}}
\newcommand{\eeqar}{\end{eqnarray}}
\newcommand{\beqars}{\begin{eqnarray*}}
\newcommand{\eeqars}{\end{eqnarray*}}

 \newcommand{\defn}{\triangleq}

 \newcommand{\ovec}[1]{\ensuremath{\boldsymbol{\Bar{#1}}}}
 \newcommand{\hvec}[1]{\ensuremath{\boldsymbol{\Hat{#1}}}}
    
 \renewcommand{\vec}[1]{\ensuremath{\boldsymbol{#1}}}
 \newcommand{\mat}[1]{\ensuremath{\begin{bmatrix}#1\end{bmatrix}}}
 \newcommand{\smallmat}[1]{\ensuremath{
        \left[\begin{smallmatrix}#1\end{smallmatrix}\right]}}

 \newcommand{\norm}[1]{\ensuremath{\| #1 \|}}
 \newcommand{\mc}[1]{\ensuremath{\mathcal{#1}}}

 \newcommand{\Real}{{\mathbb{R}}}
 \newcommand{\Complex}{{\mathbb{C}}}
 \newcommand{\Int}{{\mathbb{Z}}}

 \newcommand{\modulo}[1]{\left\langle #1 \right\rangle}

 \newcommand{\tran}{^\textsf{T}}
 \newcommand{\herm}{^\textsf{H}}
 \newcommand{\of}[1]{^{(#1)}}
 \newcommand{\oft}[1]{^{{(#1)}\textsf{T}}}
 
 \newcommand{\ofH}[1]{^{{(#1)}\textsf{H}}}
 \newcommand{\ofsq}[1]{^{{(#1)}2}}
 
\newcommand{\CNor}{\mathcal{CN}}
\newcommand{\Nor}{\mathcal{N}}

 \DeclareMathOperator{\E}{E}
 \DeclareMathOperator{\var}{var}

 \DeclareMathOperator{\tr}{tr}
 \DeclareMathOperator{\diag}{diag}

 \DeclareMathOperator{\vect}{vec}
 
 \DeclareMathOperator{\Diag}{Diag}
 \DeclareMathOperator{\Cir}{Circ}
 \DeclareMathOperator{\Conv}{Conv}
 \DeclareMathOperator*{\argmax}{arg\, max}
 \DeclareMathOperator*{\argmin}{arg\, min}

 \renewcommand{\eqref}[1]{(\ref{eq:#1})}
 
 \newcommand{\Figref}[1]{Figure~\ref{fig:#1}}
 \newcommand{\figref}[1]{Fig.~\ref{fig:#1}}
 \newcommand{\tabref}[1]{Table~\ref{tab:#1}}
 \newcommand{\secref}[1]{Section~\ref{sec:#1}} 
 
 \newcommand{\appref}[1]{Appendix~\ref{app:#1}}

 \newcommand{\textb}[1]{\textcolor{black}{#1}}

\newcommand{\putFrag}[4]{\begin{figure}[tp]
                            \begin{center}
                            #4
                            \includegraphics[width=#3in]{figures/#1.eps}
                            \end{center}
			    \caption{#2}
			    \label{fig:#1}
                          \end{figure} }
\newcommand{\twoFrag}[4]{\begin{figure}[tp]
                            \centering
                            #4
                            \includegraphics[width=1.65in]{figures/#1.eps}
			    \hfill
                            \includegraphics[width=1.65in]{figures/#2.eps}
			    \caption{#3}
			    \label{fig:#1}
                          \end{figure} }

 \newcommand{\putTable}[3]{\begin{table}[tp]
  			    \centering
     			    #3
     			    \caption{#2}
     			    \label{tab:#1}
			  \end{table} }

 \newcommand{\giv}{\,|\,}
 \newcommand{\biggiv}{\,\big|\,}
 \newcommand{\Biggiv}{\,\Big|\,}

 \newcommand{\fxnvar}[3]{\Delta_{#1{\scriptscriptstyle \rightarrow} #2}^{#3}}
 \newcommand{\varfxn}[3]{\Delta_{#1{\scriptscriptstyle \leftarrow} #2}^{#3}}
 \newcommand{\X}{\textsf{x}}

 \newcommand{\Y}{\textsf{y}}
 
 \newcommand{\Z}{\textsf{z}}

 \newcommand{\p}{\textsf{p}}
 \newcommand{\R}{\textsf{r}}
 \newcommand{\Q}{\textsf{q}}
 \newcommand{\vX}{\textsf{\textbf{\textit{X}}}}

  \newcommand{\vY}{\textsf{\textbf{\textit{y}}}} 
  \newcommand{\vZ}{\textsf{\textbf{\textit{z}}}} 
  \newcommand{\mZ}{\textsf{\textbf{\textit{Z}}}} 

 \newcommand{\const}{\text{\sf const}}

 \newcommand{\Ord}{O}
 
  \newcommand{\Br}{b} 
  \newcommand{\B}{\textsf{\Br}} 
  \newcommand{\vB}{\textsf{\textbf{\textit{\Br}}}}  
  \newcommand{\mA}{\textsf{\textbf{\textit{A}}}}  
  \newcommand{\vBr}{\vec{\Br}} 
  \newcommand{\Nb}{N_b} 
  \newcommand{\iB}{i} 
  
  \newcommand{\Cr}{c} 
   \newcommand{\C}{\textsf{\Cr}} 
  \newcommand{\vC}{\textsf{\textbf{\textit{\Cr}}}}  
  \newcommand{\mX}{\textsf{\textbf{\textit{X}}}}  
  \newcommand{\vCr}{\vec{\Cr}} 
  \newcommand{\Nc}{N_c} 
  \newcommand{\jC}{j} 
 
  \newcommand{\Np}{N_p}
  \newcommand{\Ng}{N_g}
  \newcommand{\Nz}{N_z}
  \newcommand{\Na}{N_a} 
  \newcommand{\Nphi}{N_\phi} 

  \newcommand{\zmatml}[5]{{\hat{z}}_{{\scriptscriptstyle \rightarrow}#1}^{(#2,#3)}(#4)^{#5}}
  \newcommand{\zmat}[5]{{\hat{z}}_{#1}^{(#2,#3)}(#4)^{#5}}
  \newcommand{\zmatC}[5]{{z}_{#1}^{(#2,#3) #5}} 
  \newcommand{\nod}{*}
    \newcommand{\zbar}[5]{{\bar{z}}_{#1}^{(#2,#3)}(#4)^{#5}}
    
\graphicspath{{./figures/}} 

\newcommand{\remove}[1]{}

\newcommand{\figsize}{3.0}
 
\begin{document}
\setlength{\arraycolsep}{0.4mm}
\title{Parametric Bilinear Generalized Approximate Message Passing}

\author{Jason~T.~Parker 
        and~Philip~Schniter 
\thanks{J. Parker is with the Air Force Research Laboratory, Dayton, OH 45433, e-mail: jason.parker.13@us.af.mil.  His work on this project has been supported by AFOSR Lab Task 11RY02COR.}
\thanks{P. Schniter is with the Dept. ECE, The Ohio State University, 2015 Neil Ave., Columbus OH 43210, e-mail: schniter@ece.osu.edu, phone 614.247.6488, fax 614.292.7596.  His work on this project has been supported by NSF grants IIP-0968910, CCF-1018368, CCF-1218754, and by an allocation of computing time from the Ohio Supercomputer Center.}
\thanks{Portions of this work appeared in \cite{Parker:Diss:14} and were presented at the Information Theory and Applications Workshop, La Jolla, CA, USA, February 2015.}
}

\markboth{\today}{}

\maketitle

\begin{abstract}
We propose a scheme to estimate the parameters $b_i$ and $c_j$ of the bilinear form $z_m=\sum_{i,j} b_i z_m^{(i,j)} c_j$ from noisy measurements $\{y_m\}_{m=1}^M$, where $y_m$ and $z_m$ are related through an arbitrary likelihood function and $z_m^{(i,j)}$ are known.
Our scheme is based on generalized approximate message passing (G-AMP): it treats $b_i$ and $c_j$ as random variables and $z_m^{(i,j)}$ as an i.i.d.\ Gaussian 3-way tensor in order to derive a tractable simplification of the sum-product algorithm in the large-system limit.
It generalizes previous instances of bilinear G-AMP, such as those that estimate matrices $\boldsymbol{B}$ and $\boldsymbol{C}$ from a noisy measurement of $\boldsymbol{Z}=\boldsymbol{BC}$, allowing the application of AMP methods to problems such as self-calibration, blind deconvolution, and matrix compressive sensing.
Numerical experiments confirm the accuracy and computational efficiency of the proposed approach.
\end{abstract}


\IEEEpeerreviewmaketitle

\section{Introduction}	\label{sec:intro}

\subsection{Motivation}  \label{sec:motiv}

Many problems in engineering, science, and finance can be formulated as the estimation of a structured matrix $\vec{Z}\in\Real^{M\times L}$ from a noisy (or otherwise corrupted) observation $\vec{Y}\in\Real^{M\times L}$.
For various types of structure, the problem reduces to a well-known specialized problem.
For example,
when $\vec{Z}$ has a low-rank structure and only a subset of its entries are observed (possibly in noise), the estimation of $\vec{Z}$ is known as \emph{matrix completion} (MC) \cite{Candes:PROC:10}.
When $\vec{Z}=\vec{L}+\vec{S}$ for low-rank $\vec{L}$ and sparse $\vec{S}$, 
the estimation of $\vec{L}$ and $\vec{S}$ from a (noisy) observation of $\vec{Z}$ is known as \emph{robust principal components analysis} (RPCA) \cite{Candes:JACM:11,Chandrasekaran:JO:11} or \emph{stable principle components pursuit} (SPCP) \cite{Zhou:ISIT:10}.
When $\vec{Z}=\vec{BC}$ with sparse $\vec{C}$, the problem of estimating $\vec{B}$ and $\vec{C}$ from a (noisy) observation of $\vec{Z}$ is known as \emph{dictionary learning} (DL) \cite{jtp_Rubinstein2010}.
When $\vec{Z}\!=\!\vec{BC}$ and both $\vec{B}$ and $\vec{C}$ are positive,
the problem of estimating $\vec{B},\vec{C}$ from a (noisy) observation of $\vec{Z}$ is known as \emph{nonnegative matrix factorization} (NMF) \cite{Lee:NIPS:01}.

In this paper, we propose an AMP-based approach to a more general class of structured-matrix estimation problems.
Our work is motivated by problems like the following.
\begin{enumerate}
\item
Estimate $\vec{b}$ and $\vec{C}$ from a noisy observation of%
\footnote{For clarity, we typeset matrices in bold capital, vectors in bold lowercase, and scalars in non-bold. Furthermore, we typeset random variables in san-serif font (e.g., $\mZ$) and deterministic realizations in serif font (e.g., $\vec{Z}$).}
\begin{align}
\vec{Z} 
&= \Diag(\vec{Hb})\vec{AC} 
\label{eq:SparseLift}
\end{align}
with known $\vec{H}$ and $\vec{A}$. 
This problem manifests, e.g., in 
\begin{itemize}
\item
\emph{Self-calibration} \cite{Ling:IP:15}. 
Here the columns of $\vec{C}$ are measured through a linear system, represented by the matrix $\vec{A}$, whose outputs are subject to unknown (but structured) gains of the form $\vec{Hb}$.
The goal is to simultaneously recover the signal $\vec{C}$ and the calibration parameters $\vec{b}$. 
\item
\emph{Blind circular deconvolution}: 
Here the columns of $\vec{C}$ are circularly convolved with the channel $\vec{b}$, and the goal is to simultaneously recover $\vec{C}$ and $\vec{b}$ from a noisy version of the Fourier-domain convolution outputs.\footnote{%
Recall that circular convolution between $\vec{b}$ and $\vec{c}_l$ can be written as $\vec{v}_l=\Cir(\vec{b})\vec{c}_l$, with circulant matrix $\Cir(\vec{b})=\vec{A}\herm \Diag(\sqrt{N} \vec{Ab}) \vec{A}$ for unitary discrete Fourier transform (DFT) matrix $\vec{A}$.  
The DFT of the convolution outputs is then $\vec{A}\vec{v}_l=\Diag(\sqrt{N} \vec{Ab}) \vec{A}\vec{c}_l$, matching \eqref{SparseLift}.} 
\end{itemize}

\item
Consider the more general\footnote{%
Note \eqref{SparseLift} is a special case of \eqref{STLS} with $\vec{A}\of{i}=\Diag(\vec{h}_i)\vec{A}$, where $\vec{h}_i$ denotes the $i$th column of $\vec{H}$.
} problem of estimating $\{b_i\}$ and $\vec{C}$ from a noisy observation of 
\begin{align}
\vec{Z} 
&= \mytextstyle \sum_i b_i \vec{A}\of{i} \vec{C} 
\label{eq:STLS}
\end{align}
with known $\{\vec{A}\of{i}\}$.
This problem manifests, e.g., in
\begin{itemize}
\item
\emph{Compressive sensing with matrix uncertainty} \cite{Zhu:TSP:11}.
Here, $\vec{Z}=\vec{A C}$ where $\vec{A}=\sum_i b_i \vec{A\of{i}}$ is an unknown (but structured) sensing matrix and the columns of $\vec{C}\in\Real^{N\times L}$ are sparse signals.
The goal is to simultaneously recover $\vec{C}$ and the matrix uncertainty parameters $\{b_i\}$.
\item
\emph{Joint channel-symbol estimation}.
Say a symbol stream $\{c_i\}$ is transmitted through a length-$\Nb$ convolutive channel $\{b_i\}$, where the same length-$\Ng\geq \Nb-1$ guard interval is repeated every $\Np$ samples in $\{c_i\}$.
Then the noiseless convolution outputs can be written as $\vec{Z} = \sum_i b_i \vec{A}\of{i} \vec{C}$, where $\vec{A}\of{i}=\smallmat{\vec{0}_{\Np\times (\Ng-i+1)} & \vec{I}_{\Np} & \vec{0}_{\Np\times (i-1)}}$ and where the first and last $\Ng$ rows in $\vec{C}$ are guard symbols.
The goal is to jointly estimate the channel $\{b_i\}$ and the (finite-alphabet) data symbols in $\vec{C}$.
\end{itemize}

\item
Consider the yet more general\footnote{%
\iftoggle{long}{\appref{intro} shows}{\cite{Parker:PBiGAMP} shows that} \eqref{STLS} is a special case of \eqref{MCS} with rank-one $\vec{L}$ and $\vec{S}=\vec{0}$.
} problem of estimating low-rank $\vec{L}$ and sparse $\vec{S}$ from noisy observations of 
\begin{align}
z_m
&= \tr\{\vec{\Phi}_m\tran (\vec{L}+\vec{S})\}~\text{for}~m=1,\dots,\Nz
\label{eq:MCS}
\end{align}
with known $\{\vec{\Phi}_m\}$.
This problem is sometimes known as \emph{matrix compressive sensing} (MCS),
which has applications in, e.g.,
video surveillance \cite{Waters:NIPS:11},
hyperspectral imaging \cite{Waters:NIPS:11}, 
quantum state tomography \cite{Candes:TIT:11}, 
multi-task regression \cite{Agarwal:AS:12},
and
image processing \cite{Wright:II:13}. 

\iftoggle{long}{
\color{\longcolor}
\item
Another problem of interest is the estimation of matrices $\vec{B}$ and $\vec{C}$ from a noisy observation of
\begin{align}
\vec{Z}_l
= \vec{F}_l \vec{BC} \vec{G}_l ~\text{for}~l=0,\dots,\Nz,
\label{eq:fusion}
\end{align}
with known $\{\vec{F}_l,\vec{G}_l\}$
This problem arises, e.g., in \emph{spatial-spectral data fusion super-resolution}, which aims to the hyperspectral images captured by $\Nz$ cameras \cite{Bioucas:GRSM:13}.
In this case, the matrix $\vec{BC}$ models the high-resolution spatial-spectral scene of interest: 
$\vec{B}$ is a tall positive matrix containing material spectra and $\vec{C}$ is a wide positive (and often sparse) matrix containing material abundances.
Then $\vec{G}_l$ and $\vec{F}_l$ represent the spatial and spectral blurring/downsampling operators associated with the $l$th camera, which have fast implementations.
\color{black}
}{}
\end{enumerate}

\subsection{Approach}  \label{sec:approach}

To solve structured-matrix estimation problems like those above, we start with a noiseless model of the form
\begin{align}
\vec{z}
&= \mytextstyle \sum_{\iB=0}^{\Nb} \sum_{\jC=0}^{\Nc} b_{\iB} \vec{z}\of{\iB,\jC} c_{\jC} 
\in \Real^M
\label{eq:ZPB} ,
\end{align}
where 
$b_0=1/\sqrt{\Nb}$, 
$c_0=1/\sqrt{\Nc}$, and
\textb{$\vec{z}\of{\iB,\jC}\in\Real^M~\forall \iB,\jC$} 
are known.
\textb{Note that the collection $\{\vec{z}\of{\iB,\jC}\}_{\forall i,j}$ defines a tensor of size $M\times (\Nb+1)\times (\Nc+1)$.}
We then estimate the parameters $\vec{b}=[b_1,\dots,b_{\Nb}]\tran$ and $\vec{c}=[c_1,\dots,c_{\Nc}]\tran$ 
from $\vec{y}$, a ``noisy'' observation of $\vec{z}$.
In doing so, we treat $\vBr$ and $\vCr$ as realizations of random vectors $\vB$ and $\vC$ with independent components, i.e., 
\begin{align}
p_{\vB,\vC}(\vBr,\vCr) 
&= \mytextstyle \prod_{\iB = 1}^{\Nb} p_{\B_{\iB}}(\Br_{\iB}) 
   \prod_{\jC = 1}^{\Nc} p_{\C_{\jC}}(\Cr_{\jC}), 
\label{eq:pBCPB}
\end{align}
and we assume that the likelihood function of $\vec{z}$ takes the separable form
\begin{align}
p_{\vY|\vZ}(\vec{y}\giv\vec{z}) 
&= \mytextstyle \prod_{m=1}^M p_{\Y_{m}|\Z_{m}}(y_{m}\giv z_{m}) 
\label{eq:pYgivZPB} .
\end{align}
Note that our definition of ``noisy'' is quite broad due to the generality of $p_{\Y_{m}|\Z_{m}}$.
For example, \eqref{pYgivZPB} facilitates both additive noise and nonlinear measurement models like those arising with, e.g., quantization \cite{Kamilov:TSP:12}, Poisson noise \cite{Fletcher:NIPS:11}, and phase retrieval \cite{Schniter:TSP:15}.
Note also that, since $b_0$ and $c_0$ are known, the model \eqref{ZPB} includes bilinear, linear, and constant terms%
\iftoggle{long}{, i.e., 
\color{\longcolor}
\begin{align}
\vec{z}
&= \mytextstyle 
\sum_{\iB=1}^{\Nb} \sum_{\jC=1}^{\Nc} b_{\iB} \vec{z}\of{\iB,\jC} c_{\jC} 
+c_0 \sum_{\iB=1}^{\Nb} b_{\iB} \vec{z}\of{\iB,0}
+b_0 \sum_{\jC=1}^{\Nc} \vec{z}\of{0,\jC} c_{\jC} 
\non\\&\quad \mytextstyle
+ c_0 b_0 \vec{z}\of{0,0} 
\label{eq:ZPB2} .
\end{align}
\color{black}
}{.}
In \secref{simplifications}, we demonstrate how \eqref{ZPB}-\eqref{pYgivZPB} can be instantiated to solve various structured-matrix estimation problems.

Our estimation algorithm is based on the AMP framework \cite{Montanari:Chap:12}.
\textb{%
Previously, AMP was applied to the \emph{generalized linear} problem:
``estimate i.i.d.\ $\vX$ from $\vec{y}$, a noisy realization of $\vZ=\vec{A}\vX$,''
leading to the G-AMP algorithm \cite{Rangan:ISIT:11},
and the \emph{generalized bilinear} problem:
``estimate i.i.d.\ $\mA$ and $\mX$ from $\vec{Y}$, a noisy realization of $\mZ=\mA\mX$,''
leading to the BiG-AMP algorithm \cite{Parker:TSP:14a,Parker:TSP:14b,Kabashima:14}.
In this paper, we apply AMP to estimate 
$\vB$ and $\vC$ from a noisy measurement of the \emph{parametric bilinear} output $\mZ=\vec{A}(\vB)\vec{X}(\vC)$, where $\vec{A}(\cdot)$ and $\vec{X}(\cdot)$ are matrix-valued affine linear functions. 
We write the relationship between $\vec{b}$, $\vec{c}$, and $\vec{z}\defn\vect(\vec{Z})$ more concisely as \eqref{ZPB} and coin the resulting algorithm ``\emph{Parametric BiG-AMP}'' (P-BiG-AMP).}

We also show that, using an expectation-maximization (EM) \cite{Dempster:JRSS:77} approach similar to those used in other AMP-based works \cite{Krzakala:JSM:12,Vila:TSP:13,Kamilov:TIT:14}, we can generalize our approach to the case where the parameters governing the distributions $p_{\B_i}$, $p_{\C_j}$, and $p_{\Y_m|\Z_m}$ are unknown.

\subsection{Relation to Previous Work}

We now describe related literature, starting with versions of compressive sensing (CS) under sensing-matrix uncertainty.

\iftoggle{long}{
\color{\longcolor}
Consider first the problem of \emph{single} measurement vector (SMV) CS with \emph{unstructured} matrix uncertainty, i.e., recovering the sparse vector $\vec{c}$ from a noisy observation of
$\vec{z}=(\vec{A}+\vec{B})\vec{c}$, 
where $\vec{A}$ is known and the elements of $\vec{B}$ are small i.i.d.\ perturbations \cite{Herman:JSTSP:10}.
AMP based approaches to minimum mean-squared error (MMSE) estimation were proposed in
\cite{Parker:ASIL:11,Krzakala:ICASSP:13}.
The extension to the \emph{multiple} measurement vector (MMV) case,
$\vec{Z}=(\vec{A}+\vec{B})\vec{C}$,
eliminates the need for $\vec{B}$ to be small and yields the DL problem discussed in \secref{motiv}. 
For the latter, AMP-based algorithms were proposed in \cite{Parker:TSP:14b,Kabashima:14}.
The proposed P-BiG-AMP generalizes this line of work.%
\color{black}

Next consider MMV}{
Consider first the problem of}
multiple measurement vector (MMV) CS with \emph{output gain} uncertainty, i.e., recovering $\vec{C}$ with sparse columns from a noisy observation of
$\vec{Z}=\Diag(\vec{b})\vec{AC}$,
where $\vec{A}$ is known and $\vec{b}$ is unknown.
For the case of positive $\vec{b}$ and no noise,
\cite{Gribonval:ICASSP:12} proposed a convex approach based on $\ell_1$ minimization, which was generalized to arbitrary $\vec{b}$ in \cite{Bilen:TSP:14}.
For MMSE estimation in the noisy case, a G-AMP-based approach to the MMV version was proposed in \cite{Schulke:NIPS:14}, 
and G-AMP approaches to the single measurement vector (SMV) version with coded-symbol $\vec{b}$ and constant-modulus $\vec{b}$ were proposed in \cite{Schniter:JSTSP:11} and \cite{Schniter:TSP:15}.
Our proposed P-BiG-AMP approach handles more general forms of matrix uncertainty than \cite{Schulke:NIPS:14,Schniter:JSTSP:11,Schniter:TSP:15}.

MMV CS with \emph{input gain} uncertainty, i.e., recovering possibly-sparse $\vec{C}$ from a noisy observation of
$\vec{Z}=\vec{A}\Diag(\vec{b})\vec{C}$,
where $\vec{A}$ is known and $\vec{b}$ is unknown, was considered in \cite{Kamilov:ICASSP:13}.  
There, G-AMP estimation of $\vec{C}$ was alternated with EM estimation of $\vec{b}$ using the EM-AMP framework from \cite{Kamilov:TIT:14}. 
As such, \cite{Kamilov:ICASSP:13} does not support a prior on $\vec{b}$.

A related problem is SMV CS with \emph{subspace-structured} output gain uncertainty, i.e., recovering sparse $\vec{c}$ from a noisy observation of
$\vec{z}=\Diag(\vec{Hb})\vec{Ac}$ with known $\vec{A},\vec{H}$.
This problem is perhaps better known as \emph{blind deconvolution} of sequences $\vec{b},\vec{c}$ when $\vec{H},\vec{A}$ are DFT matrices and $\vec{z}$ is the DFT-domain noiseless measurement vector.
Several convex approaches to blind deconvolution have been proposed using the ``lifting'' technique, which transforms the problem to that of recovering a rank-$1$ matrix $\vec{L}$ from a (noisy) observation of $z_m=\tr\{\vec{\Phi}\tran_m\vec{L}\}$ for $m=1,...,M$.
For example, \cite{Asif:ALL:09} 
proposed a convex relaxation that applies to linear convolution with sparse $\vec{c}$,
\cite{Ahmed:TIT:14} 
proposed a convex relaxation (with guarantees) that applies to circular convolution with non-sparse $\vec{b},\vec{c}$, 
\cite{Ling:IP:15} 
proposed a convex relaxation (with guarantees) that applies to circular convolution with sparse $\vec{c}$,
and 
\cite{Hedge:TSP:11} 
proposed alternating and greedy schemes for sparse $\vec{b},\vec{c}$.
Meanwhile, identifiability conditions were studied in \cite{Choudhary:deconv1:14,Choudhary:deconv2:15,Li:deconv1:15,Li:deconv2:15}.


For \eqref{STLS}, i.e., CS with \emph{general} matrix uncertainty,
\cite{Zhu:TSP:11} proposed an alternating minimization scheme and \cite{Choudhary:bilinear:14} showed that the problem can be convexified via lifting and then used that insight to study identifiability issues.

Finally, consider the \emph{matrix CS} problem given by \eqref{MCS}.
For generic\footnote{For the special case where each $\vec{\Phi}_m$ has a single unit-valued entry (i.e., noisy elements of $\vec{L}+\vec{S}$ are directly observed), many more schemes have been proposed 
(e.g., \cite{Candes:JACM:11,Chandrasekaran:JO:11,Zhou:ICML:11}), 
including AMP-based schemes \cite{Parker:TSP:14a,Parker:TSP:14b,Kabashima:14}.} 
$\{\vec{\Phi}_m\}$, greedy schemes were proposed in \cite{Waters:NIPS:11} and \cite{Kyrillidis:12b} and convex ones in \cite{Candes:TIT:11,Agarwal:AS:12,Wright:II:13,Aravkin:UAI:14}.

The P-BiG-AMP approach that we propose in this work supports all of the above matrix-uncertain CS, blind deconvolution, and low-rank-plus-sparse recovery models. 
Moreover, it allows arbitrary priors on $b_i$ and $c_j$, allowing the exploitation of (approximate) sparsity, constant-modulus structure, finite-alphabet structure, etc.
Furthermore, it allows a generic likelihood function of the form \eqref{pYgivZPB}, allowing non-linear measurement models like quantization, Poisson noise, phase-retrieval, etc.
Although it is non-convex and comes with no performance guarantees, it attacks the MMSE problem directly, and the empirical results in \secref{numerical} suggest that it offers better MSE recovery performance than recent convex relaxations while being computationally competitive (if not faster).

\subsection{Organization and Notation}

The remainder of this manuscript is organized as follows.
In \secref{prelim} we present preliminary material on belief propagation and AMP, 
and in \secref{PBigAmp} we derive our P-BiG-AMP algorithm.
In \secref{simplifications} we show how the implementation of P-BiG-AMP can be simplified for several problems of interest,
and in \secref{numerical} we present the results of several numerical experiments.
In \secref{conc}, we conclude.
\iftoggle{long}{}{Note that this manuscript is a shortened version of \cite{Parker:PBiGAMP}.}

\emph{Notation}:
For random variable $\X$, we use $p_{\X}(x)$ for the pdf, $\E\{\X\}$ for the mean, and $\var\{\X\}$ for the variance.
$\Nor(x;\hat{x},\nu^x)$ denotes the Gaussian pdf with mean $\hat{x}$ and variance $\nu^x$.
For a matrix $\vec{X}$, we use
$\vec{x}_{l}=[\vec{X}]_{:,l}$ to denote the $l^{th}$ column,
$x_{nl}=[\vec{X}]_{nl}$ to denote the entry in the $n^{th}$ row and $l^{th}$ column,
$\vec{X}\tran$ the transpose, 
$\vec{X}^*$ the conjugate, 
$\vec{X}\herm$ the conjugate transpose, 
$\|\vec{X}\|_F$ the Frobenius norm, 
and $\|\vec{X}\|_*$ the nuclear norm. 
For vectors $\vec{x}$, we use
$x_n=[\vec{x}]_n$ to denote the $n^{th}$ entry and
$\norm{\vec{x}}_p=(\sum_n |x_n|^p)^{1/p}$ to denote the $\ell_p$ norm.
$\Diag(\vec{x})$ is the diagonal matrix with diagonal elements $\vec{x}$,
$\Conv(\vec{x})$ is the convolution matrix with first column $\vec{x}$,
and $\Cir(\vec{x})$ is the circular convolution matrix with first column $\vec{x}$.

\section{Preliminaries} \label{sec:prelim}

\subsection{Bayesian Inference} \label{sec:Bayes}
For the model defined by \eqref{ZPB}-\eqref{pYgivZPB}, the posterior pdf is
\begin{align}
\lefteqn{
p_{\vB,\vC|\vY}(\vBr,\vCr\giv\vec{y}) 
= p_{\vY|\vB,\vC}(\vec{y}\giv\vBr,\vCr) 
   \,p_{\vB}(\vBr) 
   \,p_{\vC}(\vCr)/p_{\vY}(\vec{y}) 
}
\label{eq:BayesPB}\\
&\propto p_{\vY|\vZ}(\vec{y}\giv\vec{z}(\vBr,\vCr)) 
   \,p_{\vB}(\vBr) 
   \,p_{\vC}(\vCr) \label{eq:scalingPB}\\
&= \mytextstyle 
\textb{
 \Big(\prod_m p_{\Y_{m}|\Z_{m}\!}\big(y_{m}\biggiv z_{m}(\vBr,\vCr) \big)\Big)
 \Big(\prod_{\iB} p_{\B_{\iB}\!}(\Br_{\iB})\Big) 
 \Big(\prod_{\jC} p_{\C_{\jC}\!}(\Cr_{\jC})\Big), }
\label{eq:postPB}
\end{align}
where 
\eqref{BayesPB} used Bayes' rule and  
$\propto$ denotes equality up to a scale factor. 
This pdf can be represented using the bipartite factor graph shown in \figref{factor_graph}. 
There, the factors in \eqref{postPB} are represented by ``factor nodes'' appearing as black boxes and the random variables in \eqref{postPB} are represented by ``variable nodes'' appearing as white circles. 
Note that the observed data $\{y_{m}\}$ are treated as parameters of the $p_{\Y_{m}|\Z_{m}\!}(y_{m}|\cdot)$ factor nodes, and not as random variables. 
Although \figref{factor_graph} shows an edge between \emph{every} $\B_i$ and $p_{\Y_{m}|\Z_{m}\!}$ node pair, the edge will vanish when $z_{m}(\vB,\vC)$ does not depend on $\B_i$, and similar for $\C_j$.

\putFrag{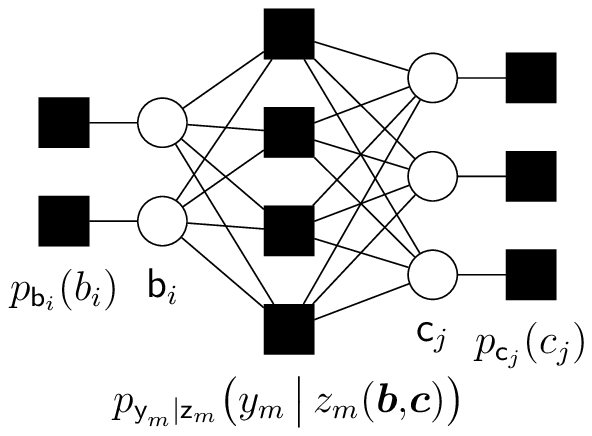}{The factor graph for parametric generalized bilinear inference under $\Nb=2$, $\Nc = 3$, and $M=4$.}{1.8}

\subsection{Loopy Belief Propagation} \label{sec:LBP}

Our goal is to compute minimum mean-squared error (MMSE) estimates of $\vec{b}$ and $\vec{c}$, i.e., the means of the marginal posteriors $p_{\B_{i}|\vY}(\cdot\giv\vec{y})$ and $p_{\C_{j}|\vY}(\cdot\giv\vec{y})$.
Since exact computation is intractable in our problem (see below), we consider approximate computation using loopy belief propagation (LBP).

In LBP, beliefs about the random variables (in the form of pdfs or log pdfs) are propagated among the nodes of the factor graph until they converge.
The standard way to compute these beliefs, known as the \emph{sum-product algorithm} (SPA) \cite{Pearl:Book:88,Kschischang:TIT:01}, says that the belief emitted by a variable node along a given edge of the graph is computed as the product of the incoming beliefs from all other edges, whereas the belief emitted by a factor node along a given edge is computed as the integral of the product of the factor associated with that node and the incoming beliefs on all other edges.
The product of all beliefs impinging on a given variable node yields the posterior pdf for that variable.
In cases where the factor graph has no loops, exact marginal posteriors result from two (i.e., forward and backward) passes of the SPA \cite{Pearl:Book:88,Kschischang:TIT:01}.
For loopy factor graphs like ours, exact inference is in general NP hard \cite{Cooper:AI:90} and so LBP does not guarantee correct posteriors.
However, it often gives good approximations \cite{Murphy:UAI:99}.

\subsection{Sum-Product Algorithm} \label{sec:SPA}
We formulate the SPA using the messages and log-posteriors specified in \tabref{logpdfPB}.
All take the form of log-pdfs with arbitrary constant offsets, which can be converted to pdfs via exponentiation and scaling.
\iftoggle{long}{
\color{\longcolor}
For example, the message $\fxnvar{m}{\iB}{\B}(t,.))$ corresponds to the pdf $\frac{1}{C}\exp(\fxnvar{m}{\iB}{\B}(t,.))$ with $C=\int_{\Br_{\iB}} \exp(\fxnvar{m}{\iB}{\B}(t,\Br_{\iB}))$.
\color{black}
}{}

Applying the SPA to the factor graph in \figref{factor_graph}, we arrive at the following update rules for the four messages in \tabref{logpdfPB}:
\begin{align}
&\fxnvar{m}{\iB}{\B}(t,\Br_{\iB}) = \mytextstyle
        \log \int_{\{\Br_{r}\}_{r \ne \iB}, \{\Cr_{k} \}_{k = 1}^{\Nc}}
        p_{\Y_{m}|\Z_{m}}\big(y_{m}\biggiv z_{m}(\vBr,\vCr) \big)
        \non\\&\quad\times \mytextstyle
        \prod_{r \ne \iB} \exp \big( \varfxn{m}{r}{\B}(t,\Br_r) \big)
        \prod_{k=1}^{\Nc} \exp\big( \varfxn{m}{k}{\C}(t,\Cr_k)\big) 
        \non\\&\quad
        + \const \label{eq:zTobPB}\\
&\fxnvar{m}{\jC}{\C}(t,\Cr_{\jC}) = \mytextstyle
        \log \int_{\{\Br_{r}\}_{r =1}^{\Nb}, \{\Cr_{k} \}_{k \ne \jC}}
        p_{\Y_{m}|\Z_{m}}\big(y_{m}\biggiv z_{m}(\vBr,\vCr) \big)
        \non\\&\quad\times \mytextstyle
        \prod_{r=1}^{\Nb} \exp \big(\varfxn{m}{r}{\B}(t,\Br_r)\big)
        \prod_{k \ne \jC} \exp \big(\varfxn{m}{k}{\C}(t,\Cr_k)\big) 
        \non\\&\quad
        + \const \label{eq:zTocPB}\\
&\varfxn{m}{\iB}{\B}(t\!+\!1,\Br_{\iB}) = \mytextstyle
        \log p_{\B_{\iB}}(\Br_{\iB}) + 
        \sum_{r \ne m} \fxnvar{r}{\iB}{\B}(t,\Br_{\iB}) + \const \label{eq:bTozPB}\\
&\varfxn{m}{\jC}{\C}(t\!+\!1,\Cr_{\jC}) = \mytextstyle
        \log p_{\C_{\jC}}(\Cr_{\jC}) + 
        \sum_{r \ne m} \fxnvar{r}{\jC}{\C}(t,\Cr_{\jC}) + \const \label{eq:cTozPB} ,
\end{align}
where $\const$ denotes a constant (w.r.t $\Br_{\iB}$ in \eqref{zTobPB} and \eqref{bTozPB} and w.r.t $\Cr_{\jC}$ in \eqref{zTocPB} and \eqref{cTozPB}).
In the sequel, we denote the mean and variance of the pdf $\frac{1}{C}\exp(\varfxn{m}{\iB}{\B}(t,.)$ by $\hat{\Br}_{m,\iB}(t)$ and $\nu^{\Br}_{m,\iB}(t)$, respectively, and we denote the mean and variance of $\frac{1}{C}\exp(\varfxn{m}{\jC}{\C}(t,.))$ by $\hat{\Cr}_{m,\jC}(t)$ and $\nu^{\Cr}_{m,\jC}(t)$.
We refer to the vectors of these statistics for a given $m$ as
$\vec{\hat{\Br}}_{m}(t),\vec{\nu}^{\Br}_{m}(t) \in \Real^{\Nb}$
and
$\vec{\hat{\Cr}}_{m}(t),\vec{\nu}^{\Cr}_{m}(t) \in \Real^{\Nc}$.
For the log-posteriors, the SPA implies
\begin{align}
\Delta_{\iB}^{\B}(t\!+\!1,\Br_{\iB}) &= \mytextstyle
        \log p_{\B_{\iB}\!}(\Br_{\iB})
        + \sum_{m} \fxnvar{m}{\iB}{\B}(t,\Br_{\iB}) + \const \label{eq:DbPB}\\
\Delta_{\jC}^{\C}(t\!+\!1,\Cr_{\jC}) &= \mytextstyle
        \log p_{\C_{\jC}\!}(\Cr_{\jC})
        + \sum_{m} \!\fxnvar{m}{\jC}{\C}(t,\Cr_{\jC}) + \const \label{eq:DcPB} 
\end{align}
and we denote the mean and variance of
$\frac{1}{C}\exp(\Delta_{\iB}^{\B}(t,.))$
by $\hat{\Br}_{\iB}(t)$ and $\nu^{\Br}_{\iB}(t)$,
and the mean and variance of
$\frac{1}{C}\exp(\Delta_{\jC}^{\C}(t,.))$
by $\hat{\Cr}_{\jC}(t)$ and $\nu^{\Cr}_{\jC}(t)$.
Finally, we denote the vectors of these statistics as
$\vec{\hat{\Br}}(t),\vec{\nu}^{\Br}(t) \in \Real^{\Nb}$
and
$\vec{\hat{\Cr}}(t),\vec{\nu}^{\Cr}(t) \in \Real^{\Nc}$.

\putTable{logpdfPB}{SPA message definitions at iteration $t\in\Int$.}
{\renewcommand{\arraystretch}{1.3}
\begin{tabular}{|r|l|} 
\hline
  $\fxnvar{m}{\iB}{\B}(t,.)$
  & SPA message from node $p_{\Y_{m}|\Z_{m}}$ to node $\B_{\iB}$\\
  $\varfxn{m}{\iB}{\B}(t,.)$
  & SPA message from node $\B_{\iB}$ to node $p_{\Y_{m}|\Z_{m}}$\\
  $\fxnvar{m}{\jC}{\C}(t,.)$
  & SPA message from node $p_{\Y_{m}|\Z_{m}}$ to node $\C_{\jC}$ \\
  $\varfxn{m}{\jC}{\C}(t,.)$
  & SPA message from node $\C_{\jC}$ to node $p_{\Y_{m}|\Z_{m}}$\\
  $\Delta_{\iB}^{\B}(t,.)$
  & SPA-approximated log posterior pdf of $\B_{\iB}$\\
  $\Delta_{\jC}^{\C}(t,.)$
  & SPA-approximated log posterior pdf of $\C_{\jC}$\\
  \textb{ $\hat{\Br}_{m,\iB}(t)$ and $\nu^{\Br}_{m,\iB}(t)$ }
  &\textb{ mean and variance of $\frac{1}{C}\exp(\varfxn{m}{\iB}{\B}(t,.))$ }\\
  \textb{ $\hat{\Cr}_{m,\jC}(t)$ and $\nu^{\Cr}_{m,\jC}(t)$ }
  &\textb{ mean and variance of $\frac{1}{C}\exp(\varfxn{m}{\jC}{\C}(t,.))$ }\\
  \textb{ $\hat{\Br}_{\iB}(t)$ and $\nu^{\Br}_{\iB}(t)$ }
  &\textb{ mean and variance of $\frac{1}{C}\exp(\Delta_{\iB}^{\B}(t,.))$ }\\
  \textb{ $\hat{\Cr}_{\jC}(t)$ and $\nu^{\Cr}_{\jC}(t)$ }
  &\textb{ mean and variance of $\frac{1}{C}\exp(\Delta_{\jC}^{\C}(t,.))$ }\\[1mm]
\hline
\end{tabular}
}

\subsection{Approximate Message Passing}

When the priors and/or likelihood are generic, as in our case, exact representation of the SPA messages becomes difficult, motivating SPA approximations. 
One such approximation technique, known as \emph{approximate message passing} (AMP) \cite{Montanari:Chap:12}, becomes applicable when the statistical model involves multiplication of the unknown vectors with large random matrices.
In this case, central-limit-theorem (CLT) and Taylor-series arguments can be used to arrive at a tractable SPA approximation that can be rigorously analyzed \cite{Javanmard:II:13}.
In the sequel, we propose an AMP-based approximation of the SPA in \secref{SPA}.

\section{Parametric BiG-AMP} \label{sec:PBigAmp}

We now derive the proposed AMP-based approximation of the SPA algorithm from \secref{SPA}, which we refer to as 
\iftoggle{long}{
\emph{parametric bilinear generalized AMP} (P-BiG-AMP).
}{
P-BiG-AMP.  
Due to space constraints, some details are omitted.
The full derivation can be found in \cite{Parker:PBiGAMP}.
}

\subsection{Randomization and Large-System Limit}

For the derivation of P-BiG-AMP, we treat $z_m\of{i,j}$ as realizations of i.i.d.\ zero-mean unit-variance Gaussian random variables $\Z_m\of{i,j}$, and we treat $\Z_m\of{i,j}, \B_i, \C_j$ as independent for all $m,i,j$.
Furthermore, we consider a large-system limit (LSL) where $M,\Nb,\Nc\to\infty$ such that $\Nb/M$ and $\Nc/M$ converge to fixed positive constants.
Without loss of generality (w.l.o.g.) we will assume that 
$\E\{\B_{\iB}^2\}$ and $\E\{\C_{\jC}^2\}$ scale as $\Ord(1/M)$. 
Given these assumptions, it is straightforward to show from \eqref{ZPB} that $\E\{\Z_m^2\}$ scales as $\Ord(1)$ \iftoggle{long}{\textl{(see \appref{scaling})}}{.}

To derive P-BiG-AMP, we will examine the SPA updates \eqref{zTobPB}-\eqref{DcPB} and drop those terms that vanish in the LSL, i.e., as $M\rightarrow\infty$.
\iftoggle{long}{
\color{\longcolor}
In doing so, we will assume that the previously assumed scalings on $\Z_m,\B_i,\C_j$ hold whether the random variables are distributed according to the priors, the SPA message pdfs \eqref{zTobPB}-\eqref{cTozPB}, or the SPA-approximated posterior pdfs \eqref{DbPB}-\eqref{DcPB}.
These assumptions lead straightforwardly to the scalings of
$\hat{z}_{m}(t)$,
$\nu^z_{m}(t)$,
$\hat{\Br}_{m,\iB}(t)$,
$\nu^{\Br}_{m,\iB}(t)$,
$\hat{\Cr}_{m,\jC}(t)$,
and $\nu^{\Cr}_{m,\jC}(t)$
specified in \tabref{termOrdersPB}.
Furthermore, we will assume that both
$\hat{\Br}_{m,\iB}(t) - \hat{\Br}_{\iB}(t)$
and
$\hat{\Cr}_{m,\jC}(t) - \hat{\Cr}_{\jC}(t)$
are $\Ord(1/M)$, which leads to the assumed scalings on the variance differences in \tabref{termOrdersPB}.
\textb{Notice that, since $\hat{b}_i(t) = O(1/\sqrt{M})$ and $\hat{c}_j(t) = O(1/\sqrt{M})$, the difference quantities $(\hat{b}_{m,i}(t)-\hat{b}_i(t))$ and $(\hat{c}_{m,j}(t)-\hat{c}_j(t))$ scale as $1/\sqrt{M}$ times the reference quantities $\hat{b}_i(t)$ and $\hat{c}_j(t)$, as in previous AMP derivations (e.g., \cite{Montanari:Chap:12,Rangan:ISIT:11,Parker:TSP:14a}).}
Other entries in \tabref{termOrdersPB} will be explained in the sequel.
\color{black}

\putTable{termOrdersPB}{P-BiG-AMP variable scalings in the large-system limit.}
{\renewcommand{\arraystretch}{1.3}
\color{\longcolor}
\begin{tabular}{||@{\:}c@{\:}|@{\:}c@{\:}||@{\:}c@{\:}|@{\:}c@{\:}||@{\:}c@{\:}|@{\:}c@{\:}||} \hline
$\hat{\Br}_{m,\iB}(t)$ & $\Ord(\frac{1}{M^{1/2}})$ 
&
$\nu^{\Br}_{m,\iB}(t)$ & $\Ord(\frac{1}{M})$ 
&
$\hat{\Br}_{m,\iB}(t) - \hat{\Br}_{\iB}(t)$ & $\Ord(\frac{1}{M})$ 
\\[0.5mm]
$\hat{\Cr}_{m,\jC}(t)$ & $\Ord(\frac{1}{M^{1/2}})$
&  
$\nu^{\Cr}_{m,\jC}(t)$ & $\Ord(\frac{1}{M})$
& 
$\hat{\Cr}_{m,\jC}(t) - \hat{\Cr}_{\jC}(t)$ & $\Ord(\frac{1}{M})$
\\
$\hat{p}_{m}(t)$ & $\Ord(1)$
&  
$\nu^p_{m}(t)$ & $\Ord(1)$
&  
$\nu^{\Br}_{m,\iB}(t) - \nu^{\Br}_{\iB}(t)$ & {$\Ord(\frac{1}{M^{3/2}})$}
\\
$\hat{z}_{m}(t)$ & $\Ord(1)$
&  
$\nu^z_{m}(t)$ & $\Ord(1)$
& 
$\nu^{\Cr}_{m,\jC}(t) - \nu^{\Cr}_{\jC}(t)$ & {$\Ord(\frac{1}{M^{3/2}})$}
\\
$\hat{s}_{m}(t)$ & $\Ord(1)$
&  
$\nu^s_{m}(t)$ & $\Ord(1)$
& 
$\nu^q_{m,\iB}(t) - \nu^q_{\iB}(t)$ & {$\Ord(\frac{1}{M^2})$}
\\
  & 
&  
  & 
& 
$\nu^r_{m,\jC}(t) - \nu^r_{\jC}(t)$
& 
{$\Ord(\frac{1}{M^2})$} 
\\
$\zmatml{m}{\iB}{\jC}{t}{}$ & $\Ord(1)$
&
$\zmatml{m}{\nod}{\jC}{t}{}$ & $\Ord(1)$
&
$\zmatml{m}{\nod}{\jC}{t}{} - \zmat{m}{\nod}{\jC}{t}{}$ & $\Ord(\frac{1}{M^{1/2}})$
\\
$\zmat{m}{\nod}{\nod}{t}{}$ & $\Ord(1)$
&
$\zmatml{m}{\iB}{\nod}{t}{}$ & $\Ord(1)$
&
$\zmatml{m}{\iB}{\nod}{t}{} - \zmat{m}{\iB}{\nod}{t}{}$ & $\Ord(\frac{1}{M^{1/2}})$
\\
$\hat{r}_{m,\jC}(t)$ & $\Ord(\frac{1}{M^{1/2}})$
&
$\nu^r_{m,\jC}(t)$   & $\Ord(\frac{1}{M})$
& 
$\hat{r}_{m,\jC}(t) - \hat{r}_{\jC}(t)$ & {$\Ord(\frac{1}{M})$}
\\
$\hat{q}_{m,\iB}(t)$ & $\Ord(\frac{1}{M^{1/2}})$
&
$\nu^q_{m,\iB}(t)$   & $\Ord(\frac{1}{M})$
& 
$\hat{q}_{m,\iB}(t) - \hat{q}_{\iB}(t)$ & {$\Ord(\frac{1}{M})$}
\\[1mm]
\hline
 \end{tabular}
\color{black}
}
}{}

\subsection{SPA message from node \texorpdfstring{$p_{\Y_{m}|\Z_{m}}$}{pYmZm} to node \texorpdfstring{$\B_{\iB}$}{Bi}} \label{sec:BmessagePB}

We begin by approximating the message defined in \eqref{zTobPB}.
First, we invoke the LSL to apply the central limit theorem (CLT) to $\Z_{m}\defn z_{m}(\vB,\vC)$, where $\vB$ and $\vC$ are distributed according to the pdfs in \eqref{zTobPB}.
(Details on the application of the CLT are given in \iftoggle{long}{\textl{\appref{CLT}}}{\cite{Parker:PBiGAMP}}.)
With the CLT, we can treat $\Z_{m}$ conditioned on $\B_{\iB}=\Br_{\iB}$ as Gaussian and thus completely characterize it by a (conditional) mean and variance. 
In particular, the conditional mean is
\begin{align}
\lefteqn{
\E\{\Z_{m} \giv \B_{\iB} = \Br_{\iB}\} 
}\nonumber\\
&= \mytextstyle \E\left\{ 
  \sum_{k,j} \B_k \C_j z_m\of{k,j} 
  + \big( b_i -\B_i \big) \sum_j \C_j z_m\of{i,j} 
\right\} \\
&= \underbrace{ \mytextstyle \sum_{k,j} \hat{b}_{m,k}(t) \hat{c}_{m,j}(t) z_m\of{k,j} }_{\displaystyle \defn \zmatml{m}{\nod}{\nod}{t}{} }
  + \big( b_i - \hat{b}_{m,i}(t) \big) 
  \underbrace{ \mytextstyle \sum_j \hat{c}_{m,j}(t) z_m\of{i,j} }_{\displaystyle \defn \zmatml{m}{\iB}{\nod}{t}{}} 
\label{eq:Zstar}\\
&= \underbrace{ \zmatml{m}{\nod}{\nod}{t}{} - \hat{b}_{m,i}(t) \zmatml{m}{\iB}{\nod}{t}{} }_{\displaystyle \defn \hat{p}_{i,m}(t)}
  + b_i \zmatml{m}{\iB}{\nod}{t}{} ,
\label{eq:ZmeanPB}
\end{align}
and it can be shown \iftoggle{long}{\textl{(see \appref{cond_var})}}{\cite{Parker:PBiGAMP}} that the conditional variance is
\begin{align}
&\var\{\Z_{m} \giv \B_{\iB} = \Br_{\iB}\} 
= \mytextstyle \nu^p_{\iB,m}(t) 
   + \Br_{\iB}^2 \sum_{\jC=1}^{\Nc} \nu^{\Cr}_{m,\jC}(t)
     z_{m}\ofsq{\iB,\jC}      
\label{eq:ZvariancePB} 
   \\&\quad \mytextstyle
   + 2\Br_{\iB} \sum_{\jC=1}^{\Nc} \nu^{\Cr}_{m,\jC}(t)
     \left( \zmatml{m}{\nod}{\jC}{t}{} z_{m}\of{\iB,\jC}
     - \hat{\Br}_{m,\iB}(t) z_{m}\ofsq{\iB,\jC}
     \right) 
\non,
\end{align}
for $\zmatml{m}{\nod}{\jC}{t}{} \defn \sum_k \hat{b}_{m,k}(t) z_m\of{k,j}$ and 
\begin{align}
\nu^p_{\iB,m}(t) 
&\defn \mytextstyle
\sum_{k \ne \iB} \nu^{\Br}_{m,k}(t) \left( \zmatml{m}{k}{\nod}{t}{2} 
+ \sum_{\jC=1}^{\Nc} \nu^{\Cr}_{m,\jC}(t) z_{m}\ofsq{k,\jC} \right) 
\non\\&\quad   \mytextstyle 
+ \sum_{\jC=1}^{\Nc} \nu^{\Cr}_{m,\jC}(t)
\Big( \zmatml{m}{\nod}{\jC}{t}{2} + \hat{\Br}_{m,\iB}(t)^2 z_{m}\ofsq{\iB,\jC}
\non\\&\quad   \mytextstyle 
- 2 \hat{\Br}_{m,\iB}(t) \zmatml{m}{\nod}{\jC}{t}{}  z_{m}\of{\iB,\jC}  \Big)
\label{eq:pvarimlPB} .
\end{align}
\iftoggle{long}{
\color{\longcolor}
We note that $\hat{p}_{i,m}(t)$ and $\nu^p_{\iB,m}(t)$ are analogous to the similarly named terms in G-AMP~\cite{Rangan:ISIT:11} and BiG-AMP~\cite{Parker:TSP:14a}. 
Since they pertain to estimates of $\Z_{m}$, they scale as $\Ord(1)$.
\color{black}
}{}

The Gaussian approximation of $\Z_{m}|_{\B_i=b_i}$ (with mean and variance above) can now be used to simplify the representation of the SPA message \eqref{zTobPB} from an $(\Nb+\Nc-1)$-dimensional integral to a one-dimensional integral:
\begin{align}
\lefteqn{ \mytextstyle
\fxnvar{m}{\iB}{\B}(t,\Br_{\iB}) 
\approx \log \int_{z_{m}} p_{\Y_{m}|\Z_{m}}\big(y_{m}\biggiv z_{m}\big) 
}\nonumber\\&\qquad\times
   \mc{N}\big(z_{m}; 
      \E\{\Z_{m} \giv \B_{\iB} = \Br_{\iB}\},
      \var\{\Z_{m} \giv \B_{\iB} = \Br_{\iB}\}
      \big)\\
&= \mytextstyle
   H_{m}\!\!\left(\hat{p}_{\iB,m}(t) +  \Br_{\iB} \zmatml{m}{\iB}{\nod}{t}{},~
   \nu^p_{\iB,m}(t) + \Br_{\iB}^2 \sum_{\jC} \nu^{\Cr}_{m,\jC}(t)
   z_{m}\ofsq{\iB,\jC} \right.
\non\\&\qquad\mytextstyle \left.
   + 2\Br_{\iB} \sum_{\jC=1}^{\Nc} \nu^{\Cr}_{m,\jC}(t)
   \left[ \zmatml{m}{\nod}{\jC}{t}{} z_{m}\of{\iB,\jC}
   - \hat{\Br}_{m,\iB}(t) z_{m}\ofsq{\iB,\jC}
   \right]
   \right) 
\nonumber\\&\quad
+ \const, \label{eq:HstepPB}
\end{align}
where we have introduced the shorthand notation
\begin{align}
H_{m}\big(\hat{q},\nu^q\big) 
&\defn \mytextstyle \log \int_z p_{\Y_{m}|\Z_{m}}(y_{m} \giv z) \,\Nor(z;\hat{q},\nu^q) 
\label{eq:HPB}.
\end{align}

We now further approximate \eqref{HstepPB}.
For this, we first introduce $\iB$-invariant versions of $\hat{p}_{i,m}(t)$ and $\nu^p_{\iB,m}(t)$:
\begin{align}
\hat{p}_{m}(t) 
&\defn \zmatml{m}{\nod}{\nod}{t}{}
\label{eq:phatPB}\\
\nu^p_{m}(t) 
&\defn \mytextstyle \sum_{\jC=1}^{\Nc} \nu^{\Cr}_{m,\jC}(t) \zmatml{m}{\nod}{\jC}{t}{2}   
       + \sum_{k=1}^{\Nb} \nu^{\Br}_{m,k}(t) \iftoggle{long}{\vast[}{\Big[} \zmatml{m}{k}{\nod}{t}{2} 
\non\\&\quad  \mytextstyle \left. 
       +  \sum_{\jC=1}^{\Nc} \nu^{\Cr}_{m,\jC}(t)
       z_{m}\ofsq{k,\jC} \right]
\label{eq:pvarPB} ,
\end{align}
noting that
\begin{align}
&\hat{p}_{m,i}(t) 
= \hat{p}_{m}(t) - \hat{\Br}_{m,\iB}(t) \zmatml{m}{\iB}{\nod}{t}{}
        \label{eq:phatiml2PB}\\
&\nu^p_{\iB,m}(t) 
= \mytextstyle \nu^p_{m}(t) 
   - \nu^{\Br}_{m,i}(t) \left[ \zmatml{m}{i}{\nod}{t}{2} 
   +  \sum_{\jC=1}^{\Nc} \nu^{\Cr}_{m,\jC}(t) 
   z_{m}\ofsq{i,\jC} \right] 
\non\\&\quad  \mytextstyle  
   + \sum_{\jC=1}^{\Nc} \nu^{\Cr}_{m,\jC}(t) \left[
   \hat{\Br}_{m,\iB}(t)^2 z_{m}\ofsq{\iB,\jC} 
   - 2 \hat{\Br}_{m,\iB}(t) \zmatml{m}{\nod}{\jC}{t}{}  
   z_{m}\of{\iB,\jC}  \right]
\label{eq:pvariml2PB} .
\end{align}
\iftoggle{long}{
\color{\longcolor}
As with $\hat{p}_{\iB,m}(t)$ and $\nu^p_{\iB,m}(t)$, the quantities $\hat{p}_{m}(t)$ and $\nu^p_{m}(t)$ are $\Ord(1)$.
\color{black}
}{}
Next, we define
\begin{align}
\zmat{m}{\iB}{\nod}{t}{} 
&\defn \mytextstyle \sum_j \hat{c}_{j}(t) z_m\of{\iB,j}
        \label{eq:zi0PB} \\
\zmat{m}{\nod}{\jC}{t}{} 
&\defn \mytextstyle \sum_i \hat{b}_{i}(t) z_m\of{i,\jC}
        \label{eq:z0jPB} \\
\zmat{m}{\nod}{\nod}{t}{} 
&\defn \mytextstyle \sum_{i,j} \hat{b}_{i}(t)\hat{c}_{j}(t) z_m\of{i,j}
        \label{eq:z00PB} ,
\end{align}
which are versions of 
$\zmatml{m}{\iB}{\nod}{t}{}, \zmatml{m}{\nod}{\jC}{t}{}, \zmatml{m}{\nod}{\nod}{t}{}$
evaluated at $\vec{\hat{\Br}}(t)$ and $\vec{\hat{\Cr}}(t)$,
the means of the SPA-approximated posteriors, rather than at $\vec{\hat{\Br}}_{m}(t)$ and $\vec{\hat{\Cr}}_{m}(t)$,
the means of the SPA messages.
\iftoggle{long}{
\color{\longcolor}
As such, the quantities in \eqref{zi0PB}-\eqref{z00PB} are also $\Ord(1)$.
\color{black}
}{}
\textb{Note that 
$\zmat{m}{\iB}{\nod}{t}{},\zmat{m}{\nod}{\jC}{t}{},z_{m}\of{\iB,\jC}$ 
can also be interpreted as as partial derivatives:}
\begin{align}
\zmat{m}{\iB}{\nod}{t}{} 
&= \mytextstyle \left.\frac{\partial}{\partial \Br_{\iB}} z_{m}(\vBr,\vCr)  
        \right|_{\displaystyle \vBr = \vec{\hat{\Br}}(t),\vCr = \vec{\hat{\Cr}}(t)}
        \label{eq:zmli0PB}\\
\zmat{m}{\nod}{\jC}{t}{} 
&= \mytextstyle \left.\frac{\partial}{\partial \Cr_{\jC}} z_{m}(\vBr,\vCr)  
        \right|_{\displaystyle \vBr = \vec{\hat{\Br}}(t),\vCr = \vec{\hat{\Cr}}(t)}
        \label{eq:zml0jPB}\\
z_{m}\of{\iB,\jC}
&= \mytextstyle \left.\frac{\partial^2}{\partial \Br_{\iB} \partial \Cr_{\jC}} z_{m}(\vBr,\vCr) 
        \right|_{\displaystyle \vBr = \vec{\hat{\Br}}(t),\vCr = \vec{\hat{\Cr}}(t)}
        \label{eq:zijPB} .
\end{align}
\iftoggle{long}{
\color{\longcolor}
\textb{Comparing \eqref{zi0PB} to \eqref{Zstar} 
and invoking the independence of $\{\C_j\}$,} 
it follows that $\big(\zmatml{m}{\iB}{\nod}{t}{} - \zmat{m}{\iB}{\nod}{t}{}\big)$ is $\Ord(1/M^{1/2})$.
Similarly it can be shown that $\big(\zmatml{m}{\nod}{\jC}{t}{} - \zmat{m}{\nod}{\jC}{t}{}\big)$ is $\Ord(1/M^{1/2})$.
\color{black}
}{}
\textb{With these new quantities, it can be shown \iftoggle{long}{\textl{(see \appref{H})}}{\cite{Parker:PBiGAMP}} that \eqref{HstepPB} can be expressed as}
\begin{align}
\lefteqn{
\fxnvar{m}{\iB}{\B}(t,\Br_{\iB}) 
= \const 
}\non\\& \mytextstyle
+ H_{m} \iftoggle{long}{\Bigg(}{\Big(}
        \hat{p}_{m}(t) + \big(\Br_{\iB} - \hat{\Br}_{\iB}(t)\big) \zmatml{m}{\iB}{\nod}{t}{} 
        + \Ord(1/M),
\label{eq:HtodiffPB} 
\\&\qquad~  \mytextstyle
        \nu^p_{m}(t) +\big(\Br_{\iB} - \hat{\Br}_{\iB}(t) \big)^2  \sum_{\jC=1}^{\Nc} \nu^{\Cr}_{m,\jC}(t)                    z_{m}\ofsq{\iB,\jC}
\non\\&\qquad  \mytextstyle
        + 2 \big( \Br_{\iB} -  \hat{\Br}_{\iB}(t) \big) \sum_{\jC=1}^{\Nc} \nu^{\Cr}_{m,\jC}(t)
        \zmat{m}{\nod}{\jC}{t}{} z_{m}\of{\iB,\jC} 
        + \Ord(1/M) 
        \iftoggle{long}{\Bigg)}{\Big)}
\non .
\end{align}

The next step is to perform a Taylor series expansion of \eqref{HtodiffPB} in $\Br_{\iB}$ about $\hat{\Br}_{\iB}(t)$.
By carefully analyzing the scaling of all terms in the expansion, and neglecting those that vanish as $M\rightarrow\infty$, it can be shown \iftoggle{long}{\textl{(see \appref{Taylor})}}{\cite{Parker:PBiGAMP}} that
\begin{align}
\lefteqn{
\fxnvar{m}{\iB}{\B}(t,\Br_{\iB})
}\label{eq:bmlToiPB}\\
&\approx \const 
  + \iftoggle{long}{\Bigg[}{\!\Big[}
    \hat{s}_{m}(t) \zmatml{m}{\iB}{\nod}{t}{} 
    + \nu^s_{m}(t)  \hat{\Br}_{\iB}(t) \zmat{m}{\iB}{\nod}{t}{2}   
\non\\& \mytextstyle \left.
    + \big(\hat{s}_{m}^2(t) - \nu^s_{m}(t) \big) 
      \sum_{\jC} \nu^{\Cr}_{\jC}(t) z_{m}\of{\iB,\jC} 
      \big(\zmat{m}{\nod}{\jC}{t}{} - \hat{\Br}_{\iB}(t) z_{m}\of{\iB,\jC} \big)
  \right] \Br_{\iB} 
\non\\& \mytextstyle
  - \frac{1}{2} \left[  \nu^s_{m}(t) \zmat{m}{\iB}{\nod}{t}{2}   
  - \big(\hat{s}_{m}^2(t) - \nu^s_{m}(t) \big) 
  \sum_{\jC} \nu^{\Cr}_{\jC}(t) z_{m}\ofsq{\iB,\jC}\right]  
  \Br_{\iB}^2 
\non ,
\end{align}
using the definitions
\begin{align}
\hat{s}_{m}(t) 
&\defn H'_{m}\big( \hat{p}_{m}(t),  \nu^p_{m}(t) \big)\label{eq:shatPB}\\
\nu^s_{m}(t) 
&\defn -H''_{m}\big( \hat{p}_{m}(t),  \nu^p_{m}(t) \big)\label{eq:nushatPB} ,
\end{align}
where $H_{m}'(\cdot,\cdot)$ and $H_{m}''(\cdot,\cdot)$ respectively denote the first and second derivative w.r.t.\ the first argument of $H_{m}(\cdot,\cdot)$.
Note that, since \eqref{bmlToiPB} is quadratic, the (exponentiated) message from $p_{\Y_{m}|\Z_{m}}$ to $\B_{\iB}$ is Gaussian in the LSL.
\iftoggle{long}{
\color{\longcolor}
Finally, since the function $H_{m}(\cdot,\cdot)$ and its partials are $\Ord(1)$, 
we conclude that $\hat{s}_{m}(t)$ and $\nu^s_{m}(t)$ are $\Ord(1)$ as well.
\color{black}
}{}

Furthermore, the derivation in \cite[App.~A]{Parker:TSP:14a} shows that \eqref{shatPB}-\eqref{nushatPB} can be rewritten as
\begin{align}
\hat{s}_{m}(t) 
&= \big( \hat{z}_{m}(t) - \hat{p}_{m}(t)  \big) / \nu^p_{m}(t) 
\label{eq:sPB} \\
\nu^s_{m}(t) 
&= \left( 1 - \nu^z_{m}(t)/\nu^p_{m}(t)\right) / \nu^p_{m}(t) 
\label{eq:nusPB} ,
\end{align}
using the conditional mean and variance
\begin{align}
\hat{z}_{m}(t) &\defn \E\{\Z_{m}\giv \p_{m}\!=\!\hat{p}_{m}(t);\nu^p_{m}(t)\} \label{eq:zhatPB}\\
\nu^z_{m}(t) &\defn \var\{\Z_{m}\giv \p_{m}\!=\!\hat{p}_{m}(t);\nu^p_{m}(t)\}, \label{eq:zvarPB} .
\end{align}
Note \eqref{zhatPB}-\eqref{zvarPB} are computed according to the pdf
\iftoggle{long}{
\color{\longcolor}
\begin{align}
\lefteqn{ p_{\Z_{m}|\p_{m}\!}\big(z_{m}\giv\hat{p}_{m}(t);\nu^p_{m}(t)\big) }\non\\
&\defn 
\mytextstyle \frac{1}{C} \, p_{\Y_{m}|\Z_{m}\!}(y_{m} \giv z_{m}) \, \Nor\big(z_{m};\hat{p}_{m}(t),\nu^p_{m}(t)\big) ,
\label{eq:pZgivYPPB}
\end{align}
with $C=\int_{z} p_{\Y_{m}|\Z_{m}\!}(y_{m} \giv z) \Nor\big(z;\hat{p}_{m}(t),\nu^p_{m}(t)\big)$,
\color{black}
}{given in (D1) of \tabref{PbigAmpScalar},}
which is P-BiG-AMP's iteration-$t$ approximation to the true marginal posterior $p_{\Z_{m}|\vY}(z_{m}|\vec{y})$.
\iftoggle{long}{
\color{\longcolor}
We note that \eqref{pZgivYPPB} can also be interpreted as the (exact) posterior pdf for $\Z_{m}$ given the likelihood $p_{\Y_{m}|\Z_{m}\!}(y_{m}|\cdot)$ from \eqref{pYgivZPB} and the prior $\Z_{m}\sim \Nor\big(\hat{p}_{m}(t),\nu^p_{m}(t)\big)$ that is implicitly adopted by iteration-$t$ P-BiG-AMP.
\color{black}
}{}

\subsection{SPA message from node \texorpdfstring{$p_{\Y_{m}|\Z_{m}}$}{pYmZm} to node \texorpdfstring{$\C_{\jC}$}{Cj}} \label{sec:CmessagePB}

Since $\Z_{m}=\sum_{\iB=0}^{\Nb} \sum_{\jC=0}^{\Nc}
\B_{\iB} z_m\of{i,j} \C_{\jC}$ implies a symmetry between $\B_\iB$ and $\C_\jC$, the procedure to approximate $\fxnvar{m}{\jC}{\C}(t,\cdot)$ is essentially the same as that to approximate $\fxnvar{m}{\iB}{\B}(t,\cdot)$ from \secref{BmessagePB}.
The end result is
\begin{align}
\lefteqn{
\fxnvar{m}{\jC}{\C}(t,\Cr_{\jC}) 
}\label{eq:cmlTojPB}\\
&\approx \const
  + \iftoggle{long}{\Bigg[}{\!\Big[}
    \hat{s}_{m}(t) \zmatml{m}{\nod}{\jC}{t}{} 
    + \nu^s_{m}(t)  \hat{\Cr}_{\jC}(t) \zmat{m}{\nod}{\jC}{t}{2}   
\non\\& \mytextstyle \left.
    + \big(\hat{s}_{m}^2(t) - \nu^s_{m}(t) \big) 
      \sum_{\iB} \nu^{\Br}_{\iB}(t) z_{m}\of{\iB,\jC}
      \Big(\zmat{m}{\iB}{\nod}{t}{} - \hat{\Cr}_{\jC}(t) z_{m}\of{\iB,\jC} \Big)
  \right] \Cr_{\jC} 
\non\\& \mytextstyle
  - \frac{1}{2} \left[  \nu^s_{m}(t) \zmat{m}{\nod}{\jC}{t}{2}   
  - \big(\hat{s}_{m}^2(t) - \nu^s_{m}(t) \big) 
  \sum_{\iB} \nu^{\Br}_{\iB}(t) z_{m}\ofsq{\iB,\jC}\right]  
  \Cr_{\jC}^2 
\non .
\end{align}

\subsection{SPA message from node \texorpdfstring{$\C_{\jC}$}{Cj} to \texorpdfstring{$p_{\Y_{m}|\Z_{m}}$}{pYmZm}}

We now turn our attention to approximating the messages flowing out of the variable nodes.
To start, we plug the approximation of $\fxnvar{m}{\jC}{\C}(t,\Cr_{\jC})$ from \eqref{cmlTojPB} into \eqref{cTozPB} and find 
\begin{align}
\lefteqn{
\varfxn{m}{\jC}{\C}(t\!+\!1,\Cr_{\jC})
} \non\\&\quad
&\approx \const + \log\big( p_{\C_{\jC}\!}(\Cr_{\jC})\Nor(\Cr_{\jC};\hat{r}_{m,\jC}(t),\nu^r_{m,\jC}(t)) \big) \label{eq:cTozFinalPB}
\end{align}
where
\begin{align}
\nu^r_{m,\jC}(t) 
&\defn \mytextstyle 
     \left[ \sum_{r \ne m} \left( \nu^s_{r}(t) \zmat{r}{\nod}{\jC}{t}{2} 
     \right.\right.
\label{eq:rvarmlPB}\\&\quad \left.\left. \mytextstyle
     - \big(\hat{s}_{r}^2(t) - \nu^s_{r}(t) \big) 
     \sum_{\iB=1}^{\Nb} \nu^{\Br}_{\iB}(t) z_{r}\ofsq{\iB,\jC}\right)
     \right]^{-1}
\non\\
\hat{r}_{m,\jC}(t) 
&\defn \mytextstyle
        \hat{\Cr}_{\jC}(t) + \nu^r_{m,\jC}(t) \sum_{r \ne m} 
                \iftoggle{long}{\Bigg(}{\Big(}
                \big(\hat{s}_{r}^2(t) - \nu^s_{r}(t) \big) 
\label{eq:rhatmlPB}\\&\quad\times \left. \mytextstyle
        \sum_{\iB=1}^{\Nb} \nu^{\Br}_{\iB}(t) 
                z_{r}\of{\iB,\jC} \zmat{r}{\iB}{\nod}{t}{}
        + \hat{s}_{r}(t) \zmatml{r}{\nod}{\jC}{t}{} \right)
\non .
\end{align}
\iftoggle{long}{
\color{\longcolor}
Since $\nu^r_{m,\jC}(t)$ is the reciprocal of a sum of $M$ terms of $\Ord(1)$, we conclude that it is $\Ord(1/M)$.
Given this and the scalings from \tabref{termOrdersPB}, we see that $\hat{r}_{m,\jC}(t)$ is $\Ord(1/M^{1/2})$.
Since $\hat{r}_{m,\jC}(t)$ can be interpreted as an estimate of $\C_{\jC}$, this scaling is anticipated.
\color{black}
}{}

The mean and variance of the pdf associated with the $\varfxn{m}{\jC}{\C}(t\!+\!1,\Cr_{\jC})$ message approximation from \eqref{cTozFinalPB} are
\begin{align}
\lefteqn{
\hat{\Cr}_{m,\jC}(t\!+\!1)
\defn \underbrace{ \mytextstyle  \frac{1}{K} \int_{\Cr} \Cr\, p_{\C_{\jC}\!}(\Cr) \Nor\big(\Cr;\hat{r}_{m,\jC}(t), \nu^r_{m,\jC}(t)\big) }_{ \displaystyle \defn g_{\C_{\jC}\!}(\hat{r}_{m,\jC}(t),\nu^r_{m,\jC}(t)) } 
\label{eq:cmljPB}}\\[-3mm]
\lefteqn{ 
\nu^{\Cr}_{m,\jC}(t\!+\!1) 
}\nonumber\\
&\defn \underbrace{ \mytextstyle \frac{1}{K} \int_{\Cr} \big|\Cr - \hat{\Cr}_{m,\jC}(t\!+\!1)\big|^2
p_{\C_{\jC}\!}(\Cr) \Nor\big(\Cr;\hat{r}_{m,\jC}(t), \nu^r_{m,\jC}(t)\big) }_{ \displaystyle \nu^r_{m,\jC}(t) \, g'_{\C_{\jC}\!}(\hat{r}_{m,\jC}(t),\nu^r_{m,\jC}(t))}
\nonumber\\[-6mm]&\label{eq:nucmljPB}
\end{align}
with $K = \int_{\Cr} p_{\C_{\jC}\!}(\Cr) \Nor\big(\Cr;\hat{r}_{m,\jC}(t), \nu^r_{m,\jC}(t)\big)$ and where $g'_{\C_{\jC}}$ denotes the derivative of $g_{\C_{\jC}}$ with respect to its first argument.
The fact that \eqref{cmljPB} and \eqref{nucmljPB} are related through a derivative was shown in \cite{Rangan:ISIT:11}.

Next we develop mean and variance approximations that do not depend on the destination node $m$.
For this, we introduce $m$-invariant versions of $\hat{r}_{m,\jC}(t)$ and $\nu^r_{m,\jC}(t)$:
\begin{align}
\nu^r_{\jC}(t) 
&\defn \mytextstyle \left[ \sum_{m} \left( \nu^s_{m}(t) \zmat{m}{\nod}{\jC}{t}{2} 
\right.\right.\label{eq:rvarPB}\\&\quad\left.\left.  \mytextstyle
                - \big(\hat{s}_{m}^2(t) - \nu^s_{m}(t) \big) \sum_{\iB=1}^{\Nb} \nu^{\Br}_{\iB}(t) z_{m}\ofsq{\iB,\jC}\right)\right]^{-1}
\non\\
\hat{r}_{\jC}(t) 
&\defn \mytextstyle  \hat{\Cr}_{\jC}(t) + 
        \nu^r_{\jC}(t) \sum_{m}  \iftoggle{long}{\Bigg(}{\Big(} \big(\hat{s}_{m}^2(t) - \nu^s_{m}(t) \big) 
\label{eq:rhatPB}\\&\quad\times\left.  \mytextstyle
\sum_{\iB=1}^{\Nb} \nu^{\Br}_{\iB}(t) 
                z_{m}\of{\iB,\jC} \zmat{m}{\iB}{\nod}{t}{}
        + \hat{s}_{m}(t) \zmatml{m}{\nod}{\jC}{t}{} \right) 
\non.
\end{align}
Comparing \eqref{rvarmlPB}-\eqref{rhatmlPB} to \eqref{rvarPB}-\eqref{rhatPB} reveals that $\big(\nu^r_{m,\jC}(t) - \nu^r_{\jC}(t)\big)$ scales as $\Ord(1/M^2)$ and that $\hat{r}_{m,\jC}(t) = \hat{r}_{\jC}(t) -\nu^r_{\jC}(t) \hat{s}_{m}(t) \zmat{m}{\nod}{\jC}{t}{} + \Ord(1/M^{3/2})$, and thus \eqref{cmljPB} implies
\begin{align}
\lefteqn{ 
\hat{\Cr}_{m,\jC}(t\!+\!1)
}\nonumber\\ 
&= g_{\C_{\jC}\!} \big( \hat{r}_{\jC}(t) -\nu^r_{\jC}(t) \hat{s}_{m}(t) \zmat{m}{\nod}{\jC}{t}{} + \Ord(1/M^{3/2}),
\non\\&\qquad
\nu^r_{\jC}(t) + \Ord(1/M^2) \big) \label{eq:chatStep1PB}\\
&= g_{\C_{\jC}\!} \big(\hat{r}_{\jC}(t) -\nu^r_{\jC}(t) \hat{s}_{m}(t) \zmat{m}{\nod}{\jC}{t}{},~\nu^r_{\jC}(t) \big) + \Ord(1/M^{3/2})\label{eq:chatStep2PB}\\
&= g_{\C_{\jC}\!}\big (\hat{r}_{\jC}(t),\nu^r_{\jC}(t) \big) 
\label{eq:chatStep3PB}\\&\quad
        - \nu^r_{\jC}(t) g'_{\C_{\jC}}\big (\hat{r}_{\jC}(t),\nu^r_{\jC}(t) \big) \hat{s}_{m}(t) \zmat{m}{\nod}{\jC}{t}{} + \Ord(1/M^{3/2}) \non\\
&= \hat{\Cr}_{\jC}(t\!+\!1) - \hat{s}_{m}(t) \zmat{m}{\nod}{\jC}{t}{} \nu^{\Cr}_{\jC}(t\!+\!1)  + \Ord(1/M^{3/2}), \label{eq:chatStep4PB}
\end{align}
where \eqref{chatStep2PB} follows by taking Taylor series expansions of \eqref{chatStep1PB} about the perturbations to the arguments;
\eqref{chatStep3PB} follows by taking a Taylor series expansion of \eqref{chatStep2PB} in the first argument about the point $\hat{r}_{\jC}(t)$; and \eqref{chatStep4PB} follows
from the definitions
\begin{align}
\hat{\Cr}_{\jC}(t\!+\!1) 
&\defn g_{\C_{\jC}\!}\big (\hat{r}_{\jC}(t),\nu^r_{\jC}(t) \big)\\
\nu^{\Cr}_{\jC}(t\!+\!1) 
&\defn \nu^r_{\jC}(t) g'_{\C_{\jC}\!}\big (\hat{r}_{\jC}(t),\nu^r_{\jC}(t) \big).
\end{align}

\subsection{SPA message from node \texorpdfstring{$\B_{\iB}$}{Bi} to \texorpdfstring{$p_{\Y_{m}|\Z_{m}}$}{pYmZm}}

Once again, due to symmetry, the derivation for $\varfxn{m}{\iB}{\B}(t\!+\!1,\Br_{\iB})$ closely parallels that for $\varfxn{m}{\jC}{\C}(t\!+\!1,\Cr_{\jC})$.
Plugging approximation \eqref{bmlToiPB} into \eqref{bTozPB}, we obtain
\begin{align}
&\varfxn{m}{\iB}{\B}(t\!+\!1,\Br_{\iB})
\approx 
\log\big( p_{\C_{\iB}}(\Br_{\iB})\Nor(\Br_{\iB};\hat{q}_{m,\iB}(t),\nu^q_{m,\iB}(t)) \big) 
\non\\&\qquad\qquad\qquad\quad 
+ \const 
\label{eq:bTozFinalPB} \\
&\nu^q_{m,\iB}(t) 
\defn  \mytextstyle \left[ \sum_{r \ne m} \left( \nu^s_{r}(t) \zmat{r}{\iB}{\nod}{t}{2}   
\right.\right.\label{eq:qvarmlPB}\\&\qquad\qquad \left.\left.  \mytextstyle
                - \big(\hat{s}_{r}^2(t) - \nu^s_{r}(t) \big) \sum_{\jC=1}^{\Nc} \nu^{\Cr}_{\jC}(t) z_{r}\ofsq{\iB,\jC}\right)\right]^{-1}
\non\\
&\hat{q}_{m,\iB}(t) 
\defn \mytextstyle \hat{\Br}_{\iB}(t)  + \nu^q_{m,\iB}(t) \sum_{r \ne m} \iftoggle{long}{\Bigg(}{\Big(} \big(\hat{s}_{r}^2(t) - \nu^s_{r}(t) \big) 
\label{eq:qhatmlPB}\\&\qquad\qquad\times \left.  \mytextstyle
\sum_{\jC=1}^{\Nc} 
\nu^{\Cr}_{\jC}(t) 
                z_{r}\of{\iB,\jC} \zmat{r}{\nod}{\jC}{t}{} 
                + \hat{s}_{r}(t) \zmatml{r}{\iB}{\nod}{t}{} \right) 
\non.
\end{align}

The mean and variance of the pdf associated with the $\varfxn{m}{\iB}{\B}(t\!+\!1,\Br_{\iB})$ approximation from \eqref{bTozFinalPB} are then
\begin{align}
\lefteqn{ 
\hat{\Br}_{m,\iB}(t\!+\!1)
\defn \underbrace{ \mytextstyle \frac{1}{K} \int_{\Br} \Br\, p_{\B_{\iB}\!}(\Br) \Nor\big(\Br;\hat{q}_{m,\iB}(t), \nu^q_{m,\iB}(t)\big)}_{\displaystyle \defn g_{\B_{\iB}\!}(\hat{q}_{m,\iB}(t),\nu^q_{m,\iB}(t)) } 
}\label{eq:bmliPB}\\[-3mm]
\lefteqn{ 
\nu^{\Br}_{m,\iB}(t\!+\!1) 
}\nonumber\\
&\defn \underbrace{ \mytextstyle \frac{1}{K} \int_{\Br} \big|\Br - \hat{\Br}_{m,\iB}(t\!+\!1)\big|^2
p_{\B_{\iB}\!}(\Br) \Nor\big(\Br;\hat{q}_{m,\iB}(t), \nu^q_{m,\iB}(t)\big) }_{ \displaystyle \nu^q_{m,\iB}(t) \, g'_{\B_{\iB}\!}(\hat{q}_{m,\iB}(t),\nu^q_{m,\iB}(t))}
\nonumber\\[-6mm]&\label{eq:nubmliPB}
\end{align}
where $K = \int_{\Br} p_{\B_{\iB}}(\Br) \Nor\big(\Br;\hat{q}_{m,\iB}(t), \nu^q_{m,\iB}(t)\big)$ and where $g'_{\B_{\iB}}$ denotes the derivative of $g_{\B_{\iB}}$ with respect to the first argument.
As before, we define the $m$-invariant quantities
\begin{align}
\nu^q_{\iB}(t) 
&\defn \mytextstyle \left[ \sum_{m} \left( \nu^s_{m}(t) \zmat{m}{\iB}{\nod}{t}{2} 
\right.\right.\label{eq:qvarPB}\\&\quad  \mytextstyle \left.\left.
                - \big(\hat{s}_{m}^2(t) - \nu^s_{m}(t) \big) \sum_{\jC=1}^{\Nc} \nu^{\Cr}_{\jC}(t) z_{m}\ofsq{\iB,\jC}\right)\right]^{-1}
\non\\
\hat{q}_{\iB}(t) 
&\defn \mytextstyle \hat{\Br}_{\iB}(t) + 
        \nu^q_{\iB}(t) \sum_{m} \left(
                \big(\hat{s}_{m}^2(t) - \nu^s_{m}(t) \big) 
\right.\label{eq:qhatPB}\\&\quad\times \mytextstyle \left.
\sum_{\jC=1}^{\Nc} \nu^{\Cr}_{\jC}(t) 
                z_{m}\of{\iB,\jC} \zmat{m}{\nod}{\jC}{t}{} 
                + \hat{s}_{m}(t) \zmatml{m}{\iB}{\nod}{t}{} \right) 
\non
\end{align}
and perform several Taylor series expansions, finally dropping terms that vanish in the LSL, to obtain
\begin{align}
\hat{\Br}_{m,\iB}(t\!+\!1)
&=\hat{\Br}_{\iB}(t\!+\!1) - \hat{s}_{m}(t) \zmat{m}{\iB}{\nod}{t}{} \nu^{\Br}_{\iB}(t\!+\!1) 
\non\\&\quad 
+ \Ord(1/M^{3/2}),\label{eq:bhatStep4PB} \\
\hat{\Br}_{\iB}(t\!+\!1) 
&\defn g_{\B_{\iB}\!}\big(\hat{q}_{\iB}(t),\nu^q_{\iB}(t) \big)\\
\nu^{\Br}_{\iB}(t\!+\!1) 
&\defn \nu^q_{\iB}(t) g'_{\B_{\iB}\!}\big(\hat{q}_{\iB}(t),\nu^q_{\iB}(t) \big).
\end{align}

\subsection{Closing the loop}

To complete the derivation of P-BiG-AMP, we use \eqref{chatStep4PB} and \eqref{bhatStep4PB} to eliminate the dependence on $m$ in the $\B_{\iB}$ and $\C_{\jC}$ estimates and on $i$ and $j$ in the $\Z_{m}$ estimates.
By plugging \eqref{chatStep4PB} and \eqref{bhatStep4PB} into the expression \eqref{phatPB} for $\hat{p}_{m}(t)$ and dropping terms that vanish in the LSL, it can be shown \iftoggle{long}{\textl{(see \appref{phat})}}{\cite{Parker:PBiGAMP}} that 
\begin{align}
\hat{p}_{m}(t) 
&\approx \mytextstyle \zmat{m}{\nod}{\nod}{t}{} - \hat{s}_{m}(t\!-\!1) \left( 
        \sum_{\iB=1}^{\Nb} \zmat{m}{\iB}{\nod}{t\!-\!1}{}\zmat{m}{\iB}{\nod}{t}{} \nu^{\Br}_{\iB}(t)  
\right. \non\\&\quad \mytextstyle  \left.
        + \sum_{\jC=1}^{\Nc} \zmat{m}{\nod}{\jC}{t\!-\!1}{}\zmat{m}{\nod}{\jC}{t}{} \nu^{\Cr}_{\jC}(t)  \right) 
\label{eq:phat2PB} .
\end{align}
Although not justified by the LSL, we also approximate
\begin{align}
\mytextstyle \sum_{\iB=1}^{\Nb} \zmat{m}{\iB}{\nod}{t\!-\!1}{} \zmat{m}{\iB}{\nod}{t}{} \nu^{\Br}_{\iB}(t)
&\approx \mytextstyle \sum_{\iB=1}^{\Nb} \zmat{m}{\iB}{\nod}{t}{2} \nu^{\Br}_{\iB}(t)  \label{eq:phat_time_approx1PB}\\
\mytextstyle \mytextstyle \sum_{\jC=1}^{\Nc} \zmat{m}{\nod}{\jC}{t\!-\!1}{} \zmat{m}{\nod}{\jC}{t}{} \nu^{\Cr}_{\jC}(t)
&\approx \mytextstyle \sum_{\jC=1}^{\Nc} \zmat{m}{\nod}{\jC}{t}{2} \nu^{\Cr}_{\jC}(t)  \label{eq:phat_time_approx2PB}
\end{align}
for the sake of algorithmic simplicity, yielding
\begin{align}
\hat{p}_{m}(t) 
&\approx \zmat{m}{\nod}{\nod}{t}{} - \hat{s}_{m}(t\!-\!1) 
        \label{eq:phat3PB}
\\&\quad \times
\underbrace{ \mytextstyle \left( 
        \sum_{\iB=1}^{\Nb} \zmat{m}{\iB}{\nod}{t}{2}\nu^{\Br}_{\iB}(t)  
        + \sum_{\jC=1}^{\Nc}
        \zmat{m}{\nod}{\jC}{t}{2}\nu^{\Cr}_{\jC}(t) \right)}_{\displaystyle \defn \bar{\nu}^p_{m}(t)},
\non
\end{align}
noting that similar approximations were made for BiG-AMP \cite{Parker:TSP:14a}, where empirical tests showed little effect.
Of course, a more complicated variant of P-BiG-AMP could be stated using \eqref{phat2PB} instead of \eqref{phat3PB}.

Equations \eqref{chatStep4PB} and \eqref{bhatStep4PB} can also be used to simplify $\nu^p_{m}(t)$.
For this, we first use the facts $\nu^{\Cr}_{m,\jC}(t) = \nu^{\Cr}_{\jC}(t) + \Ord(1/M^{3/2})$ and $\nu^{\Br}_{m,\iB}(t) = \nu^{\Br}_{\iB}(t) + \Ord(1/M^{3/2})$ to write \eqref{pvarPB} as
\begin{align}
\nu^p_{m}(t) 
&= \mytextstyle \sum_{\jC=1}^{\Nc} \nu^{\Cr}_{\jC}(t) \zmatml{m}{\nod}{\jC}{t}{2}   
   + \sum_{\iB=1}^{\Nb} \nu^{\Br}_{\iB}(t) \zmatml{m}{\iB}{\nod}{t}{2}
\label{eq:pvar2PB} \\&\quad \mytextstyle
   +  \sum_{\iB=1}^{\Nb}\sum_{\jC=1}^{\Nc} \nu^{\Br}_{\iB}(t) \nu^{\Cr}_{\jC}(t)
   z_{m}\ofsq{\iB,\jC} + \Ord(1/M^{1/2}) 
\non.
\end{align}
Then we use \eqref{chatStep4PB} with \eqref{Zstar} and \eqref{zi0PB} to write 
\begin{align}
\zmatml{m}{\iB}{\nod}{t}{} 
&= \mytextstyle \zmat{m}{\iB}{\nod}{t}{} - \hat{s}_{m}(t\!-\!1)
        \sum_{\jC=1}^{\Nc}  \zmat{m}{\nod}{\jC}{t\!-\!1}{} z_{m}\of{\iB,\jC} \nu^{\Cr}_{\jC}(t)
\non\\&\quad
        + \Ord(1/M) 
        \label{eq:zi0approxPB} ,
\end{align}
and similarly we use \eqref{bhatStep4PB} to write 
\begin{align}
\zmatml{m}{\nod}{\jC}{t}{} 
&= \mytextstyle \zmat{m}{\nod}{\jC}{t}{} - \hat{s}_{m}(t\!-\!1)
        \sum_{\iB=1}^{\Nb}  \zmat{m}{\iB}{\nod}{t\!-\!1}{} z_{m}\of{\iB,\jC} \nu^{\Br}_{\iB}(t)
\non\\&\quad
        + \Ord(1/M) \label{eq:z0japproxPB} .
\end{align}
Plugging \eqref{zi0approxPB}-\eqref{z0japproxPB} into \eqref{pvar2PB} and dropping the terms that vanish in the LSL yields \iftoggle{long}{\textl{(see \appref{pvar})}}{\cite{Parker:PBiGAMP}}
\begin{align}
\nu^p_{m}(t) 
&\approx \mytextstyle \bar{\nu}^p_{m}(t) 
        + \sum_{\iB=1}^{\Nb}\sum_{\jC=1}^{\Nc} \nu^{\Br}_{\iB}(t) \nu^{\Cr}_{\jC}(t)z_{m}\ofsq{\iB,\jC} .
        \label{eq:pvar3PB}
\end{align}

Next, we eliminate the dependence on $\zmatml{m}{\nod}{\jC}{t}{}$ from $\hat{r}_{\jC}(t)$.
Plugging \eqref{z0japproxPB} into \eqref{rhatPB} and dropping the terms that vanish in the LSL yields 
\begin{align}
\hat{r}_{\jC}(t) 
&\approx \mytextstyle \hat{\Cr}_{\jC}(t) + 
        \nu^r_{\jC}(t) \sum_{m}  \big(\hat{s}_{m}^2(t) - \nu^s_{m}(t) \big) 
\label{eq:rhat3PB}\\&\quad\times  \mytextstyle 
                \sum_{\iB=1}^{\Nb} \nu^{\Br}_{\iB}(t) 
                \zmatC{m}{\iB}{\jC}{t}{} \zmat{m}{\iB}{\nod}{t}{}
                + \nu^r_{\jC}(t) \sum_{m} \hat{s}_{m}(t) 
\non\\&\quad \times \mytextstyle
                \left(  \zmat{m}{\nod}{\jC}{t}{} - \hat{s}_{m}(t\!-\!1)
                \sum_{\iB=1}^{\Nb}  \zmat{m}{\iB}{\nod}{t\!-\!1}{} \zmatC{m}{\iB}{\jC}{t}{} \nu^{\Br}_{\iB}(t) \right)
\non , 
\end{align}
Although not justified by the LSL, we also approximate
\begin{align}
\lefteqn{ \mytextstyle
\sum_{m} \hat{s}_{m}(t)\hat{s}_{m}(t\!-\!1) \sum_{\iB=1}^{\Nb} \nu^{\Br}_{\iB}(t) \zmatC{m}{\iB}{\jC}{t}{} \zmat{m}{\iB}{\nod}{t\!-\!1}{}
}\non\\
&\approx \mytextstyle \sum_{m} \hat{s}_{m}^2(t) \sum_{\iB=1}^{\Nb} \nu^{\Br}_{\iB}(t) \zmatC{m}{\iB}{\jC}{t}{} \zmat{m}{\iB}{\nod}{t}{}
\end{align}
for the sake of algorithmic simplicity, yielding
\begin{align}
\hat{r}_{\jC}(t) 
&\approx \mytextstyle \hat{\Cr}_{\jC}(t) + 
        \nu^r_{\jC}(t) \sum_{m} \left( \hat{s}_{m}(t) \zmat{m}{\nod}{\jC}{t}{} 
\right. \non\\&\quad  \mytextstyle \left.
                - \nu^s_{m}(t)  \sum_{\iB=1}^{\Nb} \nu^{\Br}_{\iB}(t) 
                \zmatC{m}{\iB}{\jC}{t}{} \zmat{m}{\iB}{\nod}{t}{} \right)
                \label{eq:rhat4PB} ,
\end{align}
noting that a similar approximation was made for BiG-AMP~\cite{Parker:TSP:14a}.
The expression \eqref{rhat4PB} then simplifies.
Using \eqref{zi0PB} to expand $\zmat{m}{\iB}{\nod}{t}{}$, the last term in \eqref{rhat4PB} can be written as
\begin{align}
\lefteqn{ \mytextstyle
\nu^r_j(t) 
\sum_{m} \nu^s_{m}(t)  \sum_{\iB=1}^{\Nb} \nu^{\Br}_{\iB}(t) 
                \zmatC{m}{\iB}{\jC}{t}{} \zmat{m}{\iB}{\nod}{t}{}
}\non\\
&= \mytextstyle
\nu^r_j(t) 
\hat{c}_j(t) 
\sum_{\iB=1}^{\Nb} \nu^{\Br}_{\iB}(t) 
\sum_{m} \nu^s_{m}(t) \zmatC{m}{\iB}{\jC}{t}{2}  
\label{eq:rhat_complicated}\\&\quad \mytextstyle
+ \nu^r_j(t) 
\sum_{\iB=1}^{\Nb} \nu^{\Br}_{\iB}(t) 
\sum_{k\neq j} \hat{c}_k(t) 
\sum_{m} \nu^s_{m}(t) \zmatC{m}{\iB}{\jC}{t}{} \zmatC{m}{\iB}{k}{t}{} 
\non\\
&\approx \mytextstyle
\nu^r_j(t) 
\hat{c}_j(t) 
\sum_{\iB=1}^{\Nb} \nu^{\Br}_{\iB}(t) 
\sum_{m} \nu^s_{m}(t) \zmatC{m}{\iB}{\jC}{t}{2}  
\label{eq:rhat_simp} ,
\end{align}
where \eqref{rhat_simp} holds in the LSL \iftoggle{long}{\textl{(see \appref{rhat})}}{\cite{Parker:PBiGAMP}}.
Thus, \eqref{rhat4PB} reduces to
\begin{align}
\hat{r}_{\jC}(t) 
&\approx \mytextstyle \hat{\Cr}_{\jC}(t) + 
        \nu^r_{\jC}(t) \sum_{m} \hat{s}_{m}(t) \zmat{m}{\nod}{\jC}{t}{} 
\non\\&\quad \mytextstyle
        -\nu^r_{\jC}(t) \hat{c}_{\jC}(t) \sum_{m} 
                \nu^s_{m}(t)  \sum_{\iB=1}^{\Nb} \nu^{\Br}_{\iB}(t) 
                \zmatC{m}{\iB}{\jC}{t}{2} 
                \label{eq:rhatFinalPB} .
\end{align}
Similarly, we substitute \eqref{zi0approxPB} into \eqref{qhatPB} and make analogous approximations to obtain
\begin{align}
\hat{q}_{\iB}(t) 
&\approx \mytextstyle \hat{\Br}_{\iB}(t) + 
        \nu^q_{\iB}(t) \sum_{m} \hat{s}_{m}(t) \zmat{m}{\iB}{\nod}{t}{} 
\non\\&\quad \mytextstyle
                - \nu^q_{\iB}(t) \hat{\Br}_{\iB}(t) \sum_{m} \nu^s_{m}(t)  \sum_{\jC=1}^{\Nc} \nu^{\Cr}_{\jC}(t) 
                 \zmatC{m}{\iB}{\jC}{t}{2} 
                \label{eq:qhatFinalPB} .
\end{align}

Next, we simplify expressions for the variances $\nu^r_{\jC}(t)$ and $\nu^q_{\iB}(t)$.
First, it can be shown \iftoggle{long}{\textl{(see \appref{rvar})}}{\cite{Parker:PBiGAMP}} that \eqref{sPB} and \eqref{nusPB} can be used to rewrite the second half of $\nu^r_{\jC}(t)$ from \eqref{rvarPB} as 
\begin{align}
\lefteqn{ 
\mytextstyle \sum_{m}   
\big(\hat{s}_{m}^2(t) - \nu^s_{m}(t) \big) 
        \sum_{\iB=1}^{\Nb} \nu^{\Br}_{\iB}(t) \zmatC{m}{\iB}{\jC}{t}{2} 
}\label{eq:NishimoriPB}\\
&= 
\iftoggle{long}{
\sum_{m} \left(\E\left\{
\frac{\big(\Z_{m}-\hat{p}_{m}(t)\big)^2}{\nu^p_{m}(t)}
\right\} - 1 \right) 
\frac{\sum_{\iB=1}^{\Nb} \nu^{\Br}_{\iB}(t) 
\zmatC{m}{\iB}{\jC}{t}{2}}{\nu^p_{m}(t)} ,
}{
\mytextstyle 
\sum_{m} \Big(\E\Big\{
\frac{(\Z_{m}-\hat{p}_{m}(t))^2}{\nu^p_{m}(t)}
\Big\} - 1 \Big) 
\frac{\sum_{\iB=1}^{\Nb} \nu^{\Br}_{\iB}(t) 
\zmatC{m}{\iB}{\jC}{t}{2}}{\nu^p_{m}(t)} ,
}
\non
\end{align}
where the random variable $\Z_{m}$ above is distributed according to the pdf in
\iftoggle{long}{\eqref{pZgivYPPB}}{line (D1) of \tabref{PbigAmpScalar}}.
For the G-AMP algorithm, \cite[Sec.~VI.D]{Rangan:ISIT:11} clarifies that, under i.i.d priors and scalar variances, in the LSL, the true $z_{m}$ and the G-AMP iterates $\hat{p}_m(t)$ converge empirically to a pair of random variables $(\Z,\p)$ that satisfy $p_{\Z|\p}(z|\hat{p}(t))=\mc{N}(z;\hat{p}(t),\nu^p(t))$.
This suggests that \eqref{NishimoriPB} is negligible in the LSL, in which case \eqref{rvarPB} implies
\begin{align}
\nu^r_{\jC}(t) 
&\approx \mytextstyle \left( \sum_{m} \nu^s_{m}(t) \zmat{m}{\nod}{\jC}{t}{2}  \right)^{-1}.
\end{align}
A similar argument yields
\begin{align}
\nu^q_{\iB}(t) 
&\approx \mytextstyle \left( \sum_{m} \nu^s_{m}(t) \zmat{m}{\iB}{\nod}{t}{2}  \right)^{-1}.
\end{align}

The final step in the derivation of P-BiG-AMP is to approximate the SPA posterior log-pdfs in \eqref{DbPB} and \eqref{DcPB}.
Plugging \eqref{bmlToiPB} and \eqref{cmlTojPB} into these expressions, we get
\begin{align}
\Delta_{\iB}^{\B}(t\!+\!1,\Br_{\iB}) &\approx \label{eq:DbfinalPB}
         \const + \log\!\big( p_{\B_{\iB}\!}(\Br_{\iB})\Nor(\Br_{\iB};\hat{q}_{\iB}(t),\nu^q_{\iB}(t)) \big)\\
\Delta_{\jC}^{\C}(t\!+\!1,\Cr_{\jC}) &\approx \label{eq:DcfinalPB}
         \const + \log\!\big( p_{\C_{\jC}\!}(\!\Cr_{\jC})\Nor(\Cr_{\jC};\hat{r}_{\jC}(t),\nu^r_{\jC}(t)) \big) 
\end{align}
using steps similar to those used for \eqref{cTozFinalPB}.
The corresponding pdfs are given as (D2) and (D3) in 
\iftoggle{long}{\tabref{PbigAmp}}{\tabref{PbigAmpScalar}}
and represent P-BiG-AMP's iteration-$t$ approximations to the true marginal posteriors
$p_{\B_{\iB}|\vY}(\Br_{\iB}\giv\vec{y})$ and $p_{\C_{\jC}|\vY}(\Cr_{\jC}\giv\vec{y})$.
The quantities $\hat{\Br}_{\iB}(t\!+\!1)$ and $\nu^{\Br}_{\iB}(t\!+\!1)$ are then respectively defined as the mean and variance of the pdf associated with \eqref{DbfinalPB}, and $\hat{\Cr}_{\jC}(t\!+\!1)$ and $\nu^{\Cr}_{\jC}(t\!+\!1)$ are the mean and variance of the pdf associated with \eqref{DcfinalPB}.
\iftoggle{long}{
\color{\longcolor}
As such, $\hat{\Br}_{\iB}(t\!+\!1)$ represents P-BiG-AMP's approximation to the MMSE estimate of $\B_{\iB}$ and $\nu^{\Br}_{\iB}(t\!+\!1)$ represents its approximation of the corresponding MSE.
Likewise, $\hat{\Cr}_{\jC}(t\!+\!1)$ represents P-BiG-AMP's approximation to the MMSE estimate of $\C_{\jC}$ and $\nu^{\Cr}_{\jC}(t\!+\!1)$ represents its approximation of the corresponding MSE.
\color{black}
}{}
This completes the derivation of P-BiG-AMP.

\iftoggle{long}
{
\color{\longcolor}
\subsection{Algorithm Summary}    \label{sec:algSummaryPB}

\putTable{PbigAmp}{The P-BiG-AMP Algorithm 
}{\scriptsize
\color{\longcolor}
\begin{equation*}
\begin{array}{|lr@{\,}c@{\,}l@{}r|}\hline
  \multicolumn{2}{|l}{\textsf{definitions:}}&&&\\[-1mm]
  &p_{\Z_{m}|\p_{m}\!}\big(z\giv\hat{p};\nu^p\big) 
   &\defn& \frac{p_{\Y_{m}|\Z_{m}\!}(y_{m} \giv z) \, \Nor(z;\hat{p},\nu^p)}{\int_{z'}p_{\Y_{m}|\Z_{m}\!}(y_{m} \giv z') \, \Nor(z';\hat{p},\nu^p)} &\text{(D1)}\\
   
&p_{\C_{\jC}|\textsf{r}_{\jC}\!}(\Cr\giv\hat{r};\nu^r) 
	&\defn& \frac{p_{\C_{\jC}\!}(\Cr) \, \Nor(\Cr;\hat{r},\nu^r)}{\int_{\Cr'}p_{\C_{\jC}\!}(\Cr') \, \Nor(\Cr';\hat{r},\nu^r)}&\text{(D2)}\\
&p_{\B_{\iB}|\textsf{q}_{\iB}\!}(\Br\giv\hat{q};\nu^q) 
	&\defn& \frac{p_{\B_{\iB}\!}(\Br) \, \Nor(\Br;\hat{q},\nu^q)}{\int_{\Br'}p_{\B_{\iB}\!}(\Br') \, \Nor(\Br';\hat{q},\nu^q)}&\text{(D3)}\\
	
    \multicolumn{2}{|l}{\textsf{initialization:}}&&&\\
    &\forall m:
      \hat{s}_{m}(0) &=& 0  & \text{(I1)}\\
  &\forall \iB,\jC: \textsf{choose~} &
  \multicolumn{2}{l}{\hat{\Br}_{\iB}(1), \nu^{\Br}_{\iB}(1), \hat{\Cr}_{\jC}(1), \nu^{\Cr}_{\jC}(1)} &\text{(I2)}\\
  \multicolumn{2}{|l}{\textsf{for $t=1,\dots T_\textrm{max}$}}&&&\\
   &\forall m,\iB:
   \zmat{m}{\iB}{\nod}{t}{}
   &=& \sum_{\jC=0}^{\Nc} z_{m}\of{\iB,\jC} \hat{c}_{\jC}(t) & \text{(R1)}\\[0.5mm]
   & \forall m,\jC:
   \zmat{m}{\nod}{\jC}{t}{}
   &=& \sum_{\iB=0}^{\Nb} \hat{b}_{\iB}(t) z_{m}\of{\iB,\jC} & \text{(R2)}\\[0.5mm]  
   & \forall m:
   \zmat{m}{\nod}{\nod}{t}{} 
   &=& \sum_{\iB=0}^{\Nb} \hat{b}_{\iB}(t) \zmat{m}{\iB}{\nod}{t}{} 
       \text{~or~} \sum_{\jC=0}^{\Nc} \hat{c}_{\jC}(t) \zmat{m}{\nod}{\jC}{t}{}
       & \text{(R3)}\\[0.5mm]  
   &\forall m:
   \bar{\nu}^p_{m}(t)
   &=& \sum_{\iB=1}^{\Nb} \nu^{\Br}_{\iB}(t) |\zmat{m}{\iB}{\nod}{t}{}|^2   
   + \sum_{\jC=1}^{\Nc} \nu^{\Cr}_{\jC}(t) |\zmat{m}{\nod}{\jC}{t}{}|^2 & \text{(R4)}\\[0.5mm]
   &\forall m:
   \nu^p_{m}(t)
   &=& \bar{\nu}^p_{m}(t)+  \sum_{\iB=1}^{\Nb} \nu^{\Br}_{\iB}(t) \sum_{\jC=1}^{\Nc} \nu^{\Cr}_{\jC}(t)|z_{m}\of{\iB,\jC}|^2 & \text{(R5)}\\[0.5mm] 

   &\forall m: \hat{p}_{m}(t) 
   &=& \zmat{m}{\nod}{\nod}{t}{} - \hat{s}_{m}(t\!-\!1)\bar{\nu}^p_{m}(t)& \text{(R6)}\\[0.5mm] 
   &\forall m: \nu^z_{m}(t) 
   &=& \var\{\Z_{m}\giv\p_{m}\!=\!\hat{p}_{m}(t);\nu^p_{m}(t)\} & \text{(R7)}\\[0.5mm] 
   &\forall m: \hat{z}_{m}(t) 
   &=& \E\{\Z_{m}\giv\p_{m}\!=\!\hat{p}_{m}(t);\nu^p_{m}(t)\} & \text{(R8)}\\[0.5mm] 
   &\forall m: \nu^s_{m}(t) 
   &=& (1 -  \nu^z_{m}(t)/\nu^p_{m}(t) )/\nu^p_{m}(t)  & \text{(R9)}\\[0.5mm] 
   &\forall m: \hat{s}_{m}(t) 
   &=& ( \hat{z}_{m}(t) - \hat{p}_{m}(t))/\nu^p_{m}(t) & \text{(R10)}\\[0.5mm] 
   &\forall \jC: \nu^r_{\jC}(t)
   &=&  \Big( \sum_{m=1}^M \nu^s_{m}(t) |\zmat{m}{\nod}{\jC}{t}{}|^2  \Big)^{-1} & \text{(R11)}\\[1mm] 
   &\forall \jC: \hat{r}_{\jC}(t) 
   &=& \hat{\Cr}_{\jC}(t) + \nu^r_{\jC}(t) \sum_{m=1}^M \hat{s}_{m}(t) \zmat{m}{\nod}{\jC}{t}{*} &\\
   &&&~ -\nu^r_{\jC}(t)\hat{\Cr}_{\jC}(t) \sum_{m=1}^M \nu^s_{m}(t) \sum_{\iB=1}^{\Nb} \nu^{\Br}_{\iB}(t) 
	|\zmatC{m}{\iB}{\jC}{t}{}|^2 & \text{(R12)}\\[0.5mm]
   &\forall \iB: \nu^q_{\iB}(t) 
   &=&  \Big( \sum_{m=1}^M \nu^s_{m}(t) |\zmat{m}{\iB}{\nod}{t}{}|^2  \Big)^{-1} & \text{(R13)}\\[1mm] 
   &\forall \iB: \hat{q}_{\iB}(t) 
   &=& \hat{\Br}_{\iB}(t) + \nu^q_{\iB}(t) \sum_{m=1}^M \hat{s}_{m}(t)   \zmat{m}{\iB}{\nod}{t}{*} &\\
   &&&~ -\nu^q_{\iB}(t)\hat{\Br}_{\iB}(t) \sum_{m=1}^M \nu^s_{m}(t) \sum_{\jC=1}^{\Nc} \nu^{\Cr}_{\jC}(t) 
        |\zmatC{m}{\iB}{\jC}{t}{}|^2  &\text{(R14)}\\[0.5mm]
   &\forall \jC: \nu^{\Cr}_{\jC}(t\!+\!1) 
   &=& \var\{\C_{\jC}\giv\R_{\jC}\!=\!\hat{r}_{\jC}(t); \nu^r_{\jC}(t)\} & \text{(R15)}\\[0.5mm] 
   &\forall \jC: \hat{\Cr}_{\jC}(t\!+\!1) 
   &=& \E\{\C_{\jC}\giv \R_{\jC}\!=\!\hat{r}_{\jC}(t); \nu^r_{\jC}(t)\}& \text{(R16)}\\[0.5mm] 
   &\forall \iB: \nu^{\Br}_{\iB}(t\!+\!1) 
   &=& \var\{\B_{\iB}\giv\Q_{\iB}\!=\!\hat{q}_{\iB}(t); \nu^q_{\iB}(t)\}& \text{(R17)}\\[0.5mm] 
   &\forall \iB: \hat{\Br}_{\iB}(t\!+\!1) 
   &=& \E\{\B_{\iB}\giv \Q_{\iB}\!=\!\hat{q}_{\iB}(t); \nu^q_{\iB}(t)\}& \text{(R18)}\\[0.5mm] 
   
      \multicolumn{4}{|c}{\textsf{if $\sum_{m=1}^M |\zmat{m}{\nod}{\nod}{t}{} - \zmat{m}{\nod}{\nod}{t\!-\!1}{}|^2 \le \tau_\textrm{stop} \sum_{m=1}^M |\zmat{m}{\nod}{\nod}{t}{}|^2$, \textsf{stop}}}&\text{(R19)}\\

    \multicolumn{2}{|l}{\textsf{end}}&&&\\\hline
\end{array}
\end{equation*}
}

The P-BiG-AMP algorithm is summarized in \tabref{PbigAmp}. 
The version in \tabref{PbigAmp} includes a maximum number of iterations $T_\textrm{max}$, as well as a stopping condition (R19) that terminates the iterations when the change in $\zmat{m}{\nod}{\nod}{t}{}$ falls below a user-defined parameter $\tau_\textrm{stop}$. 
Noting the complex conjugates in (R12) and (R14), the algorithm also allows the use of \emph{complex-valued} quantities, in which case $\Nor$ in (D1)-(D3) would denote a circular complex Gaussian pdf.
{However, for ease of interpretation, \tabref{PbigAmp} does not include the important damping steps that will be detailed in \secref{dampPB}.}


The complexity scaling of each line in \tabref{PbigAmp} is tabulated in \tabref{compCost}
assuming that all $M\Nb\Nc$ entries in the tensor $z_m\of{i,j}$ are nonzero.
In practice, $z_m\of{i,j}$ is often sparse or implementable using a fast transformation, allowing drastic reduction in complexity, as shown in \secref{simplifications}.
Thus, \tabref{compCost} should be interpreted as ``worst-case'' complexity.

\putTable{compCost}{Worst-case complexity of P-BiG-AMP from \tabref{PbigAmp}.}
{
 \color{\longcolor}
 \renewcommand{\arraystretch}{1}
 \begin{tabular}{||@{\:}c@{\:}|@{\;}l@{\;}||@{\:}c@{\:}|@{\;}l@{\;}||@{\:}c@{\:}|@{\;}l@{\;}||} \hline
  (R1) & $\Ord(M\Nb\Nc)$ &
  (R2) & $\Ord(M\Nb\Nc)$ &
  (R3) & $\Ord(M(\Nb \!\wedge\! \Nc))$ \\\hline 
  (R4) & $\Ord(M\Nb+M\Nc)$ &
  (R5) & $\Ord(M\Nc\Nb)$ &
  (R6) & $\Ord(M)$ \\\hline
  (R7) & $\Ord(M)$ &
  (R8) & $\Ord(M)$ &
  (R9) & $\Ord(M)$ \\\hline
  (R10) & $\Ord(M)$ &
  (R11) & $\Ord(M\Nc)$ &
  (R12) & $\Ord(M\Nb\Nc)$ \\\hline
  (R13) & $\Ord(M\Nb)$ &
  (R14) & $\Ord(M\Nb\Nc)$ &
  (R15) & $\Ord(\Nc)$ \\\hline
  (R16) & $\Ord(\Nc)$ &
  (R17) & $\Ord(\Nb)$ &
  (R18) & $\Ord(\Nb)$ \\\hline
 \end{tabular}
}

\color{black}
}
{
A table summarizing the steps of the P-BiG-AMP algorithm is given in \cite{Parker:PBiGAMP} but is omitted here for reasons of space.
}

\subsection{Scalar-Variance Approximation}      \label{sec:scalar}

The P-BiG-AMP algorithm 
\iftoggle{long}
{from \tabref{PbigAmp}}
{derived above}
stores and processes variance terms $\bar{\nu}^p_m, \nu^p_m, \nu^z_m, \nu^s_m, \nu^r_j, \nu^q_i, \nu^c_j, \nu^b_i$ that depend on the indices $m,j,i$.
The use of scalar (i.e., index-invariant) variances significantly reduces its complexity.

To derive scalar-variance P-BiG-AMP, we first assume 
$\forall i: \nu^b_i(t) \approx \nu^b(t)\defn \frac{1}{\Nb}\sum_{i=1}^{\Nb} \nu^b_i(t)$ 
and
$\forall j: \nu^c_j(t) \approx \nu^c(t)\defn \frac{1}{\Nc}\sum_{j=1}^{\Nc} \nu^c_j(t)$.
Then we approximate $\bar{\nu}^p_m(t)$ as
\begin{align}
&\lefteqn{ \mytextstyle \bar{\nu}^p_m(t)
\approx \nu^{\Br}(t) \sum_{\iB=1}^{\Nb} |\zmat{m}{\iB}{\nod}{t}{}|^2   
   + \nu^{\Cr}(t) \sum_{\jC=1}^{\Nc} |\zmat{m}{\nod}{\jC}{t}{}|^2 
}
\iftoggle{long}{\\}{\non\\[-2mm]\\}
&\approx \mytextstyle
\frac{\nu^{\Br}(t)}{M}\sum_{i=1}^{\Nb}\|\hvec{z}\of{\iB,\nod}(t)\|^2 
+ \frac{\nu^{\Cr}(t)}{M}\sum_{j=1}^{\Nc}\|\hvec{z}\of{\nod,\jC}(t)\|^2 
\defn \bar{\nu}^p(t)
\label{eq:nup_bar} .
\end{align}
Similarly, $\nu^p_m(t)$ is approximated as
\begin{align}
&\lefteqn{ \nu^p_m(t)
\approx \mytextstyle \bar{\nu}^p(t)+  \nu^{\Br}(t) \nu^{\Cr}(t) \sum_{\iB=1}^{\Nb} \sum_{\jC=1}^{\Nc} |z_{m}\of{\iB,\jC}|^2 }\\
&\approx \mytextstyle \bar{\nu}^p(t)+  \frac{\nu^{\Br}(t) \nu^{\Cr}(t)}{M} \sum_{\iB=1}^{\Nb} \sum_{\jC=1}^{\Nc} \|\vec{z}\of{\iB,\jC}\|^2 
\defn \nu^p(t)
\label{eq:nup} ,
\end{align}
where $\frac{1}{M}\sum_{\iB=1}^{\Nb} \sum_{\jC=1}^{\Nc} \|\vec{z}\of{\iB,\jC}\|^2$ can be pre-computed.
Even with the above scalar-variance approximations, $\nu^s_m(t)$ is not guaranteed to be $m$-invariant.
Still, it can be approximated as such using $\nu^s(t)\defn \frac{1}{M}\sum_{m=1}^M \nu^s_m(t)$, in which case 
\begin{align}
\nu^r_j(t)
&\approx \big( \nu^s(t) \|\hvec{z}\of{\nod,j}(t)\|^2 \big)^{-1} \\
&\approx \mytextstyle \left( \nu^s(t) \frac{1}{\Nc}\sum_{j=1}^{\Nc}\|\hvec{z}\of{\nod,j}(t)\|^2 \right)^{-1}
\defn \nu^r(t) 
\label{eq:nur} \\
\hat{r}_{\jC}(t) 
&= \mytextstyle \hat{\Cr}_{\jC}(t) + \nu^r(t) \sum_{m=1}^M \hat{s}_{m}(t) \zmat{m}{\nod}{\jC}{t}{*}
\non\\&\quad \mytextstyle
-\nu^r(t)\nu^s(t) \nu^{\Br}(t) \hat{\Cr}_{\jC}(t) 
\sum_{\iB=1}^{\Nb} \|\vec{z}\of{\iB,\jC}\|^2 
\label{eq:rhat_scalar} ,
\end{align}
where $\sum_{\iB=1}^{\Nb} \|\vec{z}\of{\iB,\jC}\|^2$ can be pre-computed.
Similarly,
\begin{align}
\nu^q_i(t)
&\approx \big( \nu^s(t) \|\hvec{z}\of{i,\nod}(t)\|^2 \big)^{-1}\\
&\approx \mytextstyle \left( \nu^s(t) \frac{1}{\Nb}\sum_{i=1}^{\Nb}\|\hvec{z}\of{i,\nod}(t)\|^2 \right)^{-1}
\defn \nu^q(t) 
\label{eq:nuq} \\
\hat{q}_{\iB}(t) 
&= \mytextstyle \hat{\Br}_{\iB}(t) + \nu^q(t) \sum_{m=1}^M \hat{s}_{m}(t) \zmat{m}{\iB}{\nod}{t}{*}
\non\\&\quad \mytextstyle
-\nu^q(t)\nu^s(t) \nu^{\Cr}(t) \hat{\Br}_{\iB}(t) 
\sum_{\jC=1}^{\Nc} \|\vec{z}\of{\iB,\jC}\|^2 
\label{eq:qhat_scalar} .
\end{align}

\putTable{PbigAmpScalar}{The Scalar-Variance P-BiG-AMP Algorithm 
}{\scriptsize
\begin{equation*}
\begin{array}{|lr@{\,}c@{\,}l@{}r|}\hline
  \multicolumn{2}{|l}{\textsf{definitions:}}&&&\\[-1mm]
  &p_{\Z_{m}|\p_{m}\!}\big(z\giv\hat{p};\nu^p\big) 
   &\defn& \frac{p_{\Y_{m}|\Z_{m}\!}(y_{m} \giv z) \, \Nor(z;\hat{p},\nu^p)}{\int_{z'}p_{\Y_{m}|\Z_{m}\!}(y_{m} \giv z') \, \Nor(z';\hat{p},\nu^p)} &\text{(D1)}\\
   
&p_{\C_{\jC}|\textsf{r}_{\jC}\!}(\Cr\giv\hat{r};\nu^r) 
	&\defn& \frac{p_{\C_{\jC}\!}(\Cr) \, \Nor(\Cr;\hat{r},\nu^r)}{\int_{\Cr'}p_{\C_{\jC}\!}(\Cr') \, \Nor(\Cr';\hat{r},\nu^r)}&\text{(D2)}\\
&p_{\B_{\iB}|\textsf{q}_{\iB}\!}(\Br\giv\hat{q};\nu^q) 
	&\defn& \frac{p_{\B_{\iB}\!}(\Br) \, \Nor(\Br;\hat{q},\nu^q)}{\int_{\Br'}p_{\B_{\iB}\!}(\Br') \, \Nor(\Br';\hat{q},\nu^q)}&\text{(D3)}\\
	
    \multicolumn{2}{|l}{\textsf{initialization:}}&&&\\
    &\forall m:
      \hat{s}_{m}(0) &=& 0  & \text{(I1)}\\
  &\forall \iB,\jC: \textsf{choose~} &
  \multicolumn{2}{l}{\hat{\Br}_{\iB}(1), \nu^{\Br}(1), \hat{\Cr}_{\jC}(1), \nu^{\Cr}(1)} &\text{(I2)}\\
  \multicolumn{2}{|l}{\textsf{for $t=1,\dots T_\textrm{max}$}}&&&\\
   &\forall \iB:
   \hvec{z}\of{\iB,\nod}(t)
   &=& \sum_{\jC=0}^{\Nc} \vec{z}\of{\iB,\jC} \hat{c}_{\jC}(t) & \text{(R1)}\\[0.5mm]
   & \forall \jC:
   \hvec{z}\of{\nod,\jC}(t)
   &=& \sum_{\iB=0}^{\Nb} \hat{b}_{\iB}(t) \vec{z}\of{\iB,\jC} & \text{(R2)}\\[0.5mm]  
   & 
   \hvec{z}\of{\nod,\nod}(t) 
   &=& \sum_{\iB=0}^{\Nb} \hat{b}_{\iB}(t) \hvec{z}\of{\iB,\nod}(t) 
       \text{~or~} \sum_{\jC=0}^{\Nc} \hat{c}_{\jC}(t) \hvec{z}\of{\nod,\jC}(t)
       & \text{(R3)}\\[0.5mm]  
   &
   \bar{\nu}^p(t)
   &=& \big( \nu^{\Br}(t) \sum_{\iB=1}^{\Nb} \|\hvec{z}\of{i,*}(t)\|^2   
       &\\&&&~ 
       + \nu^{\Cr}(t) \sum_{\jC=1}^{\Nc} \|\hvec{z}\of{*,j}(t)|^2 \big)/M & \text{(R4)}\\[0.5mm]
   &
   \nu^p(t)
   &=& \bar{\nu}^p(t)+  \nu^{\Br}(t) \nu^{\Cr}(t) \sum_{\iB=1}^{\Nb} \sum_{\jC=1}^{\Nc} \|\vec{z}\of{\iB,\jC}\|^2/M & \text{(R5)}\\[0.5mm] 
   &
   \hvec{p}(t) 
   &=& \hvec{z}\of{\nod,\nod}(t) - \hvec{s}(t\!-\!1)\bar{\nu}^p(t)& \text{(R6)}\\[0.5mm] 
   &
   \nu^z(t) 
   &=& \sum_{m=1}^M \var\{\Z_{m}\giv\p_{m}\!=\!\hat{p}_{m}(t);\nu^p(t)\}/M & \text{(R7)}\\[0.5mm] 
   &\forall m: 
   \hat{z}_{m}(t) 
   &=& \E\{\Z_{m}\giv\p_{m}\!=\!\hat{p}_{m}(t);\nu^p(t)\} & \text{(R8)}\\[0.5mm]
   &
   \nu^s(t) 
   &=& (1 -  \nu^z(t)/\nu^p(t) )/\nu^p(t)  & \text{(R9)}\\[0.5mm] 
   &
   \hvec{s}(t) 
   &=& ( \hvec{z}(t) - \hvec{p}(t))/\nu^p(t) & \text{(R10)}\\[0.5mm] 
   &
   \nu^r(t)
   &=&  \Big( \nu^s(t) \sum_{j=1}^{\Nc}\|\hvec{z}\of{\nod,\jC}(t)\|^2/\Nc \Big)^{-1} & \text{(R11)}\\[1mm] 
   &\forall \jC: 
   \hat{r}_{\jC}(t) 
   &=& \Big(1 - \nu^r(t) \nu^s(t) \nu^{\Br}(t) 
        \sum_{\iB=1}^{\Nb} \|\vec{z}\of{\iB,\jC}\|^2 \Big)\hat{\Cr}_{\jC}(t) 
       &\\&&&~ 
       + \nu^r(t) \hvec{z}\ofH{\nod,\jC}(t) \hvec{s}(t) & \text{(R12)}\\[0.5mm]
   &
   \nu^q(t) 
   &=&  \Big( \nu^s(t) \sum_{i=1}^{\Nb}\|\hvec{z}\of{\iB,\nod}(t)\|^2/\Nb \Big)^{-1} & \text{(R13)}\\[1mm] 
   &\forall \iB: 
   \hat{q}_{\iB}(t) 
   &=& \Big(1 - \nu^q(t) \nu^s(t) \nu^{\Cr}(t) 
        \sum_{\jC=1}^{\Nc} \|\vec{z}\of{\iB,\jC}\|^2 \Big)\hat{\Br}_{\iB}(t) 
       &\\&&&~ 
       + \nu^q(t) \hvec{z}\ofH{\iB,\nod}(t) \hvec{s}(t) & \text{(R14)}\\[0.5mm]
   &
   \nu^{\Cr}(t\!+\!1) 
   &=& \sum_{\jC=1}^{\Nc} \var\{\C_{\jC}\giv\R_{\jC}\!=\!\hat{r}_{\jC}(t); \nu^r_{\jC}(t)\}/\Nc & \text{(R15)}\\[0.5mm] 
   &\forall \jC: 
   \hat{\Cr}_{\jC}(t\!+\!1) 
   &=& \E\{\C_{\jC}\giv \R_{\jC}\!=\!\hat{r}_{\jC}(t); \nu^r_{\jC}(t)\}& \text{(R16)}\\[0.5mm] 
   &
   \nu^{\Br}(t\!+\!1) 
   &=& \sum_{\iB=1}^{\Nb} \var\{\B_{\iB}\giv\Q_{\iB}\!=\!\hat{q}_{\iB}(t); \nu^q_{\iB}(t)\}/\Nb& \text{(R17)}\\[0.5mm] 
   &\forall \iB: \hat{\Br}_{\iB}(t\!+\!1) 
   &=& \E\{\B_{\iB}\giv \Q_{\iB}\!=\!\hat{q}_{\iB}(t); \nu^q_{\iB}(t)\}& \text{(R18)}\\[0.5mm] 
   
      \multicolumn{4}{|c}{\textsf{if $\|\hvec{z}\of{\nod,\nod}(t) - \hvec{z}\of{\nod,\nod}(t\!-\!1)\|^2 \le \tau_\textrm{stop} \|\hvec{z}\of{\nod,\nod}(t)\|^2$, \textsf{stop}}}&\text{(R19)}\\

    \multicolumn{2}{|l}{\textsf{end}}&&&\\\hline
\end{array}
\end{equation*}
}

The scalar-variance P-BiG-AMP algorithm is summarized in \tabref{PbigAmpScalar}.
\iftoggle{long}
{
The complexity scaling of each line in \tabref{PbigAmpScalar} is tabulated in \tabref{compCostScalar}.
Like with \tabref{compCost}, the values in \tabref{compCostScalar} should be interpreted as ``worst-case.''
}
{
Noting the Hermitian transposes in lines (R12) and (R14), the algorithm allows the use of complex-valued quantities, in which case $\Nor$ in (D1)-(D3) becomes a circular complex Gaussian pdf.
The complexity scaling of each line in \tabref{PbigAmpScalar} is tabulated in \tabref{compCostScalar}
assuming that all $M\Nb\Nc$ entries in the tensor $z_m\of{i,j}$ are nonzero.
In practice, $z_m\of{i,j}$ is often sparse or implementable using a fast transformation, allowing drastic reduction in complexity, as shown in \secref{simplifications}.
Thus, \tabref{compCostScalar} should be interpreted as ``worst-case'' complexity.
}

\putTable{compCostScalar}{Worst-case complexity of scalar-variance P-BiG-AMP.}
{
\renewcommand{\arraystretch}{1}
\begin{tabular}{||@{\:}c@{\:}|@{\;}l@{\;}||@{\:}c@{\:}|@{\;}l@{\;}||@{\:}c@{\:}|@{\;}l@{\;}||} \hline
  (R1) & $\Ord(M\Nb\Nc)$ &
  (R2) & $\Ord(M\Nb\Nc)$ &
  (R3) & $\Ord(M(\Nb \!\wedge\! \Nc))$ \\\hline 
  (R4) & $\Ord(1)$ &
  (R5) & $\Ord(1)$ &
  (R6) & $\Ord(M)$ \\\hline
  (R7) & $\Ord(M)$ &
  (R8) & $\Ord(M)$ &
  (R9) & $\Ord(M)$ \\\hline
  (R10) & $\Ord(M)$ &
  (R11) & $\Ord(1)$ &
  (R12) & $\Ord(M\Nc)$ \\\hline
  (R13) & $\Ord(1)$ &
  (R14) & $\Ord(M\Nb)$ &
  (R15) & $\Ord(\Nc)$ \\\hline
  (R16) & $\Ord(\Nc)$ &
  (R17) & $\Ord(\Nb)$ &
  (R18) & $\Ord(\Nb)$ \\\hline
 \end{tabular}
}

\subsection{Damping}    \label{sec:dampPB}
\iftoggle{long}{
\color{\longcolor}
Damping has been applied to both G-AMP~\cite{Rangan:ISIT:14} and BiG-AMP~\cite{Parker:TSP:14a} to prevent divergence.
Essentially, damping (or ``relaxation'' in the optimization literature) slows the evolution of the algorithm's state variables.
For G-AMP, damping yields provable local-convergence guarantees with arbitrary matrices~\cite{Rangan:ISIT:14} while, for BiG-AMP, damping has been shown to be very effective through an extensive empirical study~\cite{Parker:TSP:14b}.

Motivated by these successes, we adopt a similar damping scheme for P-BiG-AMP.
In particular, we use the iteration-$t$ damping factor $\beta(t) \in [0,1]$ to slow the evolution of certain variables, namely, $\bar{\nu}^p_{m}$, $\nu^p_{m}$, $\nu^s_{m}$, $\hat{s}_{m}$, $\hat{\Br}_{\iB}$, and $\hat{\Cr}_{\jC}$.
To do this, we replace steps (R4), (R5), (R4), and (R10) in \tabref{PbigAmp} with 
\begin{align}
\bar{\nu}^p_{m}(t)
&=  \beta(t) \bigg( \sum_{\iB=1}^{\Nb} |\zmat{m}{\iB}{\nod}{t}{}|^2\nu^{\Br}_{\iB}(t)   
   + \sum_{\jC=1}^{\Nc} |\zmat{m}{\nod}{\jC}{t}{}|^2\nu^{\Cr}_{\jC}(t) \bigg)
\non\\&\quad
+(1 - \beta(t))\bar{\nu}^p_{m}(t-1) \\
\nu^p_{m}(t)
&= \beta(t)  \bigg(\bar{\nu}^p_{m}(t)+  \sum_{\iB=1}^{\Nb}\sum_{\jC=1}^{\Nc} \nu^{\Br}_{\iB}(t) \nu^{\Cr}_{\jC}(t)|\zmat{m}{\iB}{\jC}{t}{}|^2  \bigg)
\non\\&\quad
+(1 - \beta(t)) \nu^p_{m}(t-1)\\
\nu^s_{m}(t) 
&= \beta(t)\big(( 1 - \nu^z_{m}(t)/\nu^p_{m}(t) )/\nu^p_{m}(t) \big)
\non\\&\quad
+(1-\beta(t)) \nu^s_{m}(t\!-\!1)\\
\hat{s}_{m}(t) 
&= \beta(t) \big( \hat{z}_{m}(t) - \hat{p}_{m}(t))/\nu^p_{m}(t)\big)
\non\\&\quad 
+(1-\beta(t)) \hat{s}_{m}(t\!-\!1) ,
\end{align}
and we insert the following lines between (R10) and (R11):
\begin{align}
\bar{b}_{\iB}(t) 
&= \beta(t)\hat{b}_{\iB}(t) + (1-\beta(t))\bar{b}_{\iB}(t-1)\\
\bar{c}_{\jC}(t) 
&= \beta(t)\hat{c}_{\jC}(t) + (1-\beta(t))\bar{c}_{\jC}(t-1)\\
\zbar{m}{i}{\nod}{t}{}
   &= \sum_{j=0}^{\Nc} z\of{i,j}_m \bar{c}_j(t)\\
\zbar{m}{\nod}{\jC}{t}{}
   &= \sum_{i=0}^{\Nb} \bar{b}_i(t) z\of{i,j}_m .
\end{align}
The quantities $\zbar{m}{\iB}{\nod}{t}{}$ and $\zbar{m}{\nod}{\jC}{t}{}$ are then used in steps (R11)-(R14), but not in (R4)-(R6), in place of the versions computed in steps (R1)-(R2).
Similarly, the newly created state variables $\bar{b}_{\iB}(t)$ and $\bar{c}_{\jC}(t)$ are used only to compute $\zbar{m}{\iB}{\nod}{t}{}$ and $\zbar{m}{\nod}{\jC}{t}{}$.
Note that, when $\beta(t)\!=\!1$, the damping has no effect, whereas when $\beta(t)\!=\!0$, all quantities become frozen in $t$. 
Although these modifications pertain to the full P-BiG-AMP algorithm from \tabref{PbigAmp}, similar damping steps can be applied to the scalar-variance version from \tabref{PbigAmpScalar}.

\subsubsection{Adaptive Damping}
Because damping slows the convergence of the algorithm, we would like to damp only as much as needed to prevent divergence, i.e., to \emph{adapt} the damping.
An \emph{adaptive} damping scheme for G-AMP was described in \cite{Vila:ICASSP:15} and a similar one was described for BiG-AMP in \cite{Parker:TSP:14a}.
Both are based on monitoring an appropriate cost $J(t)$ and applying more damping when the cost increases or less when the cost is decreasing.
The same approach can be used for P-BiG-AMP.
For example, extending the approach used for BiG-AMP \cite{Parker:TSP:14a} would lead to the cost
\begin{align}
\hat{J}(t) &= 
\sum_{\jC} D\Big( p_{\C_{\jC}|\R_{\jC}}\big(\cdot\biggiv\hat{r}_{\jC}(t);\nu_{\jC}^r(t)\big) \Big\|\, p_{\C_{\jC}}(\cdot) \Big) 		\label{eq:costPBiGAMPPB} 
\\&\quad+
\sum_{\iB} D\Big( p_{\B_{\iB}|\Q_{\iB}}\big(\cdot\biggiv\hat{q}_{\iB}(t);\nu_{\iB}^q(t)\big) \Big\|\, p_{\B_{\iB}}(\cdot) \Big) 
\non\\&\quad-
\sum_{m} \E_{\Z_{m}\sim\mathcal{N}(\zmat{m}{\nod}{\nod}{t}{};\nu^p_{m}(t))}\big\{ \log p_{\Y_{m}|\Z_{m}}(y_{m} \giv \Z_{m}) \big\}.	\non
\end{align}
Meanwhile, the Bethe-free-energy approach used in \cite{Kabashima:14,Vila:ICASSP:15} offers a more principled, yet more complex, alternative.
Intuitively, the first term in \eqref{costPBiGAMPPB} penalizes the deviation between the (P-BiG-AMP approximated) posterior and the assumed prior on $\vC$, 
the second penalizes the deviation between the (P-BiG-AMP approximated) posterior and the assumed prior on $\vB$, and 
the third term rewards highly likely estimates $\vZ$.

For adaptive damping, we adopt the approach used for both G-AMP and BiG-AMP in the public domain GAMPmatlab implementation~\cite{GAMPmatlab}.
In particular,
if the current cost $J(t)$ is not smaller than the largest cost in the most recent \texttt{stepWindow} iterations, then the ``step'' is declared unsuccessful, the damping factor $\beta(t)$ is reduced by the factor \texttt{stepDec}, and the step is attempted again.
These attempts continue until either the cost criterion decreases or the damping factor reaches \texttt{stepMin}, at which point the step is considered successful, or the iteration count exceeds $T_{\max}$ or the damping factor reaches \texttt{stepTol}, at which point the algorithm terminates. 
Otherwise, the step is declared successful, and the damping factor is increased by the factor \texttt{stepInc} up to a maximum allowed value \texttt{stepMax}. 

\color{black}
}{

Damping has been applied to both G-AMP~\cite{Rangan:ISIT:14} and BiG-AMP~\cite{Parker:TSP:14a} to prevent divergence.
Essentially, damping (or ``relaxation'' in the optimization literature) slows the evolution of the algorithm's state variables.
For G-AMP, damping yields provable local-convergence guarantees with arbitrary matrices~\cite{Rangan:ISIT:14} while, for BiG-AMP, damping has been shown to be very effective through an extensive empirical study~\cite{Parker:TSP:14b}.
Furthermore, \emph{adaptive} damping has been proposed for G-AMP with the goal of preventing divergence with minimal impact on runtime \cite{Vila:ICASSP:15}.
The same damping strategies can be straightforwardly applied to P-BiG-AMP.
For details, we direct the interested reader to \cite{Parker:PBiGAMP} and the public domain GAMPmatlab implementation \cite{GAMPmatlab}. 
}

\subsection{Tuning of the Prior and Likelihood} \label{sec:EMPB}
To run P-BiG-AMP, one must specify the priors and likelihood in lines (D1)-(D3) of \iftoggle{long}{\tabref{PbigAmp} and }{}\tabref{PbigAmpScalar}. 
Although a reasonable family of distributions may be dictated by the application, the specific parameters of the distributions must often be tuned in practice.
\iftoggle{long}{
\color{\longcolor}
Building on the approach developed to address this challenge for G-AMP~\cite{Vila:TSP:13}, which was extended successfully to BiG-AMP in \cite{Parker:TSP:14a}, we outline a methodology that takes a given set of P-BiG-AMP priors $\{p_{\B_{\iB}}(\cdot;\vec{\theta}),p_{\C_{\jC}}(\cdot;\vec{\theta}),p_{\Y_{m}|\Z_{m}}(y_{m}|\cdot;\vec{\theta})\}_{\forall m,n,l}$ and tunes the vector $\vec{\theta}$ using an expectation-maximization (EM) \cite{Dempster:JRSS:77} based approach, with the goal of maximizing its likelihood, i.e., finding $\hvec{\theta}\defn \argmax_{\vec{\theta}}p_{\vY}(\vec{y};\vec{\theta})$. 

Taking $\vB$, $\vC$, and $\vZ$ to be the hidden variables, the EM recursion can be written as \cite{Dempster:JRSS:77}
\begin{align}
\vec{\hat{\theta}}^{k+1} 
&= \argmax_{\vec{\theta}} \E\Big\{\log p_{\vB,\vC,\vZ,\vY}(\vB,\vC,\vZ,\vY;\vec{\theta})\Biggiv \vec{y}; \hvec{\theta}^k\Big\} \nonumber \\
&= \argmax_{\vec{\theta}} \bigg\{
	\sum_{\iB} \E\Big\{\log p_{\B_{\iB}}(\B_{\iB};\vec{\theta}) \Biggiv \vec{y}; \vec{\hat{\theta}}^k \Big\}	
	\label{eq:EMPB}
	\\&\quad
	+ \sum_{\jC} \E\Big\{\log p_{\C_{\jC}}(\C_{\jC};\vec{\theta}) \Biggiv \vec{y}; \vec{\hat{\theta}}^k\Big\}
	\nonumber\\&\quad
	+ \sum_{m} \E\Big\{\log p_{\Y_{m}|\Z_{m}}(y_{m}\giv \Z_{m};\vec{\theta}) \Biggiv \vec{y}; \vec{\hat{\theta}}^k \Big\}
	\bigg\}	\nonumber
\end{align}
where for \eqref{EMPB} we used the fact
$p_{\vB,\vC,\vZ,\vY}(\vB,\vC,\vZ,\vY;\vec{\theta})
= p_{\vB}(\vB;\vec{\theta}) p_{\vC}(\vC;\vec{\theta})
p_{\vY|\vZ}(\vY|\vZ;\vec{\theta}) \,\Dirac_{\vZ-\vec{z}(\vB,\vC)}$
and the separability of $p_{\vB}$, $p_{\vC}$, and $p_{\vY|\vZ}$. 
As can be seen from \eqref{EMPB}, knowledge of the marginal posteriors $\{p_{\B_{\iB}|\vY}, p_{\C_{\jC}|\vY}, p_{\Z_{m}|\vY}\}_{\forall \iB,\jC,m}$ is sufficient to compute the EM update.
Since the exact marginal posteriors are too difficult to compute, we employ the iteration-$t$ approximations produced by P-BiG-AMP, i.e., 
\begin{align}
p_{\B_{\iB}|\vY}(b_{\iB}\giv\vec{y}) 
&\approx p_{\B_{\iB}|\Q_{\iB}}\big(b_{\iB}\giv\hat{q}_{\iB}(t);\nu^q_{\iB}(t)\big) \\
p_{\C_{\jC}|\vY}(c_{\jC}\giv\vec{y})
&\approx p_{\C_{\jC}|\R_{\jC}}\big(c_{\jC}\giv\hat{r}_{\jC}(t);\nu^r_{\jC}(t)\big) \\
p_{\Z_{m}|\vY}(z_{m}\giv\vec{y}) 
&\approx p_{\Z_{m}|\p_{m}}\big(z_{m}\giv\hat{p}_{m}(t);\nu^p_{m}(t)\big) ,
\end{align}
for suitably large $t$, where the distributions above are defined in (D1)-(D3) of \tabref{PbigAmp}.
In addition, we adopt the ``incremental'' update strategy from \cite{Neal:Jordan:98}, where the maximization over $\vec{\theta}$ is performed one element at a time while holding the others fixed.
The remaining details are analogous to the G-AMP case, for which we refer the interested reader to \cite{Vila:TSP:13}.

\color{black}
}{
To address this issue, we propose to apply the expectation-maximization (EM) \cite{Dempster:JRSS:77} based approach developed for G-AMP~\cite{Vila:TSP:13} and later extended to BiG-AMP in \cite{Parker:TSP:14a}. 
In this approach, the user specifies the families of P-BiG-AMP priors $\{p_{\B_{\iB}}(\cdot;\vec{\theta}),p_{\C_{\jC}}(\cdot;\vec{\theta}),p_{\Y_{m}|\Z_{m}}(y_{m}|\cdot;\vec{\theta})\}_{\forall m,n,l}$ and the EM algorithm is used to (locally) maximize the likelihood of the parameter vector $\vec{\theta}$, i.e., to find $\hvec{\theta}\defn \argmax_{\vec{\theta}}p_{\vY}(\vec{y};\vec{\theta})$. 
If the families themselves are unknown, it may suffice to use Gaussian-mixture models and learn their parameters \cite{Vila:TSP:13}.
Since the details are nearly identical to the BiG-AMP case, we omit them and refer the reader to \cite{Parker:PBiGAMP}. 

}

\section{Example Parameterizations}    \label{sec:simplifications}

P-BiG-AMP was summarized and derived in \secref{PBigAmp} for generic parameterizations $\vec{z}\of{i,j}$ in \eqref{ZPB}. 
A naive implementation, which treats every $z_m\of{i,j}$ as nonzero, would lead to the worst-case complexities stated in\iftoggle{long}{ \tabref{compCost} (or \tabref{compCostScalar} under the scalar-variance approximation).}{ \tabref{compCostScalar}.}
In practice, however, $\{z_m\of{i,j}\}$ is often sparse or implementable using a fast transformation, in which case the implementation can be dramatically simplified.
We now describe several examples of structured $\vec{z}\of{i,j}$, detailing the computations needed for the essential scalar-variance P-BiG-AMP quantities
$\hvec{z}\of{*,*}(t)$, 
$\sum_{i=1}^{\Nb}\|\hvec{z}\of{i,*}(t)\|^2$,
$\sum_{j=1}^{\Nc}\|\hvec{z}\of{*,j}(t)\|^2$,
$\{\hvec{z}\ofH{i,*}(t)\hvec{s}(t)\}_{i=1}^{\Nb}$ and
$\{\hvec{z}\ofH{*,j}(t)\hvec{s}(t)\}_{j=1}^{\Nc}$.

\subsection{Multi-snapshot Structure} \label{sec:multi}

With multi-snapshot structure, the noiseless outputs become
\begin{align}
\vec{Z} 
&= \mytextstyle \sum_{i=0}^{\Nb} b_i \vec{A}\of{i} \vec{C} \text{~with known~} \{\vec{A}\of{i}\}
\label{eq:Z_multi} ,
\end{align}
where $\vec{Z}\in\Complex^{K\times L}$ and $\vec{C}\in\Complex^{N\times L}$ for\footnote{When $L=1$, \eqref{Z_multi} reduces to the general parameterization \eqref{ZPB}.}
$L>1$.
Thus we have 
$\vec{A}\of{i}\in\Complex^{K\times N}$, 
$M=KL$, and $\Nc=NL$.
Defining
$\vec{z}\defn\vect(\vec{Z})$ and 
$\vec{c}\defn\vect(\vec{C})$, 
we find
\begin{align}
\vec{z}
&= \mytextstyle \sum_{i=0}^{\Nb} b_i \big(\vec{I}_L \otimes \vec{A}\of{i}\big) \vec{c} ,
\label{eq:z_multi} 
\end{align}
which implies that
\begin{align}
\vec{z}\of{i,j}
&= \big[ \vec{I}_L \otimes \vec{A}\of{i} \big]_{:,j} 
\label{eq:zij_multi}\\
\hvec{z}\of{i,*}(t)
&=\vect\big( \vec{A}\of{i}\hvec{C}(t) \big)
\label{eq:zi*_multi}\\
\hvec{z}\of{*,j}(t)
&=\big[ \vec{I}_L \otimes \hvec{A}(t) \big]_{:,j}
\label{eq:z*j_multi} \\
\hvec{z}\of{*,*}(t)
&=\mytextstyle \sum_{i=0}^{\Nb}\hat{b}_i(t)\vect\big( \vec{A}\of{i}\hvec{C}(t) \big)
=\vect\big( \hvec{A}(t)\hvec{C}(t) \big)
\label{eq:z**_multi} \\
\hvec{A}(t)
&\defn \mytextstyle \sum_{i=0}^{\Nb} \hat{b}_i(t) \vec{A}\of{i}
\label{eq:Ahat_multi} ,
\end{align}
where 
$[\vec{X}]_{:,j}$ denotes the $j$th column of $\vec{X}$ and
$\hvec{C}(t)\in\Complex^{N\times L}$ is a reshaping of $\hvec{c}(t)$. 
Note that \eqref{zij_multi}-\eqref{z*j_multi} follow directly from \eqref{z_multi} via the derivative interpretations \eqref{zmli0PB}-\eqref{zijPB}.


From the above expressions, it can be readily shown that
\begin{align}
\mytextstyle
\sum_{i=1}^{\Nb} \|\hvec{z}\of{i,*}(t)\|^2
&= \mytextstyle \sum_{i=1}^{\Nb}\big\|\vec{A}\of{i}\hvec{C}(t)\big\|_F^2 
= \tr\big(\vec{\Gamma}\hvec{C}(t)\hvec{C}(t)\herm\big)
\label{eq:sumzi*2_multi} \\
&
\mytextstyle
\sum_{j=1}^{\Nc} \|\hvec{z}\of{*,j}(t)\|^2
= L\|\hvec{A}(t)\|_F^2 
\label{eq:sumz*j2_multi} 
\end{align}
with pre-computed 
\begin{align}
\vec{\Gamma}
&\defn 
\mytextstyle 
\sum_{i=1}^{\Nb} \vec{A}\ofH{i}\vec{A}\of{i} 
\label{eq:Gamma_multi} .
\end{align}
\iftoggle{long}{
\color{\longcolor}
The following quantities can also be pre-computed:
\begin{align}
\sum_{i=1}^{\Nb} \|\vec{z}\of{i,j}\|^2 
&= \sum_{i=1}^{\Nb} \|\vec{a}_{\modulo{j-1}_N+1}\of{i}\|^2\\
\sum_{j=1}^{\Nc} \|\vec{z}\of{i,j}\|^2 
&= L\|\vec{A}\of{i}\|_F^2\\ 
\sum_{i=1}^{\Nb} \sum_{j=1}^{\Nc} \|\vec{z}\of{i,j}\|^2 
&= L\tr(\vec{\Gamma}) .
\end{align}
\color{black}
}
{}
Furthermore, under the scalar variance approximation, 
\begin{align}
\hvec{R}(t)
&= \big(1-\nu^r(t)\nu^s(t)\nu^b(t) \vec{D}^r\big) \hvec{C}(t) 
\non\\&\quad
   + \nu^r(t) \hvec{A}\herm(t)\hvec{S}(t) 
\label{eq:rhat_multi}\\
\hvec{q}(t)
&= \big(1-\nu^q(t)\nu^s(t)\nu^c(t)\vec{D}^q\big)\hvec{b}(t)
\non\\&\quad
   + \nu^q(t) \mat{\vect\big(\vec{A}\of{1}\hvec{C}(t)\big)\herm\\[-2mm]
                   \vdots\\
                   \vect\big(\vec{A}\of{\Nb}\hvec{C}(t)\big)\herm} 
                   \hvec{s}(t)
\label{eq:qhat_multi} ,
\end{align}
with the following pre-computed using 
$\vec{a}_n\of{i} \defn [\vec{A}\of{i}]_{:,n}$:
\begin{align}
\vec{D}^r 
&\defn \textstyle \diag\big\{\sum_{i=1}^{\Nb}\|\vec{a}\of{i}_1\|^2,\dots,\sum_{i=1}^{\Nb}\|\vec{a}\of{i}_N\|^2\big\} 
\label{eq:Dr_multi}\\
\vec{D}^q 
&\defn L \diag\big\{\|\vec{A}\of{1}\|_F^2,\dots,\|\vec{A}\of{\Nb}\|_F^2\} 
\label{eq:Dq_multi}
\end{align}
Note that \eqref{z**_multi}-\eqref{Dq_multi} specify the essential quantities needed for the implementation of scalar-variance P-BiG-AMP.
We discuss the complexity of these steps for two cases below.

First, suppose w.l.o.g.\ that each $\vec{A}\of{i}$ has $\Na\leq KN$ nonzero elements, with possibly different supports among $\{\vec{A}\of{i}\}$.
This implies that $\hvec{A}(t)$ has at most $\min(\Nb\Na,KN)$ nonzero elements.
It then follows that
\eqref{z**_multi} consumes $\min(\Nb\Na,KN)L$ multiplies,
\eqref{Ahat_multi} consumes $\Nb\Na$,
\eqref{sumzi*2_multi} consumes $L\min(\Nb(\Na+K),N^2)$ and
\eqref{sumz*j2_multi} consumes $\min(\Nb\Na,KN)$ multiplies. 
Furthermore, 
\eqref{rhat_multi} consumes $\approx \min(\Nb\Na,KN)L$ multiplies and 
\eqref{qhat_multi} consumes $\approx \Nb L(\Na+K)$.
In total, 
$\Ord(\min(\Nb\Na,KN)L
      +\Nb\Na L + \Nb KL
      +L\min(\Nb(\Na+K),N^2))$ 
multiplies are consumed.
For illustration, suppose that $\Nb\Na<KN$ and $\Nb\Na<N^2$.
Then $\Ord(NL+\Nb L(\Na+K))$ multiplies are consumed,
in contrast to $\Ord(M\Nb\Nc)=\Ord(KNL^2\Nb)$ for the general case.

Now suppose w.l.o.g.\ that, for a given $\vec{b}$, the multiplication of $\vec{A}(\vec{b})$ by a vector $\vec{x}$ can be accomplished implicitly using $\Na$ multiplies.
For example, $\Na=\Ord(N\log N)$ in the case of an FFT.
Then
\eqref{z**_multi} consumes $\Na L$ multiplies,
\eqref{sumzi*2_multi} consumes $KL$ (using $\{\vec{A}\of{i}\hvec{C}(t)\}$ computed for $\hvec{q}(t)$), and
\eqref{sumz*j2_multi} can be approximated using $\Ord(\Na)$ multiplies. 
Furthermore, 
\eqref{rhat_multi} consumes $\approx (N+\Na)L$ multiplies and 
\eqref{qhat_multi} consumes $\approx \Nb L(\Na+K)$.
In total, 
$\Ord(L(N + \Nb\Na+\Nb K))$ 
multiplies are consumed,
in contrast to $\Ord(M\Nb\Nc)=\Ord(KNL^2\Nb)$ for the general case.

\subsection{Low-Rank Structure} \label{sec:lr}

With low-rank signal structure, the noiseless outputs become
\begin{align}
z_m
&= \tr\big(\vec{\Phi}_m\herm\vec{B}\tran\vec{C}\big),~ m=1,\dots,M,
\label{eq:zm_lr} 
\end{align}
with known $\{\vec{\Phi}_m\}$, where
$\vec{B}\in\Complex^{N\times K}$,
$\vec{C}\in\Complex^{N\times L}$
for\footnote{When $N=1$, \eqref{zm_lr} reduces to the general parameterization \eqref{ZPB}.} $N>1$.
Thus we have $\vec{\Phi}_m\in\Complex^{K\times L}$, $\Nb=NK$, and $\Nc=NL$.
Defining $\vec{\phi}_m\defn\vect(\vec{\Phi}_m)$, 
$\vec{b}\defn\vect(\vec{B})$, and $\vec{c}\defn\vect(\vec{C})$, 
\begin{align}
z_m
&= \vec{\phi}_m\herm \vect(\vec{B}\tran\vec{C}) 
= \vec{b}\tran \big(\vec{\Phi}^*_m \otimes \vec{I}_N \big)\vec{c} 
\label{eq:zm_lr_bc}\\
&= \vect\big(\vec{B}\vec{\Phi}_m^*\big)\tran \vec{c} 
\label{eq:zm_lr_c}\\
&= \vect\big(\vec{C}\vec{\Phi}_m\herm\big)\tran \vec{b} 
\label{eq:zm_lr_b}
\end{align}
from which the derivative interpretations \eqref{zmli0PB}-\eqref{zijPB} imply
\begin{align}
&\vec{z}\of{i,j} 
\!=\! \mat{[\vec{\Phi}_1\otimes\vec{I}_N]_{i,j} \\[-2mm]
         \vdots\\
        [\vec{\Phi}_M\otimes\vec{I}_N]_{i,j} } \!\!,
~\hvec{z}\of{*,*}(t)
\!=\! \mat{\tr\big(\vec{\Phi}_1\herm\hvec{B}(t)\tran\hvec{C}(t)\big) \\[-2mm]
         \vdots\\
        \tr\big(\vec{\Phi}_M\herm\hvec{B}(t)\tran\hvec{C}(t)\big) } 
\label{eq:z**_lr}\\
&\hvec{z}\of{i,*}(t)
\!=\!\! \mat{ \vect\!\big(\hvec{C}(t)\vec{\Phi}_1\herm\big)\tran \\[-2mm]
         \vdots\\ 
         \vect\!\big(\hvec{C}(t)\vec{\Phi}_M\herm\big)\tran }_{:,i}\hspace{-4mm},
~\hvec{z}\of{*,j}(t)
\!=\!\! \mat{ \vect\!\big(\hvec{B}(t)\vec{\Phi}_1^*\big)\tran \\[-2mm]
         \vdots\\
         \vect\!\big(\hvec{B}(t)\vec{\Phi}_M^*\big)\tran }_{:,j}
\label{eq:zi**j_lr} \hspace{-4mm}.
\\[-8mm]\non
\end{align}

From the above expressions, it can be readily shown that
\begin{subequations}
\begin{align}
\mytextstyle
\sum_{i=1}^{\Nb} \|\hvec{z}\of{i,*}(t)\|^2
&= \mytextstyle \sum_{m=1}^M \|\vec{\Phi}_m\hvec{C}(t)\herm\|_F^2
\label{eq:sumzi*2_lr1} \\
&= \tr\big(\vec{\Gamma}_1\hvec{C}(t)\herm\hvec{C}(t)\big)
\label{eq:sumzi*2_lr2} 
\end{align}
\label{eq:sumzi*2_lr} 
\end{subequations}
\vspace*{-\belowdisplayshortskip}
\vspace*{-\belowdisplayshortskip}
\begin{subequations}
\begin{align}
\mytextstyle
\sum_{j=1}^{\Nc} \|\hvec{z}\of{*,j}(t)\|^2
&= \mytextstyle \sum_{m=1}^M \|\hvec{B}(t)^*\vec{\Phi}_m\|_F^2
\label{eq:sumz*j2_lr1} \\
&= \tr\big(\vec{\Gamma}_2\hvec{B}(t)\tran\hvec{B}(t)^*\big)
\label{eq:sumz*j2_lr2} 
\end{align}
\label{eq:sumz*j2_lr} 
\end{subequations}
with pre-computed 
\begin{align}
\mytextstyle 
\vec{\Gamma}_1
\defn \sum_{m=1}^{M} \vec{\Phi}_m\herm\vec{\Phi}_m,
&\quad \mytextstyle
\vec{\Gamma}_2
\defn \sum_{m=1}^{M} \vec{\Phi}_m\vec{\Phi}_m\herm  
\label{eq:Gamma_r} .
\end{align}
\iftoggle{long}{
\color{\longcolor}
The following quantities can also be pre-computed:
\begin{align}
\sum_{i=1}^{\Nb} \|\vec{z}\of{i,j}\|^2 
&= [\vec{\Gamma}_1\otimes \vec{I}_N]_{jj} \\
\sum_{j=1}^{\Nc} \|\vec{z}\of{i,j}\|^2 
&= [\vec{\Gamma}_2\otimes \vec{I}_N]_{ii} \\
\sum_{i=1}^{\Nb} \sum_{j=1}^{\Nc} \|\vec{z}\of{i,j}\|^2 
&= N\tr(\vec{\Gamma}_1) = N\tr(\vec{\Gamma}_2) .
\end{align}
\color{black}
}
{}
Furthermore, under the scalar variance approximation,
\iftoggle{long}{
\color{\longcolor}
\begin{align}
\hvec{r}(t)
&= \big(1-\nu^r(t)\nu^s(t)\nu^b(t) 
   [\Diag\vec{\Gamma}_1\otimes \vec{I}_N]\big) \hvec{c}(t) 
\\&\quad
   + \nu^r(t) 
     \mat{ \vect\big(\hvec{B}(t)^*\vec{\Phi}_1\big) &
           ,\dots, &
           \vect\big(\hvec{B}(t)^*\vec{\Phi}_M\big) } \hvec{s}(t) 
\non\\
\hvec{q}(t)
&= \big(1-\nu^q(t)\nu^s(t)\nu^c(t)
   [\Diag\vec{\Gamma}_2\otimes \vec{I}_N]\big) \hvec{b}(t)
\\&\quad
   + \nu^q(t) 
     \mat{ \vect\big(\hvec{C}(t)^*\vec{\Phi}_1\tran\big) &
           ,\dots, &
           \vect\big(\hvec{C}(t)^*\vec{\Phi}_M\tran\big) } \hvec{s}(t) 
\non
\end{align}
and so
\color{black}
}
{}
\begin{align}
\hvec{R}(t)
&= \hvec{C}(t) \big(\vec{I}_L-\nu^r(t)\nu^s(t)\nu^b(t)\Diag\vec{\Gamma}_1\big)
\non\\&\quad
\mytextstyle
   + \nu^r(t) \hvec{B}(t)^*\left(\sum_{m=1}^M \hat{s}_m(t)\vec{\Phi}_m\right)
\label{eq:rhat_lr}\\
\hvec{Q}(t)
&= \hvec{B}(t)\big(\vec{I}_K-\nu^q(t)\nu^s(t)\nu^c(t)\Diag\vec{\Gamma}_2\big) 
\non\\&\quad
\mytextstyle
   + \nu^q(t) \hvec{C}(t)^*\left(\sum_{m=1}^M \hat{s}_m(t)\vec{\Phi}_m\tran\right) .
\label{eq:qhat_lr}
\end{align}
Note that \eqref{z**_lr}-\eqref{qhat_lr} specify the essential quantities needed for the implementation of scalar-variance P-BiG-AMP.
We discuss the complexity of these steps below.


Suppose w.l.o.g.\ that $\vec{\Phi}_m$ has $\Nphi\leq KL$ nonzero entries, with possibly different supports among $\{\vec{\Phi}_m\}$.
This implies that $\sum_m \hat{s}_m(t)\vec{\Phi}_m$ has at most $\min(KL,M\Nphi)$ nonzero elements.
It then follows that 
$\hvec{z}\of{*,*}(t)$ from \eqref{z**_lr} consumes $NKL+M\Nphi$ multiplies,
\eqref{sumzi*2_lr} consumes $\approx N\min\{L^2,M(\Nphi+K)\}$, and
\eqref{sumz*j2_lr} consumes $\approx N\min\{K^2,M(\Nphi+L)\}$.
Furthermore, 
\eqref{rhat_lr} consumes $NL+N\min(KL,M\Nphi)+M\Nphi$ multiplies and 
\eqref{qhat_lr} consumes $NK+N\min(KL,M\Nphi)$.
In total, 
$\Ord(N\min(L^2,M(\Nphi+K))
     +N\min(K^2,M(\Nphi+L))
     +NKL
     +M\Nphi)$ 
multiplies are consumed. 
For illustration, suppose that $\Nphi<K,L$ and $M<K,L$.
Then $\Ord(NKL)$ multiplies are consumed, in contrast to $\Ord(M\Nb\Nc)=\Ord(MN^2KL)$ in the general case.

\subsection{Matrix-product Structure} \label{sec:triv}

A special case of \eqref{Z_multi} and \eqref{zm_lr} is when
\begin{align}
\vec{Z} 
&= \vec{B}\vec{C}
\label{eq:Z_triv}
\end{align}
which occurs, e.g., in applications such as MC, RPCA, DL, and NMF, as discussed in \secref{motiv}.
In particular, \eqref{Z_multi} reduces to \eqref{Z_triv} when $\Nb=KN$ and $\vect(\vec{A}\of{i})=[\vec{I}]_{:,i}$, and \eqref{zm_lr} reduces to \eqref{Z_triv} when $M=KL$ and $\vect(\vec{\Phi}_m)=[\vec{I}]_{:,m}$.
It can be verified \cite{Parker:Diss:14} that, under \eqref{Z_triv}, P-BiG-AMP reduces to BiG-AMP from \cite{Parker:TSP:14a}.

\subsection{Low-Rank plus Sparse Structure} \label{sec:lrps}

Recall \eqref{MCS}, the problem of recovering a ``low-rank plus sparse'' matrix.
Writing the low-rank component as $\vec{L}=\vec{B}\tran\vec{C}_1$ with $\vec{B}\in\Complex^{N\times K}$, $\vec{C}_1\in\Complex^{N\times L}$, and $N<\min\{K,L\}$, we can invoke \eqref{zm_lr_bc} to get
\begin{align}
z_m
&= \vec{b}\tran\big(\vec{\Phi}^*_m\otimes\vec{I}_N\big)\vec{c}_1 + \vec{\phi}_m\herm\vec{c}_2,~ m=1,\dots,M,
\label{eq:zm_lrs} 
\end{align}
with 
$b_0\defn 1$ (recall \secref{approach}), $\vec{b}\defn\vect(\vec{B})$, $\vec{c}_1\defn\vect(\vec{C})$, $\vec{c}_2\defn\vect(\vec{S})$ (recall $\vec{S}$ was the sparse matrix from \eqref{MCS}), and $\vec{c}=[\vec{c}_1\tran,\vec{c}_2\tran]\tran$.

Note that the structure of the first term of \eqref{zm_lrs} can be exploited through \eqref{z**_lr}-\eqref{zi**j_lr}, as discussed in \secref{lr}. 
Meanwhile, straightforward computational simplifications of the second term in \eqref{zm_lrs} result when $\vec{\phi}_m\herm$ is sparse.
But care must be taken in applying the scalar-variance approximation in this case: it may be advantageous to use different scalar variances for $\vC_1$ and $\vC_2$ (e.g., $\nu^r_1,\nu^c_1$ and $\nu^r_2,\nu^c_2$).

\section{Numerical Experiments}   \label{sec:numerical}
We now present the results of several numerical experiments that test the performance of P-BiG-AMP and EM-P-BiG-AMP in various applications.
In most cases, we quantify recovery performance using
NMSE$(\hvec{b})\defn \norm{\vec{b}-\hvec{b}}_2^2/\norm{\vec{b}}_2^2$ and
NMSE$(\hvec{c})\defn \norm{\vec{c}-\hvec{c}}_2^2/\norm{\vec{c}}_2^2$.
Matlab code for P-BiG-AMP and EM-P-BiG-AMP can be found in~\cite{GAMPmatlab}.

\subsection{I.i.d.\ Gaussian Model} \label{sec:affine}

First, we examine the performance of P-BiG-AMP in the case of i.i.d.\ Gaussian $z_m\of{i,j}$, as assumed for its derivation. 
In particular, 
$\{z_m\of{i,j}\}$ were drawn i.i.d.\ $\CNor(0,1)$, 
$\vec{b}=[b_1,\dots,b_{\Nb}]\tran$ were drawn Bernoulli-$\CNor(0,1)$ with sparsity rate $\xi^b$,
and
$\vec{c}=[c_1,\dots,c_{\Nc}]\tran$ were drawn Bernoulli-$\CNor(0,\nu^c)$ with sparsity rate $\xi^c$.
We then attempted to recover $\vec{b}$ and $\vec{c}$ from $M$ noiseless measurements of the form \eqref{ZPB} under $b_0=0$ and $c_0=0$.
For our experiment, we used $\Nb=\Nc=100$ and $\nu^c=1$, and we varied both the sparsity rate $\xi^b=\xi^c=K/100$ and the number of measurements $M$.


We tested the performance of both P-BiG-AMP, which assumed oracle knowledge of all distributional parameters, and EM-P-BiG-AMP, which estimated the parameters $\vec{\theta} \defn [\nu^c, \xi^b, \xi^c]\tran$ as well as the additive white Gaussian noise (AWGN) variance.\footnote{EM-P-BiG-AMP was not told that the measurements were noiseless.\label{foot:noiseless}}
\Figref{bilinear/phase_P-BiG-AMP} shows the empirical success rate for both algorithms, averaged over $50$ independent problem realizations, as a function of the sparsity $K$ and the number of measurements $M$. 
Here, we declare a ``success'' when both NMSE$(\hvec{b})<-60$~dB and NMSE$(\hvec{c})<-60$~dB.
The figure shows that both P-BiG-AMP and EM-P-BiG-AMP gave sharp phase transitions.
Moreover, their phase transitions are very close to the counting bound ``$M \ge 2K$,'' shown by the red line in \figref{bilinear/phase_P-BiG-AMP}. 

\twoFrag{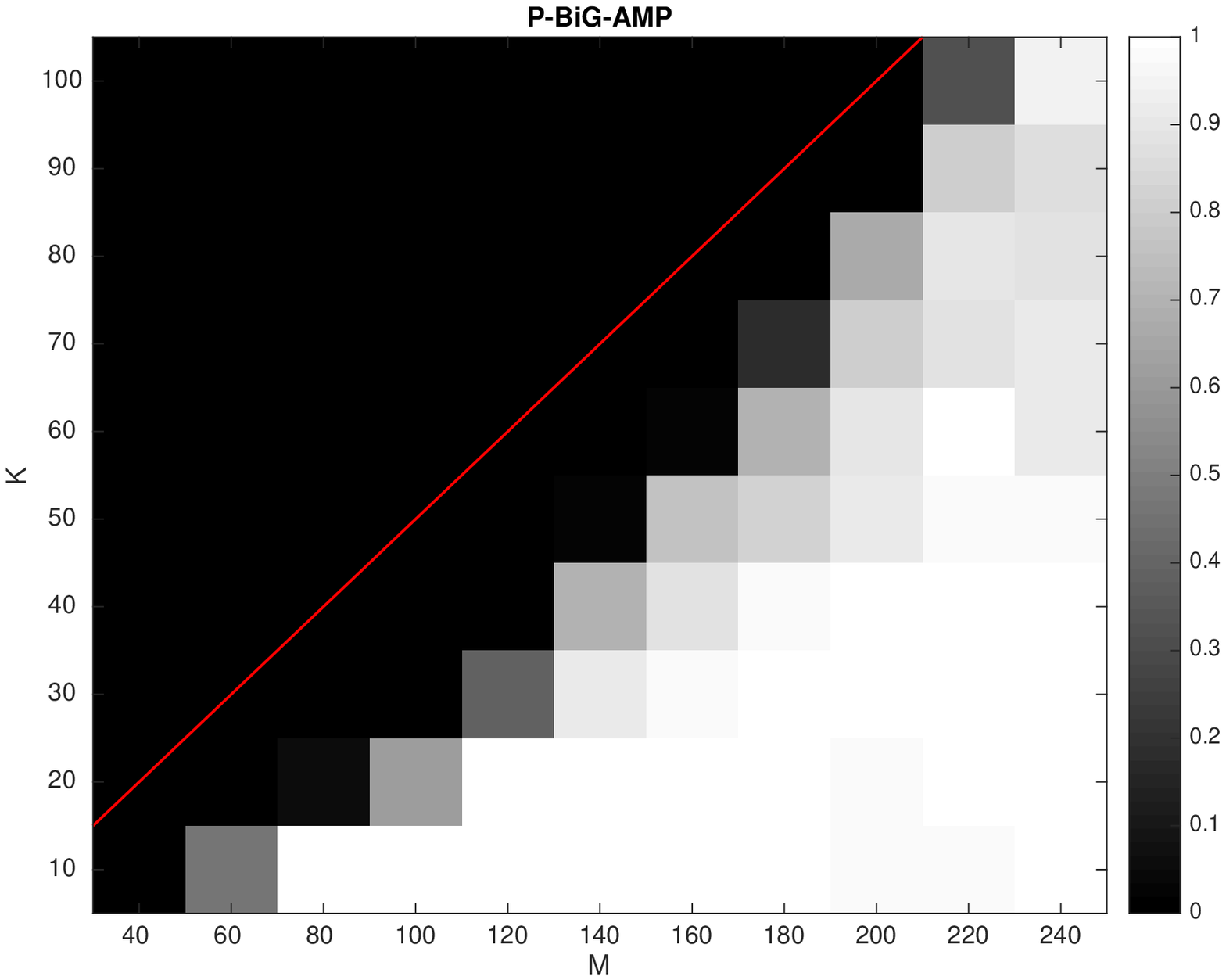}{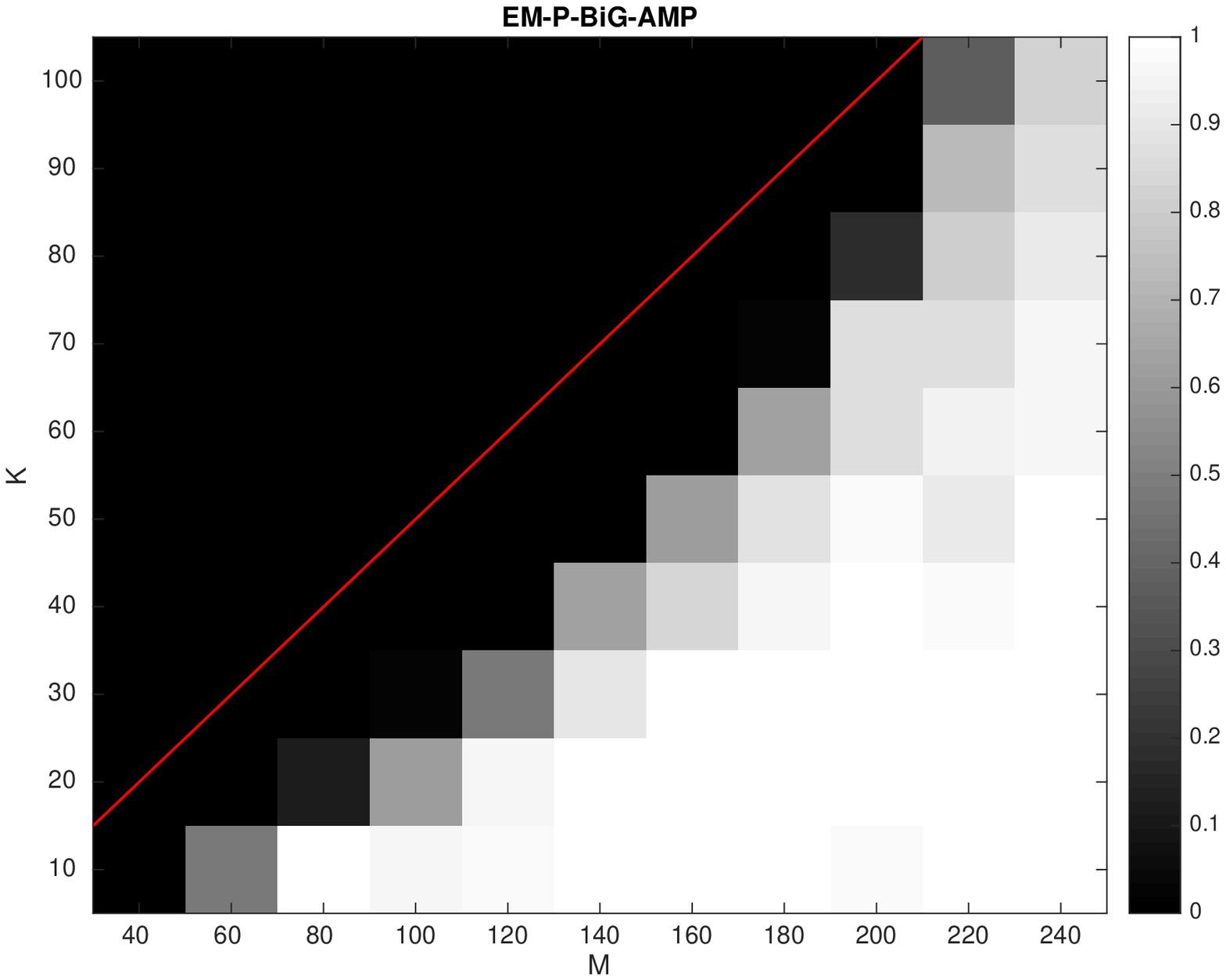}
{Empirical success rate for noiseless sparse signal recovery under the i.i.d.\ parametric bilinear model \eqref{ZPB} as a function of the number of measurements $M$ and the signal sparsity $K$.  Success rates were averaged over $50$ independent realizations.  Points above the red curve are infeasible due to counting bound, as described in the text.}
{
\psfrag{M}[t][][0.8]{\sf measurements $M$}
\psfrag{K}[][][0.8]{\sf sparsity $K$}
\psfrag{P-BiG-AMP}[b][B][0.8]{\sf P-BiG-AMP}
\psfrag{EM-P-BiG-AMP}[b][B][0.8]{\sf EM-P-BiG-AMP}
}

\subsection{Self Calibration}

We now consider the self calibration problem described in \secref{motiv}.
In particular, we consider the noiseless single measurement vector (SMV) version, where the goal is to jointly recover 
the $K$-sparse signal $\vec{c} \in \Real^{\Nc}$ 
and calibration parameters $\vec{b} \in \Real^{\Nb}$ 
from $M$ noiseless measurements of the form 
$\vec{z} = \Diag(\vec{Hb})\vec{Ac}$, where $\vec{H}$ and $\vec{A}$ are known.  
For our experiment, we mimic the setup used for \cite[Figure 1]{Ling:IP:15}. 
Thus, we set $\Nc = 256$ and $M = 128$,
we chose $\vec{H}$ as the first $\Nb$ columns of a $M$-point unitary DFT matrix,
and we drew the entries of $\vec{A}$ as i.i.d.\ $\Nor(0,1)$. 
Furthermore, we drew $K$-sparse $\vec{c}$ with i.i.d.\ $\Nor(0,\nu^c)$ non-zero elements chosen uniformly at random, and we drew $\vec{b}$ as i.i.d.\ $\Nor(0,1)$.

We compared the performance of EM-P-BiG-AMP to SparseLift~\cite{Ling:IP:15}, a recently proposed convex relaxation, using CVX for the implementation.
EM-P-BiG-AMP modeled $\vec{c}$ as Bernoulli-$\Nor(0,\nu^c)$ and learned $\nu^c$, the sparsity rate $\xi$, and the AWGN variance.\footnote{See footnote~\ref{foot:noiseless}.}
\Figref{calibration/phase_EM-P-BiG-AMP} shows empirical success rate as a function of signal sparsity $K$ and number of calibration parameters $\Nb$.
As in \cite{Ling:IP:15}, we considered NMSE~$\defn \|\vec{bc}\tran-\hvec{b}\hvec{c}\tran\|_F^2/\|\vec{bc}\tran\|_F^2$, and we declared ``success'' when NMSE~$<-60$~dB.
\Figref{calibration/phase_EM-P-BiG-AMP} shows that EM-P-BiG-AMP's success region was much larger than SparseLift's,\footnote{The SparseLift results in \figref{calibration/phase_EM-P-BiG-AMP} agree with those in \cite[Figure 1]{Ling:IP:15}.}
although it was not close to the counting bound $M \ge \Nb + K$, which lives just outside the boundaries of the figure.
Still, the shape of EM-P-BiG-AMP's empirical phase-transition suggests successful recovery when $M\gtrsim \alpha_1 (\Nb+K)$ for some $\alpha_1$, in contrast with SparseLift's empirical and theoretical \cite{Ling:IP:15} success condition of $M\gtrsim \alpha_2 \Nb K$ for some $\alpha_2$.

\twoFrag{calibration/phase_EM-P-BiG-AMP}{calibration/phase_SparseLift}
{Empirical success rate for \emph{noiseless self-calibration} as a function of the number of calibration parameters $\Nb$ and the signal sparsity $K$.
Results are averaged over $10$ independent realizations.}{
\psfrag{Nb}[b][b][0.8]{\sf \# of parameters $\Nb$}
\psfrag{K}[t][t][0.8]{\sf signal sparsity $K$}
\psfrag{SparseLift}[b][b][0.8]{\sf SparseLift}
\psfrag{EM-P-BiG-AMP}[b][b][0.8]{\sf EM-P-BiG-AMP}
}

\subsection{Noisy CS with Parametric Matrix Uncertainty} \label{sec:STLS}

Next we consider noisy compressive sensing with parametric matrix uncertainty, as described in \secref{motiv}.
Our goal is to recover a single, $K$-sparse, $\Nc$-length signal $\vec{c}$ from measurements $\vec{y}=(\vec{A}\of{0}+\sum_{i=1}^{\Nb} b_i\vec{A}\of{i}) \vec{c} + \vec{w}\in\Real^M$, where $\vec{b}=[b_1,...,b_{\Nb}]\tran$ are unknown calibration parameters and $\vec{w}$ is AWGN.
For our experiment, $\Nc=256$, $K=10$,
$\vec{c}$ had i.i.d.\ $\Nor(0,\nu^c)$ non-zero elements chosen uniformly at random with $\nu^c=1$,
$\vec{b}$ was i.i.d.\ $\mc{N}(0,\nu^b)$ with $\nu^b=1$,
$\vec{A}\of{0}$ was i.i.d.\ $\mc{N}(0,10)$,
and 
$\{\vec{A}\of{i}\}_{i=1}^{10}$ was i.i.d.\ $\mc{N}(0,1)$. 
The noise variance $\nu^w$ was adjusted to achieve an SNR $\defn\|\vec{y}-\vec{w}\|_2^2/\|\vec{w}\|_2^2$ of $40$~dB.

We compared P-BiG-AMP and EM-P-BiG-AMP to i) the MMSE oracle that knows $\vec{c}$, ii) the MMSE oracle that knows $\vec{b}$ and support($\vec{c}$), and iii) the WSS-TLS approach from \cite{Zhu:TSP:11}%
, which aims to solve the non-convex optimization problem
\begin{align}
(\hvec{b},\hvec{c}) 
&= \argmin_{\vec{b},\vec{c}} 
\Bigg\|\Big( \vec{A}\of{0} + \sum_{\iB = 1}^{\Nb} \Br_{\iB} \vec{A}\of{\iB} \Big)\vCr - \vec{y}\Bigg\|_2^2
\nonumber\\&\quad
+ \nu^w \norm{\vBr}_2^2 + \lambda \norm{\vCr}_1 
\end{align}
via alternating minimization.
For WSS-TLS, we used oracle knowledge of $\nu^w$, oracle tuning of the regularization parameter $\lambda$, and code from the authors' website (with a trivial modification to facilitate arbitrary $\vec{A}\of{i}$).  
P-BiG-AMP used a Bernoulli-Gaussian prior with sparsity rate $\xi=K/\Nc$ and perfect knowledge of $\nu^c$ and $\nu^w$, whereas EM-P-BiG-AMP learned the statistics $[\xi,\nu^c,\nu^w]\tran\defn \vec{\theta}$ from the observed data.
\Figref{trivial_slice_B_errors} shows that, for estimation of both $\vec{b}$ and $\vec{c}$, P-BiG-AMP gave near-oracle NMSE performance for $M/N\geq 0.2$.
Meanwhile, EM-P-BiG-AMP performed only slightly worse than P-BiG-AMP.
In contrast, the NMSE performance of WSS-TLS was about $10$~dB worse than P-BiG-AMP, and its ``phase transition'' occurred later, at $M/N=0.3$.

\twoFrag{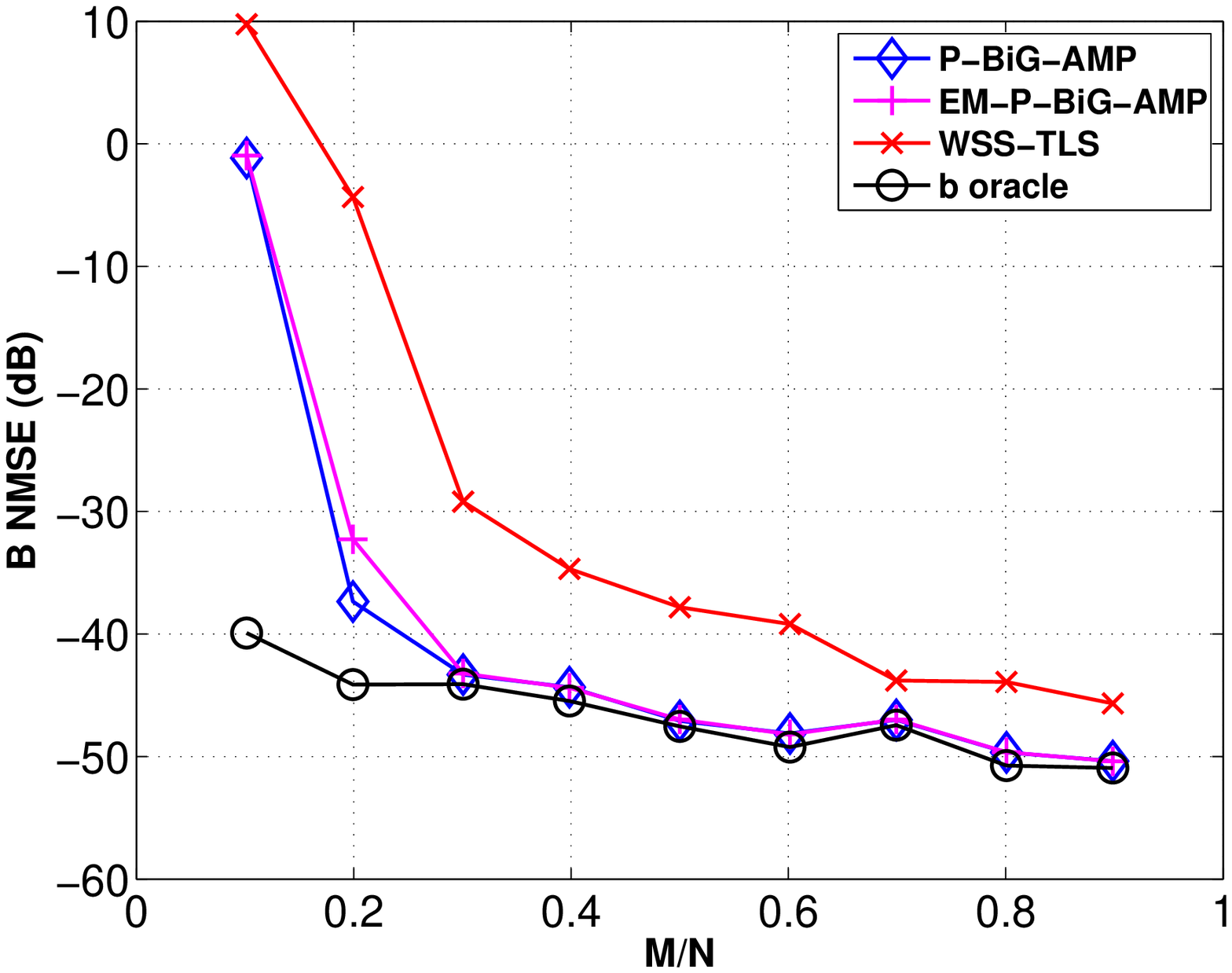}{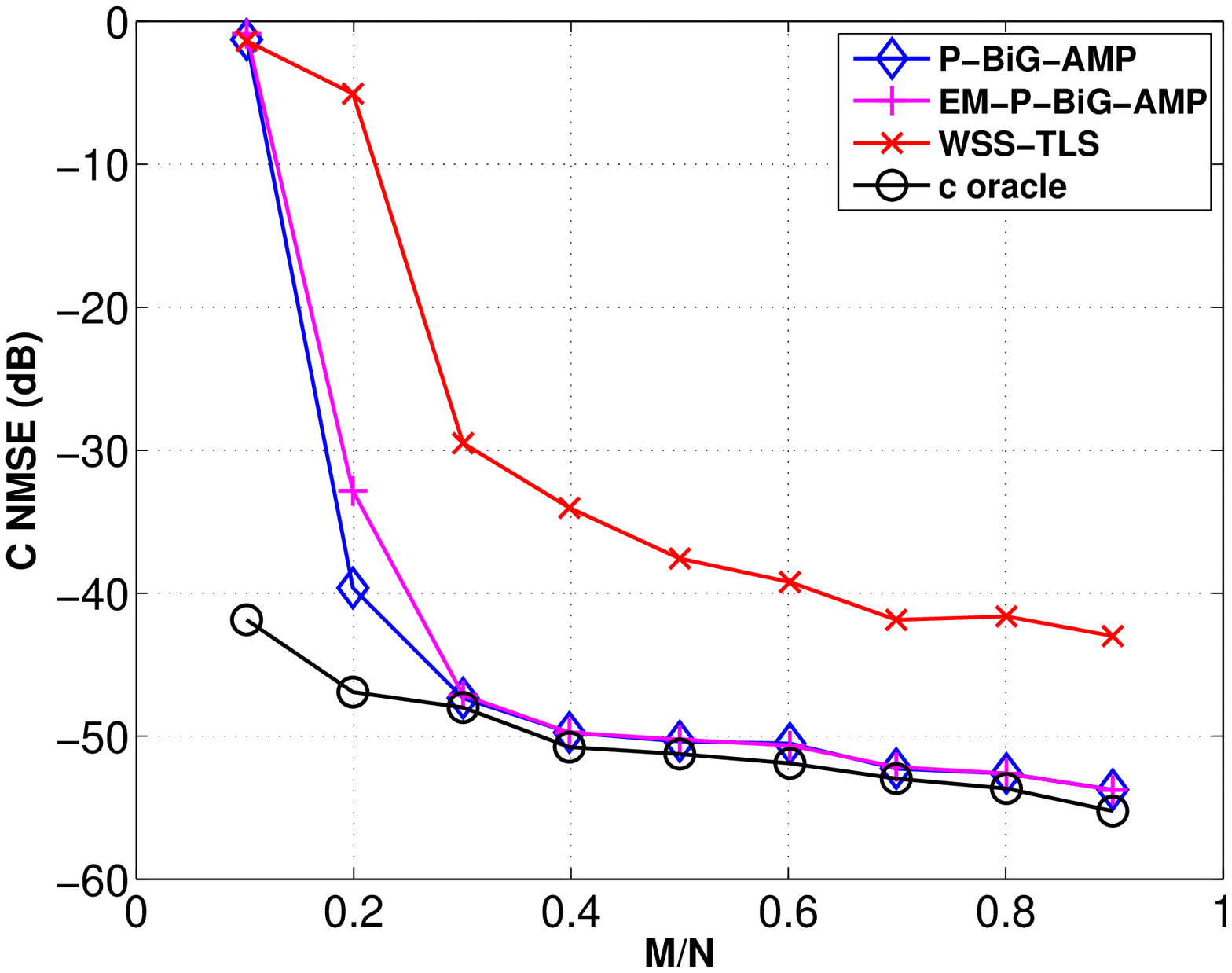}
{Parameter estimation NMSE (left) and signal estimation NMSE (right) versus sampling ratio $M/N$ for \emph{CS with parametric matrix uncertainty}.  Results are averaged over $10$ independent realizations.}
{
\psfrag{M/N}[t][t][0.8]{\sf sampling ratio $M/N$}
\psfrag{B NMSE (dB)}[b][b][0.8]{\sf NMSE ($\hvec{b}$) [dB]}
\psfrag{C NMSE (dB)}[b][b][0.8]{\sf NMSE ($\hvec{c}$) [dB]}
\psfrag{P-BiG-AMP}[l][l][0.4]{\sf PBiGAMP}
\psfrag{EM-P-BiG-AMP}[l][l][0.4]{\sf EM-PBiGAMP}
\psfrag{WSS-TLS}[l][l][0.4]{\sf WSS-TLS}
\psfrag{b oracle}[l][l][0.4]{\sf oracle}
\psfrag{c oracle}[l][l][0.4]{\sf oracle}
}

\subsection{Totally Blind Deconvolution} \label{sec:blind}

We now consider recovering an unknown signal $c_i$ and channel $b_i$ from noisy observations $y_i=z_i+w_i$ of their \emph{linear} convolution \textb{$\{z_i\}=\{b_i\}\ast \{c_i\}$}, where $w_i\sim \text{i.i.d.~}\mc{N}(0,\nu^w)$.
In particular, we consider the case of ``totally blind deconvolution'' from \cite{Manton:SCL:03}, where the signal contains zero-valued guard intervals of duration $\Ng \geq \Nb-1$ and period $\Np>\Ng$, guaranteeing identifiability.
Recalling the discussion of \emph{joint channel-symbol estimation} in \secref{motiv}, we see that a zero-valued guard allows the convolution outputs to be organized as $\vec{Z}=\Conv(\vec{b})\vec{C}$, where $\Conv(\vec{b})\in\Real^{\Np\times (\Np-\Ng)}$ is the linear convolution matrix with first column $\vec{b}$. 
For our experiment, we used an i.i.d.\ $\CNor(0,1)$ channel $\vec{b}$, and two cases of i.i.d.\ signal $\vec{c}$: 
Gaussian $c_j\sim \CNor(0,1)$ 
and 
equiprobable QPSK (i.e., $c_j\in\{1,j,-1,-j\}$).
Also, we used guard period $\Np=256$, guard duration $\Ng=64$, channel length $\Nb=63$, and $L=3$ signal periods.  

\begin{figure}[tp]
 \newcommand{\sz}{0.53}
 \psfrag{h_nmsedB}[t][t][0.8]{\sf NMSE($\hvec{b}$) [dB]} 
 \psfrag{s_nmsedB}[b][b][0.8]{\sf NMSE($\hvec{c}$) [dB]}
 \psfrag{SER}[][b][0.8]{\sf SER($\hvec{c}$)}
 \psfrag{cross_relation_CG}[l][l][\sz]{\sf cross-relation Gauss}
 \psfrag{cross_relation_FA}[l][l][\sz]{\sf cross-relation QPSK}
 \psfrag{PBiGAMP_CG}[l][l][\sz]{\sf \hspace{-1.3mm} P-BiG-AMP Gauss}
 \psfrag{PBiGAMP_FA}[l][l][\sz]{\sf \hspace{-1.3mm} P-BiG-AMP QPSK}
 \psfrag{genie_CG}[l][l][\sz]{\sf oracle Gauss}
 \psfrag{genie_FA}[l][l][\sz]{\sf oracle QPSK}
 \psfrag{SNR dB}[t][t][0.8]{\sf SNR [dB]}
 \begin{center}
   \iftoggle{long}{ 
     \includegraphics[width=3.5in, trim=-30bp 0 -23bp 0,clip]{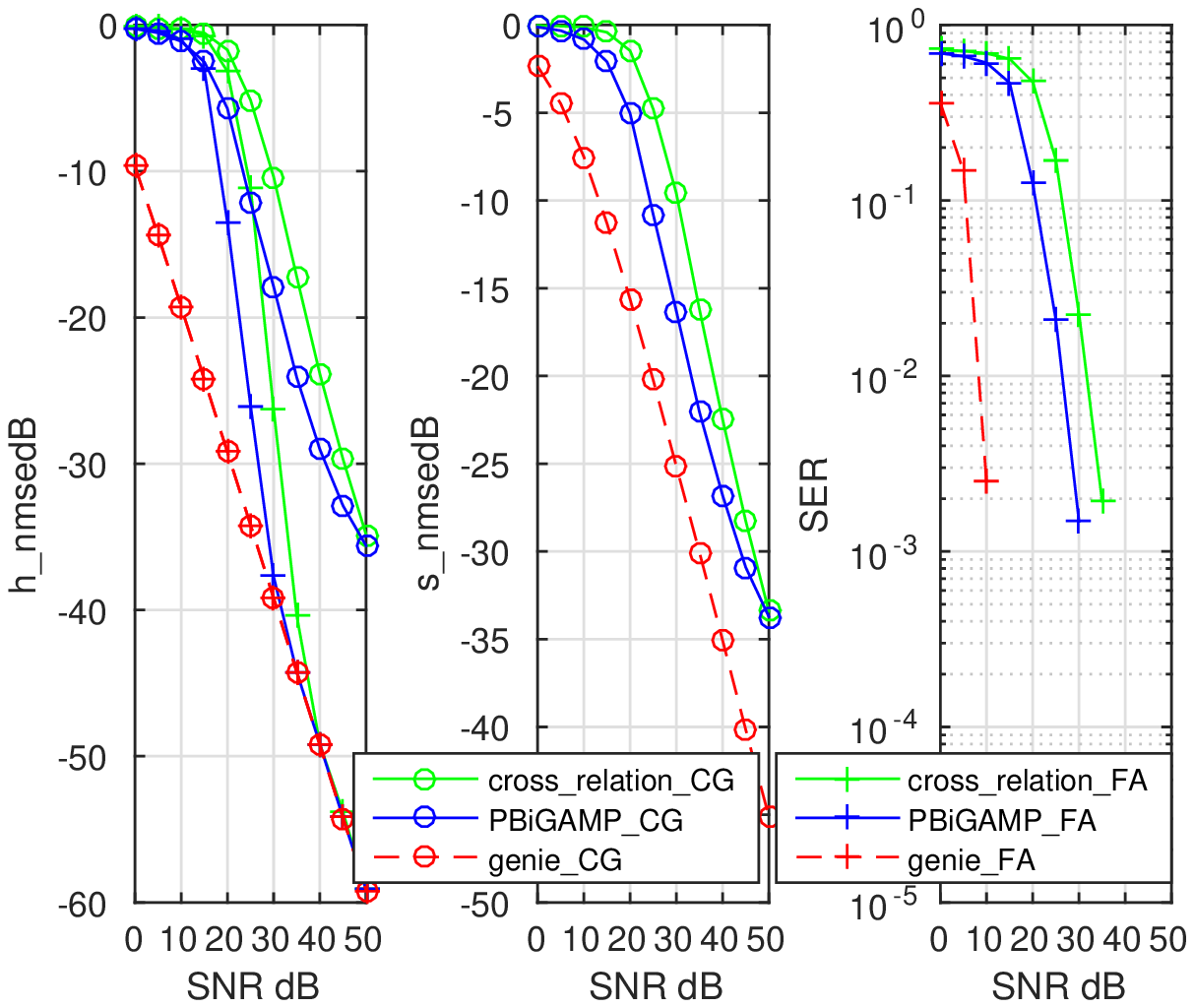}
   }{ 
     \includegraphics[width=\figsize in]{figures/blind/blind.eps}
   } 
 \end{center}
 \caption{Channel estimation NMSE (left), Gaussian-symbol estimation 
        NMSE (center), and QPSK symbol error rate (right) versus SNR 
        for \emph{totally blind deconvolution}.  Results are averaged 
        over $500$ independent realizations.}
 \label{fig:blind/blind}
\end{figure}

We compared P-BiG-AMP to i) the known-symbol and known-channel MMSE oracles and ii) the cross-relation (CR) method \cite{Hua:TSP:96}, which is known to perform close to the Cramer-Rao lower bound \cite{Hua:TSP:96}.
In particular, we used CR for blind symbol estimation, then (in the QPSK case) de-rotated and quantized the blind symbol estimates, and finally performed maximum-likelihood channel estimation assuming perfect (quantized) symbols.
\Figref{blind/blind} shows that, with both Gaussian and QPSK symbols, P-BiG-AMP outperformed the CR method by about $5$~dB in the SNR domain.
Moreover, by exploiting the QPSK constellation, both methods were able to achieve oracle-grade NMSE$(\hvec{b})$ at high SNR.

\subsection{Matrix Compressive Sensing}

Finally, we consider the problem of \emph{matrix compressive sensing}, as described in \secref{motiv} and further discussed in \secref{lrps}. 
Our goal was to jointly recover a low rank matrix $\vec{L} = \vec{B}\tran\vec{C}_1 \in \Complex^{100\times 100}$ and a sparse outlier matrix $\vec{S} = \vec{C}_2 \in \Complex^{100\times 100}$ from $M$ noiseless linear measurements of their sum, i.e., $\{z_m\}_{m=1}^M$ in \eqref{MCS}.
For our experiment, the sparse outliers were drawn with amplitudes uniformly distributed on $[-10,10]$ and uniform random phases, similar to \cite[Figure 2]{Wright:II:13}. 
But unlike \cite[Figure 2]{Wright:II:13}, the sensing matrices $\{\vec{\Phi}_m\}$ were sparse, with $K=50$ i.i.d.\ $\CNor(0,1)$ non-zero entries drawn uniformly at random.

We compare the recovery performance of EM-P-BiG-AMP to the convex formulation known as \emph{compressive principal components pursuit} (CPCP) \cite{Wright:II:13}, i.e.,
\begin{align}
\argmin_{\vec{L},\vec{S}} \norm{\vec{L}}_* + \lambda\norm{\vec{S}}_1 \text{ s.t. } z_m = \tr\{\vec{\Phi}_m\tran (\vec{L}+\vec{S})\}~\forall m,
\label{eq:CPCP}
\end{align}
which we solved with TFOCS using a continuation scheme.
In accordance with \cite[Theorem~2.1]{Wright:II:13}, we used $\lambda = 1/10$ in \eqref{CPCP}. 
EM-P-BiG-AMP learned 
the variance of the entries in $\vec{C}_1$, 
the sparsity and non-zero variance of $\vec{C}_2$, 
and the additive AWGN variance.\footnote{See footnote~\ref{foot:noiseless}.} 
Although EM-P-BiG-AMP was given knowledge of the true rank $R$, we note that an unknown rank could be accurately estimated using the scheme proposed for BiG-AMP in \cite[Sec.~V-B2]{Parker:TSP:14a} and tested for the RPCA application in \cite[Sec.~III-F2]{Parker:TSP:14b}.

\begin{figure}[tp]
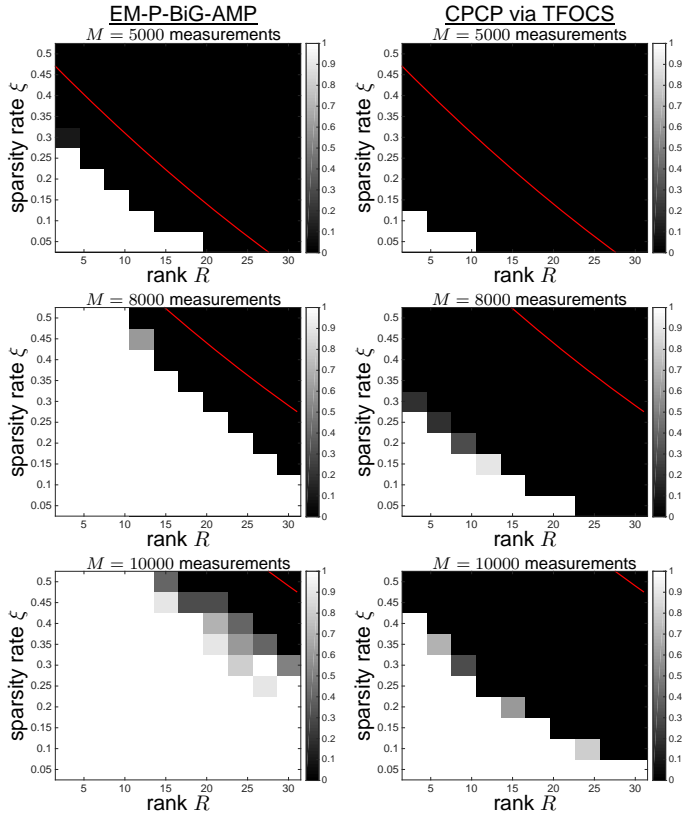

\centering
\psfrag{Lam}[b][b][0.8]{\sf sparsity rate $\xi$}
\psfrag{R}[t][t][0.8]{\sf rank $R$}
\psfrag{TFOCS5}[b][b][0.8]{\sf \begin{tabular}{c}\underline{CPCP via TFOCS}\\[-1mm]\footnotesize $M=5000$ measurements\\[-2mm]\end{tabular}}
\psfrag{TFOCS8}[][B][0.8]{\sf \footnotesize $M=8000$ measurements}
\psfrag{TFOCS10}[][B][0.8]{\sf \footnotesize $M=10000$ measurements}
\psfrag{EM-P-BiG-AMP5}[b][b][0.8]{\sf \begin{tabular}{c}\underline{EM-P-BiG-AMP}\\[-1mm]\footnotesize $M=5000$ measurements\\[-2mm]\end{tabular}}
\psfrag{EM-P-BiG-AMP8}[][B][0.8]{\sf \footnotesize $M=8000$ measurements}
\psfrag{EM-P-BiG-AMP10}[][B][0.8]{\sf \footnotesize $M=10000$ measurements}
\vspace{4mm}
\begin{tabular}{cccc}
\includegraphics[width=1.65in]{figures/lrps/phase_EM-P-BiG-AMP_M_5000.eps}&
\includegraphics[width=1.65in]{figures/lrps/phase_TFOCS_M_5000.eps}\\
\includegraphics[width=1.65in]{figures/lrps/phase_EM-P-BiG-AMP_M_8000.eps}&
\includegraphics[width=1.65in]{figures/lrps/phase_TFOCS_M_8000.eps}\\
\includegraphics[width=1.65in]{figures/lrps/phase_EM-P-BiG-AMP_M_10000.eps}&
\includegraphics[width=1.65in]{figures/lrps/phase_TFOCS_M_10000.eps}
\end{tabular}
\caption{Empirical success rate for noiseless \emph{matrix compressive sensing} as a function of rank $R$ and outlier sparsity rate $\xi$ for $M = 5000$ (top), $M = 8000$ (middle), and $M = 10000$ (bottom) measurements.
The left column shows EM-P-BiG-AMP and the right column shows CPCP solved using TFOCS.
All results are averaged over $10$ independent realizations. 
Points above the red curve are infeasible due to the counting bound, as described in the text.}
 \label{fig:phaseMatrixCS}
\end{figure}

\Figref{phaseMatrixCS} shows the empirical success rate 
of EM-P-BiG-AMP and CPCP versus  $R$ (i.e., the rank of $\vec{L}$) and $\xi=K/100^2$ (i.e., the sparsity rate of $\vec{S}$) for three fixed values of $M$ (i.e., the number of measurements).
Each point is the average of $10$ independent trials, with success defined as $\|\vec{L}-\hvec{L}\|_F^2/\|\vec{L}\|_F^2 < -60$~dB.
\Figref{phaseMatrixCS} shows that, for the three tested values of $M$, EM-P-BiG-AMP exhibited a sharp phase-transition that was significantly better than that of CPCP.\footnote{The CPCP results in \figref{phaseMatrixCS} are in close agreement with those in \cite[Figure 2]{Wright:II:13}, even though the latter correspond to real-valued and dense $\vec{\Phi}_m$.} 
In fact, EM-P-BiG-AMP's phase transition is not far from the counting bound $M \ge R(200 - R) + \xi 100^2$, shown by the red curves in \figref{phaseMatrixCS}.

\begin{figure}[tp]
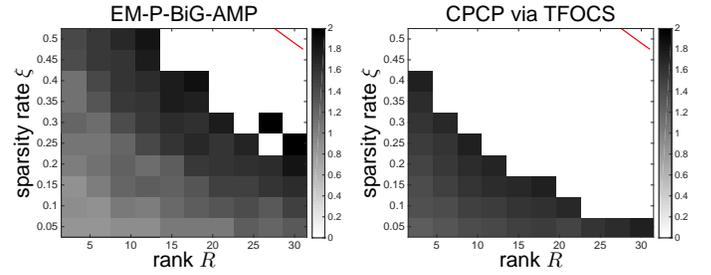

\centering
\psfrag{Lam}[b][b][0.8]{\sf sparsity rate $\xi$}
\psfrag{R}[t][t][0.8]{\sf rank $R$}
\psfrag{TFOCS}[b][b][0.8]{\sf CPCP via TFOCS}
\psfrag{EM-P-BiG-AMP}[b][b][0.8]{\sf EM-P-BiG-AMP}
\begin{tabular}{cccc}
\includegraphics[width=1.65in]{figures/lrps/phase_time_EM-P-BiG-AMP_M_10000.eps}&
\includegraphics[width=1.65in]{figures/lrps/phase_time_TFOCS_M_10000.eps}
\end{tabular}
\caption{$\log_{10}(\text{average runtime})$, in seconds, for noiseless \emph{matrix compressive sensing} as a function of rank $R$ and outlier sparsity rate $\xi$ for $M = 10000$ measurements.
Runtimes were averaged over $10$ successful trials; locations $(R,\xi)$ with any unsuccessful trials are shown in white.
}
\label{fig:phaseMatrixCSTiming}
\end{figure}

\Figref{phaseMatrixCSTiming} shows the corresponding $\log_{10}(\text{average runtime})$ versus rank $R$ and sparsity rate $\xi$ at $M=10000$ measurements. 
Runtimes were averaged over $10$ \emph{successful} trials; locations $(R,\xi)$ with any unsuccessful trials are shown in white.
The figure shows that EM-P-BiG-AMP's average runtimes were faster TFOCS's throughout the region that both algorithms were successful.
The runtimes for other values of $M$ (not shown) were similar.

\section{Conclusion}    \label{sec:conc}

We proposed P-BiG-AMP, a scheme to estimate the parameters $\vec{b}=[b_1,\dots,b_{\Nb}]\tran$ and $\vec{c}=[c_1,\dots,c_{\Nc}]\tran$ of the parametric bilinear form $z_m=\sum_{i=0}^{\Nb}\sum_{j=0}^{\Nc} b_i z_m^{(i,j)} c_j$ from noisy measurements $\{y_m\}_{m=1}^M$, where $y_m$ and $z_m$ are related through an arbitrary likelihood function and $z_m^{(i,j)}, b_0, c_0$ are known.
Our approach treats $b_i$ and $c_j$ as random variables and $z_m^{(i,j)}$ as an i.i.d.\ Gaussian tensor in order to derive a tractable simplification of the sum-product algorithm in the large-system limit, generalizing the bilinear AMP algorithms in \cite{Parker:TSP:14a,Kabashima:14}. 
We also proposed an EM extension that learns the statistical parameters of the priors on $b_i$, $c_j$, and $y_m|z_m$.
Numerical experiments suggest that our schemes yield significantly better phase transitions than several recently proposed convex and non-convex approaches to self-calibration, blind deconvolution, CS under matrix uncertainty, and matrix CS, while being competitive (or faster) in runtime.

\iftoggle{long}{
\color{\longcolor}
\section{Acknowledgement} 
The authors thank Yan Shou for help in creating \figref{blind/blind}.
\color{black}}{}

\appendices

\color{\longcolor}
\iftoggle{long}{\section{On the relation between \eqref{STLS} and \eqref{MCS}}
\label{app:intro}

Here we show that \eqref{STLS} is a special case of \eqref{MCS}.
From \eqref{STLS},
\begin{align}
z_{ml} 
&= \sum_{i=1}^{\Nb} b_i \vec{a}\oft{i}_m \vec{c}_l 
= \underbrace{[b_1,\dots,b_{\Nb}]}_{\displaystyle \defn \vec{b}\tran} \underbrace{ \mat{\vec{a}\oft{1}_m \\\vdots\\\vec{a}\oft{\Nb}_m } }_{\displaystyle \defn \ovec{A}_m} \vec{c}_l \\
&= \tr\big\{\ovec{A}_m \vec{c}_l \vec{b}\tran\big\} ,
\label{eq:zml}
\end{align}
where $\vec{a}\oft{i}_{m}$ denotes the $m$th row of $\vec{A}\of{i}$ and $\vec{c}_l$ denotes the $l$th column of $\vec{C}\textb{\in\Real^{N\times L}}$.
Then defining $\vec{e}_l$ as the $l$th column of \textb{$\vec{I}_L$} and $\vec{c}\defn\vect(\vec{C})$, we can write
\begin{align}
\ovec{A}_m \vec{c}_l
&= \underbrace{ (\vec{e}_l\tran \otimes \ovec{A}_m) }_{\displaystyle \defn \vec{\Phi}_{ml}\tran } \vec{c} .
\label{eq:kron}
\end{align}
Plugging \eqref{kron} into \eqref{zml} yields
\begin{align}
z_{ml} 
&= \tr\big\{\vec{\Phi}_{ml}\tran \vec{L}\big\}  
\end{align}
with $\vec{L}\defn \vec{c} \vec{b}\tran$, a rank-one matrix.
Thus \eqref{STLS} is equivalent to \eqref{MCS} with rank-one $\vec{L}$ and $\vec{S}=\vec{0}$.
}{}
\iftoggle{long}{\section{Scaling of \texorpdfstring{\MakeLowercase{$\E\{\Z_m^2\}$}}{E\{z2\}}} \label{app:scaling}

From \eqref{ZPB} we have
\begin{align}
\lefteqn{
\E\big\{ \Z_m^2 \big\}
= \E\left\{ \left[ 
        \sum_{i=0}^{\Nb} \sum_{j=0}^{\Nc} 
        \B_i \Z_m\of{\iB,\jC} \C_j 
        \right]^2 \right\} 
}\\
&= \sum_{i=0}^{\Nb} \sum_{j=0}^{\Nc} 
        \sum_{i'=0}^{\Nb} \sum_{j'=0}^{\Nc} 
        \E\left\{ 
        \B_i \B_{i'} 
        \C_j \C_{j'}
        \Z_m\of{i,j} \Z_m\of{i',j'} 
        \right\} \\
&= \sum_{i=0}^{\Nb} \sum_{j=0}^{\Nc} 
        \E\{\B_i^2\}
        \E\{\C_j^2\}
        \E\{\Z_m\ofsq{i,j}\} \\
&= \Ord(1)
\end{align}
since it was assumed that $\E\{\Z_m\ofsq{i,j}\}=1$, that both $\E\{\B_i^2\}$ and $\E\{\C_j^2\}$ scale as $\Ord(1/M)$, and that both $\Nc/M$ and $\Nb/M$ scale as $\Ord(1)$.
}{}
\iftoggle{long}{\section{Central Limit Theorem} \label{app:CLT}

To apply the CLT, we first expand
\begin{align}
\Z_{m}
&= \mytextstyle \sum_{i=0}^{\Nb} \sum_{j=0}^{\Nc} \B_i \C_j z_{m}\of{i,j} 
= \vB\tran \vec{Z}_{m} \vC  
        \label{eq:bZcPB}\\
&= -\hvec{b}_{m}(t)\tran\vec{Z}_{m}\hvec{c}_{m}(t)
        + \hvec{b}_{m}(t)\tran \vec{Z}_{m}\vC + \vB\tran\vec{Z}_{m}\hvec{c}_{m}(t) \non\\
        &\qquad+ \big(\vB-\hvec{b}_{m}(t)\big)\tran\vec{Z}_{m}\big(\vC-\hvec{c}_{m}(t)\big) 
        \label{eq:bZc2PB} ,
\end{align}
where the matrix $\vec{Z}_{m}$ is constructed elementwise as $[\vec{Z}_{m}]_{ij}=z_{m}\of{i,j}$ and
for \eqref{bZc2PB} we recall that $\hvec{b}_{m}(t)$ is the mean of random vector $\vB$ and $\hvec{c}_{m}(t)$ is the mean of random vector $\vC$ under the distributions in \eqref{zTobPB}.
Examining the terms in \eqref{bZc2PB}, we see that the first is an $\Ord(1)$ constant, while
the second and third are dense linear combinations of independent random variables that also scale as $\Ord(1)$.
As such, the second and third terms obey the CLT, each converging in distribution to a Gaussian as $M\rightarrow\infty$.
The last term in \eqref{bZc2PB} can be written as a quadratic form in independent zero-mean random variables:
\begin{align}
\lefteqn{
\big(\vB-\hvec{b}_{m}(t)\big)\tran\vec{Z}_{m}\big(\vC-\hvec{c}_{m}(t)\big) 
}\nonumber\\
&= \mat{\vB-\hvec{b}_{m}(t)\\\vC-\hvec{c}_{m}(t)}\tran 
                        \mat{&\frac{1}{2}\vec{Z}_{m}\\\frac{1}{2}\vec{Z}_{m}\tran&} 
                        \mat{\vB-\hvec{b}_{m}(t)\\\vC-\hvec{c}_{m}(t)} 
        \label{eq:bZc3PB} .
\end{align}
It is shown in \cite{jtp_de-Jong1987} that, for sufficiently dense $\vec{Z}_{m}$, the quadratic form in \eqref{bZc2PB} converges in distribution to a zero-mean Gaussian as $M\rightarrow\infty$.
Thus, in the LSL, $\Z_{m}$ equals a constant plus three Gaussian random variables, and thus $\Z_{m}$ is Gaussian.

}{}
\iftoggle{long}{\section{Derivation of Conditional Variance} \label{app:cond_var}

In this appendix, we derive the variance expression \eqref{pvarimlPB}.
For ease of presentation, we supress the subscript $m$ and iteration count $t$.
We begin by writing
\begin{align}
\var\{\Z \giv \B_i = b_i\} 
&= \E\left\{ \Z^2 \giv \B_i = b_i \right\} 
        - \E\left\{ \Z \giv \B_i = b_i \right\}^2 .
\label{eq:zvar_defn}
\end{align}
The first term in \eqref{zvar_defn} can be expanded as
\begin{align}
\lefteqn{
\E\left\{ \Z^2 \giv \B_i = b_i \right\} 
}\\
&= \E\Bigg\{ \Bigg[ 
  \sum_{k\neq i} \sum_j \B_k \C_j z\of{k,j} + b_i \sum_j \C_j z\of{i,j} 
  \Bigg]^2 \Bigg\} \\
&= \E\Bigg\{ \Bigg[ 
  \sum_{k\neq i} \sum_j \B_k \C_j z\of{k,j} \Bigg]^2 \Bigg\}
\non\\&\quad
  + 2 b_i \E\Bigg\{ 
  \sum_{k\neq i} \sum_{j} \B_k \C_j z\of{k,j} 
        \sum_{j'} \C_j' z\of{i,j'} \Bigg\} 
\non\\&\quad
  + b_i^2 \E\Bigg\{ \Bigg[
  \sum_j \C_j z\of{i,j} \Bigg]^2 \Bigg\} .
  \label{eq:z2}
\end{align}
We now analyze the three terms in \eqref{z2}.

The first term in \eqref{z2} can be evaluated as follows.
\begin{align}
\lefteqn{
\E\Bigg\{ \Bigg[ 
  \sum_{k\neq i} \sum_j \B_k \C_j z\of{k,j} \Bigg]^2 \Bigg\}
}\non\\
&= \E\Bigg\{ \Bigg[ 
  \sum_{k\neq i} \sum_j \Big( 
  (\B_k-\hat{b}_{k}) (\C_j-\hat{c}_{j}) 
  + \hat{b}_k(\C_j-\hat{c}_j) 
\non\\&\quad 
  + (\B_k-\hat{b}_k)\hat{c}_j + \hat{b}_k\hat{c}_j 
  \Big) z\of{k,j} 
  \Bigg]^2 \Bigg\} \\
&= \sum_{k\neq i} \sum_j \nu^b_k \nu^c_j z\ofsq{k,j}
        + \sum_j \nu^c_j 
                \Bigg[ \sum_{k\neq i} \hat{b}_k z\of{k,j} \Bigg]^2
\non\\&\quad 
        +  \sum_{k\neq i} \nu^b_k 
                \Bigg[ \sum_j \hat{c}_j z\of{k,j} \Bigg]^2
   + \Bigg[ \sum_{k\neq i} \sum_j \hat{b}_k\hat{c}_j z\of{k,j}\Bigg]^2 \\
&= \sum_{k\neq i} \sum_j \nu^b_k \nu^c_j z\ofsq{k,j}
        + \sum_j \nu^c_j 
                \Big[ \hat{z}\of{*,j} -\hat{b}_i \textb{z\of{i,j}} \Big]^2
\non\\&\quad 
        +  \sum_{k\neq i} \nu^b_k \hat{z}\ofsq{k,*}
   + \Bigg[ \sum_{k\neq i} \hat{b}_k \hat{z}\of{k,*}\Bigg]^2 .
\label{eq:z2_1}
\end{align}
The second term in \eqref{z2} then becomes
\begin{align}
\lefteqn{
  2 b_i \E\Bigg\{
  \sum_{k\neq i} \sum_{j} \B_k \C_j z\of{k,j} 
        \sum_{j'} \C_j' z\of{i,j'} \Bigg\} 
}\non\\
&= 2 b_i \E\Bigg\{ 
  \sum_{k\neq i} \sum_{j} 
  \Big[ (\B_k-\hat{b}_k) (\C_j-\hat{c}_j) 
        + \hat{b}_k(\C_j-\hat{c}_j) 
\non\\&\quad 
        + (\B_k-\hat{b}_k)\hat{c}_j 
        + \hat{b}_k\hat{c}_j 
  \Big] z\of{k,j} 
  \sum_{j'} \Big[ (\C_{j'}-\hat{c}_{j'}) 
        + \hat{c}_{j'}\Big] 
        z\of{i,j'} \Bigg\} \\
&= 2 b_i \E\Bigg\{ 
  \sum_{k\neq i} \sum_{j} 
  \Big[ (\B_k-\hat{b}_k) (\C_j-\hat{c}_j) 
        + \hat{b}_k(\C_j-\hat{c}_j) 
\non\\&\quad 
        + (\B_k-\hat{b}_k)\hat{c}_j 
        + \hat{b}_k\hat{c}_j 
  \Big] z\of{k,j} 
  \sum_{j'} (\C_{j'}-\hat{c}_{j'}) 
        z\of{i,j'} \Bigg\} 
\non\\&\quad 
  + 2 b_i \E\Bigg\{ 
  \sum_{k\neq i} \sum_{j} 
  \Big[ (\B_k-\hat{b}_k) (\C_j-\hat{c}_j) 
        + \hat{b}_k(\C_j-\hat{c}_j) 
\non\\&\quad 
        + (\B_k-\hat{b}_k)\hat{c}_j 
        + \hat{b}_k\hat{c}_j 
  \Big] z\of{k,j} 
  \sum_{j'} \hat{c}_{j'} z\of{i,j'} \Bigg\} .
\end{align}
Continuing,
\begin{align}
\lefteqn{
  2 b_i \E\Bigg\{
  \sum_{k\neq i} \sum_{j} \B_k \C_j z\of{k,j} 
        \sum_{j'} \C_j' z\of{i,j'} \Bigg\} 
}\non\\
&= 2 b_i 
  \sum_{k\neq i} \sum_{j} 
  \hat{b}_k \nu^c_j z\of{k,j}z\of{i,j}
\non\\&\quad 
  + 2 b_i \sum_{k\neq i} \sum_{j} 
        \hat{b}_k\hat{c}_j z\of{k,j} \sum_{j'} \hat{c}_{j'} z\of{i,j'} \\
&= 2 b_i \sum_{j} 
        \Bigg( 
        \sum_{k} \hat{b}_k z\of{k,j} 
        - \hat{b}_i z\of{i,j} 
        \Bigg) z\of{i,j} \nu^c_j 
\non\\&\quad 
  + 2 b_i \Bigg( 
        \sum_{k} \sum_{j} \hat{b}_k\hat{c}_j z\of{k,j} 
        - \hat{b}_i \sum_{j} \hat{c}_j z\of{i,j} 
        \Bigg) 
        \sum_{j'} \hat{c}_{j'} z\of{i,j'} \\
&= 2 b_i \sum_{j} 
        \left( \hat{z}\of{*,j}
        - \hat{b}_i z\of{i,j} 
        \right) z\of{i,j} \nu^c_j 
\non\\&\quad 
  + 2 b_i \left( \hat{z}\of{*,*} - \hat{b}_i \hat{z}\of{i,*} \right) 
        \hat{z}\of{i,*} .
\label{eq:z2_2}
\end{align}
Finally, the third term in \eqref{z2} becomes
\begin{align}
\lefteqn{
b_i^2 \E\Bigg\{ \Bigg[ \sum_j \C_j z\of{i,j} \Bigg]^2 \Bigg\}
}\\
&= b_i^2 \E\Bigg\{ \Bigg[ \sum_j \Big[ (\C_j-\hat{c}_j) + \hat{c}_j \Big] z\of{i,j} \Bigg]^2 \Bigg\} \\
&= b_i^2 \sum_j \nu^c_j z\ofsq{i,j}
   + b_i^2 \Bigg[ \sum_j \hat{c}_j z\of{i,j} \Bigg]^2 \\
&= b_i^2 \Bigg( \sum_j \nu^c_j z\ofsq{i,j} + \hat{z}\ofsq{i,*} \Bigg) .
\label{eq:z2_3}
\end{align}

Next, we analyze the second term in \eqref{zvar_defn}.
Using the expression for $\E\left\{ \Z \giv \B_i = b_i \right\}$ from \eqref{ZmeanPB}, we have
\begin{align}
\lefteqn{
-\E\left\{ \Z \giv \B_i = b_i \right\}^2
= -\left[ \big(\hat{z}\of{*,*} - \hat{b}_i\hat{z}\of{i,*}\big) 
        +b_i \hat{z}\of{i,*} \right]^2 
}\\
&= -\Big[ \hat{z}\of{*,*} - \hat{b}_i \hat{z}\of{i,*} \Big]^2
        - 2b_i \hat{z}\of{i,*} \big( \hat{z}\of{*,*} 
                - \hat{b}_i \hat{z}\of{i,*} \big)
\non\\&\quad
        - b_i^2 \hat{z}\ofsq{i,*} \\
&= -\Bigg[ \sum_{k\neq i} \hat{b}_k \hat{z}\of{k,*} \Bigg]^2
        - 2b_i \hat{z}\of{i,*} \big( \hat{z}\of{*,*} 
                - \hat{b}_i \hat{z}\of{i,*} \big)
        - b_i^2 \hat{z}\ofsq{i,*} .
\label{eq:z2_5}
\end{align}

Finally, from \eqref{zvar_defn} and \eqref{z2}, we know that $\var\{\Z \giv \B_i = b_i\}$ equals the sum of \eqref{z2_1}, \eqref{z2_2}, \eqref{z2_3}, and \eqref{z2_5}.
Adding them together and dropping the terms that cancel, we find that 
\begin{align}
\lefteqn{ 
\var\{\Z \giv \B_i = b_i\} 
}\non\\
&= \sum_{k\neq i} \sum_j \nu^b_k \nu^c_j z\ofsq{k,j}
        + \sum_j \nu^c_j 
                \Big[ \hat{z}\of{*,j} -\hat{b}_i z\of{k,j} \Big]^2
\non\\&\quad
        +  \sum_{k\neq i} \nu^b_k \hat{z}\ofsq{k,*} 
        + 2 b_i \sum_{j} 
        \left( \hat{z}\of{*,j}
        - \hat{b}_i z\of{i,j} 
        \right) z\of{i,j} \nu^c_j 
\non\\&\quad
  + b_i^2 \sum_j \nu^c_j z\ofsq{i,j} .
\label{eq:zvar}
\end{align}
The sum of the first three terms in \eqref{zvar} can then be rearranged to form
\begin{align}
\nu^p
&\defn \sum_{k\neq i} \nu^b_k \Bigg[ \sum_j \nu^c_j z\ofsq{k,j}
                + \hat{z}\ofsq{k,*} \Bigg] 
\non\\&\quad
        + \sum_j \nu^c_j \Big[ 
        \hat{z}\ofsq{*,j} 
        -2 \hat{b}_i \hat{z}\of{*,j} z\of{k,j} 
        +\hat{b}_i^2 z\ofsq{k,j} 
        \Big] .
\end{align}

}{}
\iftoggle{long}{\section{Derivation of \eqref{HtodiffPB}} \label{app:H}

In this appendix we derive equation \eqref{HtodiffPB}.
Using \eqref{phatPB} and \eqref{pvarPB}, we write the $H_{m}(\cdot)$ term in \eqref{HstepPB} as
\begin{align}
&H_{m}\Bigg(
  \hat{p}_{i,m}(t) + \Br_{\iB} \zmatml{m}{\iB}{\nod}{t}{},~~
  \nu^p_{i,m}(t) + \Br_{\iB}^2 \sum_{\jC=1}^{\Nc} \nu^{\Cr}_{m,\jC}(t)
  \zmatC{m}{\iB}{\jC}{t}{2}      
\non\\&\qquad
  + 2\Br_{\iB} \sum_{\jC=1}^{\Nc} \nu^{\Cr}_{m,\jC}(t)
  \left[ \zmatml{m}{\nod}{\jC}{t}{} \zmatC{m}{\iB}{\jC}{t}{}
  - \hat{\Br}_{m,\iB}(t) \zmatC{m}{\iB}{\jC}{t}{2}
  \right]
  \Bigg)\non\\
&= H_{m} \Bigg(\hat{p}_{m}(t) 
   + \big(\Br_{\iB} - \hat{\Br}_{m,\iB}(t)\big) \zmatml{m}{\iB}{\nod}{t}{},
\\&\qquad
   \nu^p_{m}(t) +\big(\Br_{\iB} - \hat{\Br}_{m,\iB}(t) \big)^2  
   \sum_{\jC=1}^{\Nc} \nu^{\Cr}_{m,\jC}(t) \zmatC{m}{\iB}{\jC}{t}{2}
\non\\&\qquad
   + 2 \big( \Br_{\iB} -  \hat{\Br}_{m,\iB}(t) \big) 
   \sum_{\jC=1}^{\Nc} \nu^{\Cr}_{m,\jC}(t)
   \zmatml{m}{\nod}{\jC}{t}{} \zmatC{m}{\iB}{\jC}{t}{} 
\non\\&\qquad
   -  \nu^{\Br}_{m,\iB}(t) \Bigg[\zmatml{m}{\iB}{\nod}{t}{2} + 
   \sum_{\jC=1}^{\Nc}  \nu^{\Cr}_{m,\jC}(t) \zmatC{m}{\iB}{\jC}{t}{2}   
   \Bigg] \Bigg)
\non\\
&= H_{m} \Bigg(\hat{p}_{m}(t) + \big(\Br_{\iB} 
   - \hat{\Br}_{\iB}(t)\big) \zmatml{m}{\iB}{\nod}{t}{} 
   + \Ord(1/M),
\\&\qquad
   \nu^p_{m}(t) +\big(\Br_{\iB} 
   - \hat{\Br}_{\iB}(t) \big)^2  \sum_{\jC=1}^{\Nc} \nu^{\Cr}_{m,\jC}(t)
   \zmatC{m}{\iB}{\jC}{t}{2}
\non\\&\qquad
   + 2 \big( \Br_{\iB} -  \hat{\Br}_{\iB}(t) \big) 
   \sum_{\jC=1}^{\Nc} \nu^{\Cr}_{m,\jC}(t)
   \zmat{m}{\nod}{\jC}{t}{} \zmatC{m}{\iB}{\jC}{t}{} 
   + \Ord(1/M) \Bigg)  
\non .
\end{align}
}{}
\iftoggle{long}{\section{Taylor Series Expansion} \label{app:Taylor}

In this appendix, we perform a Taylor series expansion of \eqref{HtodiffPB} and analyze the result in the LSL to obtain \eqref{bmlToiPB}.

We start by calculating the first two derivatives of the $H_{m}(\cdot)$ term from \eqref{HtodiffPB} w.r.t.\ $\Br_{\iB}$.
From \eqref{HtodiffPB}, we find that
\begin{align}
\frac{\partial H_{m}}{\partial \Br_{\iB}} 
&= \zmatml{m}{\iB}{\nod}{t}{}  H'_{m}
        + \Bigg(  2\big(\Br_{\iB} - \hat{\Br}_{\iB}(t) \big)  
                \sum_{\jC=1}^{\Nc} \nu^{\Cr}_{m,\jC}(t) \zmatC{m}{\iB}{\jC}{t}{2}
\non\\&\quad
                + 2  \sum_{\jC=1}^{\Nc} \nu^{\Cr}_{m,\jC}(t)
                          \zmat{m}{\nod}{\jC}{t}{} \zmatC{m}{\iB}{\jC}{t}{} \Bigg) \dot{H}_{m},
        \label{eq:Hderiv1aPB}
\end{align}
where $H_{m}'$ denotes the derivative of $H_m(\cdot,\cdot)$ w.r.t.\ the first argument and $\dot{H}_{m}$ denotes the derivative w.r.t.\ the second argument, supressing their arguments for brevity.
Equation \eqref{Hderiv1aPB} then implies
\begin{align}
\frac{\partial H_{m}}{\partial \Br_{\iB}}\Bigg|_{\Br_{\iB}=\hat{\Br_{\iB}}(t)} 
&= \zmatml{m}{\iB}{\nod}{t}{}  H'_{m}
\label{eq:Hderiv1PB}\\&\quad
                + \Bigg( 2  \sum_{\jC=1}^{\Nc} \nu^{\Cr}_{m,\jC}(t)
                          \zmat{m}{\nod}{\jC}{t}{} \zmatC{m}{\iB}{\jC}{t}{} \Bigg) \dot{H}_{m} ,
\non
\end{align}
and
\begin{align}
\lefteqn{
\frac{\partial^2 H_{m}}{\partial \Br_{\iB}^2} 
= \zmatml{m}{\iB}{\nod}{t}{2}  H''_{m}
        + \Bigg(  2\big(\Br_{\iB} - \hat{\Br}_{\iB}(t) \big)  
                \sum_{\jC=1}^{\Nc} \nu^{\Cr}_{m,\jC}(t) \zmatC{m}{\iB}{\jC}{t}{2}
}\non\\&\quad
                + 2  \sum_{\jC=1}^{\Nc} \nu^{\Cr}_{m,\jC}(t)
                          \zmat{m}{\nod}{\jC}{t}{} \zmatC{m}{\iB}{\jC}{t}{} \Bigg)
                                \zmatml{m}{\iB}{\nod}{t}{} \dot{H}'_{m}
        \non\\&\quad
        +  \Bigg( 2\sum_{\jC=1}^{\Nc} \nu^{\Cr}_{m,\jC}(t) \zmatC{m}{\iB}{\jC}{t}{2} \Bigg)
                \dot{H}_{m}
        \non\\&\quad
        + \Bigg[  2\big(\Br_{\iB} - \hat{\Br}_{\iB}(t) \big)  
                \sum_{\jC=1}^{\Nc} \nu^{\Cr}_{m,\jC}(t) \zmatC{m}{\iB}{\jC}{t}{2}
        \non\\&\qquad
                + 2  \sum_{\jC=1}^{\Nc} \nu^{\Cr}_{m,\jC}(t)
                          \zmat{m}{\nod}{\jC}{t}{} \zmatC{m}{\iB}{\jC}{t}{} \Bigg]
        \non\\&\qquad
        \times \Bigg[
        \zmatml{m}{\iB}{\nod}{t}{}  \dot{H}'_{m}
        + \Bigg(  2\big(\Br_{\iB} - \hat{\Br}_{\iB}(t) \big)  
                \sum_{\jC=1}^{\Nc} \nu^{\Cr}_{m,\jC}(t) \zmatC{m}{\iB}{\jC}{t}{2}
                \non\\&\qquad\qquad
                + 2  \sum_{\jC=1}^{\Nc} \nu^{\Cr}_{m,\jC}(t)
                          \zmat{m}{\nod}{\jC}{t}{} \zmatC{m}{\iB}{\jC}{t}{} \Bigg) \ddot{H}_{m}
                \Bigg]
\end{align}
which implies
\begin{align}
\lefteqn{
\frac{\partial^2 H_{m}}{\partial \Br_{\iB}^2} \Bigg|_{\Br_{\iB}=\hat{\Br_{\iB}}(t)}
}\non\\
&= \zmatml{m}{\iB}{\nod}{t}{2}  H''_{m}
        + \Bigg( 4 \sum_{\jC=1}^{\Nc} \nu^{\Cr}_{m,\jC}(t)
                          \zmat{m}{\nod}{\jC}{t}{} \zmatC{m}{\iB}{\jC}{t}{} \Bigg)
                                \zmatml{m}{\iB}{\nod}{t}{} \dot{H}'_{m}
        \non\\&\quad
        +  \Bigg( 2\sum_{\jC=1}^{\Nc} \nu^{\Cr}_{m,\jC}(t) \zmatC{m}{\iB}{\jC}{t}{2} \Bigg)
                \dot{H}_{m}
        \non\\&\quad
        + \Bigg(  2  \sum_{\jC=1}^{\Nc} \nu^{\Cr}_{m,\jC}(t)
                          \zmat{m}{\nod}{\jC}{t}{} \zmatC{m}{\iB}{\jC}{t}{} \Bigg)^2 \ddot{H}_{m} ,
        \label{eq:Hderiv2PB} .
\end{align}
The Taylor series expansion of \eqref{HstepPB} can then be stated as
\begin{align}
\lefteqn{
\fxnvar{m}{\iB}{\B}(t,\Br_{\iB}) 
}\non\\
&\approx \const
        + H_{m}\big(\hat{p}_{m}(t) + \Ord(1/M), \nu^p_{m}(t) + \Ord(1/M)\big)
        \non\\&\quad
        + \big(\Br_{\iB} - \hat{\Br}_{\iB}(t)\big) \Bigg[
                \zmatml{m}{\iB}{\nod}{t}{} 
        \non\\&\qquad\quad\times
        H'_{m}\big(\hat{p}_{m}(t) + \Ord(1/M), \nu^p_{m}(t) + \Ord(1/M)\big)
        \non\\&\qquad\quad
                + 2 \Bigg( \sum_{\jC=1}^{\Nc} \nu^{\Cr}_{m,\jC}(t) \zmat{m}{\nod}{\jC}{t}{} \zmatC{m}{\iB}{\jC}{t}{}\Bigg) 
        \non\\&\qquad\quad\times
         \dot{H}_{m}\big(\hat{p}_{m}(t) + \Ord(1/M), \nu^p_{m}(t) + \Ord(1/M)\big) \Bigg]
        \non\\&\quad
        + \frac{1}{2} \big(\Br_{\iB} - \hat{\Br}_{\iB}(t)\big)^2 \Bigg[
                \zmatml{m}{\iB}{\nod}{t}{2} 
        \non\\&\qquad\quad\times
        H''_{m}\big(\hat{p}_{m}(t) + \Ord(1/M), \nu^p_{m}(t) + \Ord(1/M)\big) 
        \non\\&\qquad\quad
                + \Bigg( 2 \sum_{\jC=1}^{\Nc} \nu^{\Cr}_{m,\jC}(t) \zmatC{m}{\iB}{\jC}{t}{2}\Bigg)  
                \non\\&\qquad\quad\quad\times
                        \dot{H}_{m}\big(\hat{p}_{m}(t) + \Ord(1/M), \nu^p_{m}(t) + \Ord(1/M)\big) \Bigg]
        \non\\&\quad
        + \Ord(1/M^{3/2}),
        \label{eq:bTaylorToApproxPB}
\end{align}
where the second and fourth terms in \eqref{Hderiv2PB} were absorbed into the $O(1/M^{3/2})$ term in \eqref{bTaylorToApproxPB} using the facts that
\begin{align}
\Bigg(\!4\sum_{\jC=1}^{\Nc} \nu^{\Cr}_{m,\jC}(t) \zmat{m}{\nod}{\jC}{t}{} \zmatC{m}{\iB}{\jC}{t}{}\Bigg)\zmatml{m}{\iB}{\nod}{t}{} &= \Ord(1/M^{1/2})\\
\Bigg(2\sum_{\jC=1}^{\Nc} \nu^{\Cr}_{m,\jC}(t) \zmatC{m}{\iB}{\jC}{t}{2}\Bigg) &= \Ord(1)\\ 
\Bigg(2\sum_{\jC=1}^{\Nc} \nu^{\Cr}_{m,\jC}(t) \zmat{m}{\nod}{\jC}{t}{} \zmatC{m}{\iB}{\jC}{t}{}\Bigg)^2 &= \Ord(1/M).
\end{align}
which follow from the $\Ord(1/M)$ scaling of $\nu^{\Cr}_{m,\jC}(t)$, as well as from the facts that $\big(\Br_{\iB} - \hat{\Br}_{\iB}(t)\big)^2$ is $\Ord(1/M)$ and the function $H_{m}$ and its partials are $\Ord(1)$.

Note that the second-order expansion term in \eqref{bTaylorToApproxPB} is $\Ord(1/M)$.
We will now approximate \eqref{bTaylorToApproxPB} by dropping terms that vanish relative to the latter as $M \to \infty$.
First, we replace $\zmatml{m}{\iB}{\nod}{t}{}$ with $\zmat{m}{\iB}{\nod}{t}{}$ in the quadratic term in \eqref{bTaylorToApproxPB}, since $\big(\zmatml{m}{\iB}{\nod}{t}{}-\zmat{m}{\iB}{\nod}{t}{}\big)$ is $\Ord(1/M^{1/2})$, which gets reduced to $\Ord(1/M^{3/2})$ via scaling by $\big(\Br_{\iB} - \hat{\Br}_{\iB}(t)\big)^2$.
Note that we cannot make a similar replacement in the linear term in \eqref{bTaylorToApproxPB}, because the $\big(\Br_{\iB} - \hat{\Br}_{\iB}(t)\big)$ scaling is not enough to render the difference negligible.
Next, we replace $\nu^{\Cr}_{m,\jC}(t)$ with $\nu^{\Cr}_{\jC}(t)$ throughout \eqref{bTaylorToApproxPB}, since the difference is $\Ord(1/M^{3/2})$.
Finally, as established in~\cite{Parker:TSP:14a}, the $\Ord(1/M)$ perturbations inside the $H_{m}$ derivatives can be dropped because they have an $O(1/M^{3/2})$ effect on the overall message.
With these approximations, and absorbing $\Br_{\iB}$-invariant terms into the $\const$, we obtain \eqref{bmlToiPB}:
\begin{align}
\lefteqn{
\fxnvar{m}{\iB}{\B}(t,\Br_{\iB}) 
}\non\\
&\approx \const 
        + \Bigg[\hat{s}_{m}(t) \zmatml{m}{\iB}{\nod}{t}{} 
                + \nu^s_{m}(t)  \hat{\Br}_{\iB}(t) \zmat{m}{\iB}{\nod}{t}{2}   
\non\\&
                + \big(\hat{s}_{m}^2(t) - \nu^s_{m}(t) \big) \sum_{\jC=1}^{\Nc} \nu^{\Cr}_{\jC}(t) 
                \zmatC{m}{\iB}{\jC}{t}{} \Big(\zmat{m}{\nod}{\jC}{t}{}  - \hat{\Br}_{\iB}(t) \zmatC{m}{\iB}{\jC}{t}{}   \Big)
          \Bigg] \Br_{\iB} 
\non\\&
        - \frac{1}{2} \Bigg[  \nu^s_{m}(t) \zmat{m}{\iB}{\nod}{t}{2}   
                - \big(\hat{s}_{m}^2(t) - \nu^s_{m}(t) \big) 
                \sum_{\jC=1}^{\Nc} \nu^{\Cr}_{\jC}(t) \zmatC{m}{\iB}{\jC}{t}{2}\Bigg]  \Br_{\iB}^2 ,
\non
\end{align}
via the definitions of $\hat{s}_{m}(t)$ and $\nu^s_{m}(t)$ from \eqref{shatPB}-\eqref{nushatPB}
and the following relationship established in~\cite{Parker:TSP:14a}:
\begin{align}
\dot{H}_{m}\big(q,\nu^q\big)
&= \frac{1}{2}\left[H'_{m}\big(q,\nu^q\big)^2 +  H''_{m}\big(q,\nu^q\big) \right] .
\end{align}
}{}
\iftoggle{long}{\section{Derivation of \eqref{phat2PB}} \label{app:phat}

In this appendix, we show how \eqref{phat2PB} results in the LSL.
From \eqref{phatPB} and \eqref{Zstar}, we have
\begin{align}
\hat{p}_{m}(t)
&= \sum_{k,j} \hat{b}_{m,k}(t) \hat{c}_{m,j}(t) z_m\of{k,j} .
\end{align}
Plugging \eqref{chatStep4PB} and \eqref{bhatStep4PB} into the previous equation gives
\begin{align}
\lefteqn{
\hat{p}_{m}(t)
}\non\\
&= \sum_{i,j} 
  \Big( \hat{b}_i(t) - \hat{s}_{m}(t\!-\!1) \hat{z}_m\of{i,*}(t\!-\!1) \nu^b_i(t) + \Ord(1/M^{3/2}) \Big)
\non\\&\quad
  \Big( \hat{c}_j(t) - \hat{s}_{m}(t\!-\!1) \hat{z}_m\of{*,j}(t\!-\!1) \nu^c_j(t) + \Ord(1/M^{3/2}) \Big)
  z_m\of{i,j} \\
&= \sum_{i,j} \hat{b}_i(t) \hat{c}_j(t) z_m\of{i,j}
\non\\&\quad
   - \hat{s}_{m}(t\!-\!1) \sum_{i} \nu^b_i(t)\hat{z}_m\of{i,*}(t\!-\!1) 
        \sum_j \hat{c}_j(t) z_m\of{i,j}
\non\\&\quad
   - \hat{s}_{m}(t\!-\!1) \sum_{j} \nu^c_j(t)\hat{z}_m\of{*,j}(t\!-\!1) 
        \sum_i \hat{b}_i(t) z_m\of{i,j}
\non\\&\quad
   + \hat{s}_{m}(t\!-\!1)^2 \sum_{i,j} 
        \hat{z}_m\of{i,*}(t\!-\!1) 
        \hat{z}_m\of{*,j}(t\!-\!1) 
        \nu^b_i(t) \nu^c_j(t) z_m\of{i,j}
\non\\&\quad
   + \Ord(1/M) 
\label{eq:phat_tmp} \\
&= \hat{z}\of{*,*}(t) - \hat{s}_{m}(t\!-\!1) 
        \Bigg( \sum_{i} \nu^b_i(t)\hat{z}_m\of{i,*}(t\!-\!1) \hat{z}\of{i,*}(t)
\non\\&\quad
   + \sum_{j} \nu^c_j(t)\hat{z}_m\of{*,j}(t\!-\!1) \hat{z}\of{*,j}(t) \Bigg)
   + \Ord(1/M)
\label{eq:phat_tmp2} .
\end{align}
since the second-to-last term in \eqref{phat_tmp} is $\Ord(1/M)$.
Because the first two terms in \eqref{phat_tmp2} are $\Ord(1)$, the $\Ord(1/M)$ term in \eqref{phat_tmp2} vanishes in the LSL, resulting in \eqref{phat2PB}.
}{}
\iftoggle{long}{\section{Derivation of \eqref{pvar3PB}} \label{app:pvar}

In this appendix, we derive \eqref{pvar3PB}.
Plugging \eqref{zi0approxPB} and \eqref{z0japproxPB} into \eqref{pvar2PB} gives
\begin{align}
\lefteqn{
\nu^p_{m}(t) 
}\non\\
&= \sum_{\jC=1}^{\Nc} \nu^{\Cr}_{\jC}(t) \Bigg(\zmat{m}{\nod}{\jC}{t}{} - \hat{s}_{m}(t\!-\!1)
                \sum_{\iB=1}^{\Nb}  \zmat{m}{\iB}{\nod}{t\!-\!1}{} \zmatC{m}{\iB}{\jC}{t}{} 
                \nu^{\Br}_{\iB}(t)\Bigg)^2      
        \non\\&
        + \sum_{\iB=1}^{\Nb} \nu^{\Br}_{\iB}(t) \Bigg(\zmat{m}{\iB}{\nod}{t}{} - \hat{s}_{m}(t\!-\!1)
                \sum_{\jC=1}^{\Nc}  \zmat{m}{\nod}{\jC}{t\!-\!1}{} \zmatC{m}{\iB}{\jC}{t}{} 
                \nu^{\Cr}_{\jC}(t)\Bigg)^2
        \non\\&
        +  \sum_{\iB=1}^{\Nb}\sum_{\jC=1}^{\Nc} \nu^{\Br}_{\iB}(t) \nu^{\Cr}_{\jC}(t)
                \zmatC{m}{\iB}{\jC}{t}{2} + \Ord(1/M^{1/2}) .
\end{align}
Using the definition of $\bar{\nu}^p_{m}(t)$ from \eqref{phat3PB},
\begin{align}
\lefteqn{
\nu^p_{m}(t) 
}\non\\
&= \bar{\nu}^p_{m}(t) 
        + \sum_{\iB=1}^{\Nb}\sum_{\jC=1}^{\Nc} \nu^{\Br}_{\iB}(t) \nu^{\Cr}_{\jC}(t)\zmatC{m}{\iB}{\jC}{t}{2} 
\non\\&\quad
        -2\hat{s}_{m}(t\!-\!1)         \Bigg[\sum_{\jC=1}^{\Nc} \nu^{\Cr}_{\jC}(t) \zmat{m}{\nod}{\jC}{t}{} 
                \sum_{\iB=1}^{\Nb}  \zmat{m}{\iB}{\nod}{t\!-\!1}{} \zmatC{m}{\iB}{\jC}{t}{} \nu^{\Br}_{\iB}(t) 
                \non\\&\quad\quad        + \sum_{\iB=1}^{\Nb} \nu^{\Br}_{\iB}(t) \zmat{m}{\iB}{\nod}{t}{}
                \sum_{\jC=1}^{\Nc}  \zmat{m}{\nod}{\jC}{t\!-\!1}{} \zmatC{m}{\iB}{\jC}{t}{} \nu^{\Cr}_{\jC}(t)         \Bigg]
\non\\&\quad
        + \hat{s}_{m}^2(t\!-\!1)
        \Bigg[\sum_{\jC=1}^{\Nc} \nu^{\Cr}_{\jC}(t) 
                \Bigg( \sum_{\iB=1}^{\Nb}  \zmat{m}{\iB}{\nod}{t\!-\!1}{} 
                        \zmatC{m}{\iB}{\jC}{t}{} \nu^{\Br}_{\iB}(t) \Bigg)^2 
                        \non\\&\quad\quad
        + \sum_{\iB=1}^{\Nb} \nu^{\Br}_{\iB}(t) 
                 \Bigg( \sum_{\jC=1}^{\Nc}  \zmat{m}{\nod}{\jC}{t\!-\!1}{} 
                        \zmatC{m}{\iB}{\jC}{t}{} \nu^{\Cr}_{\jC}(t) \Bigg)^2 \Bigg] 
        \non\\&\quad
        + \Ord(1/M^{1/2}) \\
&\approx \bar{\nu}^p_{m}(t) 
        + \sum_{\iB=1}^{\Nb}\sum_{\jC=1}^{\Nc} \nu^{\Br}_{\iB}(t) \nu^{\Cr}_{\jC}(t)\zmatC{m}{\iB}{\jC}{t}{2} ,
\end{align}
where in the last step we retained only the $\Ord(1)$ terms, since the others vanish in the LSL.
}{}
\iftoggle{long}{\section{Derivation of \eqref{rhat_simp}} \label{app:rhat}

In this appendix, we derive \eqref{rhat_simp}.
Treating $z_m\of{i,j}$ as i.i.d.\ zero-mean unit-variance Gaussian, the mean-squared value of the first term in \eqref{rhat_complicated} is (suppressing the SPA iteration $t$ for brevity)
\begin{align}
\lefteqn{
\E\Bigg\{ \Bigg| 
\nu^r_j \hat{c}_j \sum_{i=1}^{\Nb} \nu^b_i \sum_{m=1}^M \nu^s_m \Z_{m}\ofsq{i,j}
\Bigg|^2 \Bigg\}
}\\
&= \hat{c}_j^2 (\nu^r_j)^2 \sum_{i}\sum_{i'}\sum_{m}\sum_{m'} (\nu^b_i)^2 (\nu^s_m)^2 \E\big\{ \Z_{m}\ofsq{i,j} \Z_{m'}\ofsq{i',j} \big\} 
\\
&= \Ord(1/M)
\non
\end{align}
since $(\nu^r_j)^2=\Ord(1/M^2)$, $\hat{c}_j^2=\Ord(1/M)$, $(\nu^b_i)^2=\Ord(1/M^2)$, $(\nu^s_m)^2=\Ord(1)$ and 
\begin{align}
\lefteqn{
\E\big\{ \Z_{m}\ofsq{i,j} \Z_{m'}\ofsq{i',j} \big\}
}\label{eq:gaussian}\\
&= \begin{cases}
\E\big\{\Z_m^{(i,j)4}\big\} =3\big[\E\big\{\Z_m\ofsq{i,j}\big\}\big]^2 
& \text{if~$(i,m)=(i',m')$} \\
\big[ \E\big\{\Z_m\ofsq{i,j}\big\} \big]^2 & \text{if~$(i,m)\neq (i',m')$} \\
\end{cases} 
\non\\
&= \Ord(1) ,
\end{align}
where in \eqref{gaussian} we used the fact that $\E\{\Z^4\}=3[\E\{\Z^2\}]^2$ for Gaussian $\Z$.
Meanwhile, the mean-squared value of the second term in \eqref{rhat_complicated} can be shown to be
\begin{align}
\lefteqn{
\E\Bigg\{ \Bigg| 
\nu^r_j \sum_{k\neq j} \hat{c}_k \sum_{i=1}^{\Nb} \nu^b_i \sum_{m=1}^M \nu^s_m \Z_{m}\of{i,j}\Z_{m}\of{i,k} 
\Bigg|^2 \Bigg\}
}\non\\
&= (\nu^r_j)^2 \sum_{k\neq j} \sum_{i}\sum_{m} \hat{c}_k^2 (\nu^b_i)^2 (\nu^s_m)^2 \E\big\{ \Z_{m}\ofsq{i,j} \} \E\big\{ \Z_{m}\ofsq{i,k} \big\} \\
&= \Ord(1/M^2) .
\end{align}
Thus, we see that the second term in \eqref{rhat_complicated} vanishes relative to the first as $M\rightarrow\infty$.
}{}
\iftoggle{long}{\section{Derivation of \eqref{NishimoriPB}} \label{app:rvar}

In this appendix, we derive \eqref{NishimoriPB}.
Plugging \eqref{sPB} and \eqref{nusPB} into the second half of $\nu^r_{\jC}(t)$ from \eqref{rvarPB}, we find
\begin{align}
&\sum_{m} \big(\hat{s}_{m}^2(t) - \nu^s_{m}(t) \big) 
        \sum_{\iB=1}^{\Nb} \nu^{\Br}_{\iB}(t) \zmatC{m}{\iB}{\jC}{t}{2}\\
&= \sum_{m} \Bigg[\left(\frac{\hat{z}_{m}(t)-\hat{p}_{m}(t)}{\nu^p_{m}(t)}\right)^2 
        - \frac{1}{\nu^p_{m}(t)}\left(1-\frac{\nu^z_{m}(t)}{\nu^p_{m}(t)}\right) \Bigg] 
\non\\&\quad\times
        \sum_{\iB=1}^{\Nb} \nu^{\Br}_{\iB}(t) \zmatC{m}{\iB}{\jC}{t}{2} \\
&= \sum_{m} \Bigg(\frac{\big(\hat{z}_{m}(t)-\hat{p}_{m}(t)\big)^2+\nu^z_{m}(t)}{\nu^p_{m}(t)} - 1 \Bigg) 
        \frac{\sum_{\iB=1}^{\Nb} \nu^{\Br}_{\iB}(t) \zmatC{m}{\iB}{\jC}{t}{2}}{\nu^p_{m}(t)} \\
&= \sum_{m} \Bigg(\E\left\{\frac{\big(\Z_{m}-\hat{p}_{m}(t)\big)^2}{\nu^p_{m}(t)}\right\} - 1 \Bigg) 
        \frac{\sum_{\iB=1}^{\Nb} \nu^{\Br}_{\iB}(t) \zmatC{m}{\iB}{\jC}{t}{2}}{\nu^p_{m}(t)} ,
\end{align}
where the random variable $\Z_{m}$ above is distributed according to the pdf in \eqref{pZgivYPPB}.
}{}
\color{black}


\bibliographystyle{IEEEtran}
\bibliography{parker,macros_abbrev,books,blind,sparse,comm,misc,machine,multicarrier,hsi,phase}

\begin{thebibliography}{10}
\providecommand{\url}[1]{#1}
\csname url@samestyle\endcsname
\providecommand{\newblock}{\relax}
\providecommand{\bibinfo}[2]{#2}
\providecommand{\BIBentrySTDinterwordspacing}{\spaceskip=0pt\relax}
\providecommand{\BIBentryALTinterwordstretchfactor}{4}
\providecommand{\BIBentryALTinterwordspacing}{\spaceskip=\fontdimen2\font plus
\BIBentryALTinterwordstretchfactor\fontdimen3\font minus
  \fontdimen4\font\relax}
\providecommand{\BIBforeignlanguage}[2]{{%
\expandafter\ifx\csname l@#1\endcsname\relax
\typeout{** WARNING: IEEEtran.bst: No hyphenation pattern has been}%
\typeout{** loaded for the language `#1'. Using the pattern for}%
\typeout{** the default language instead.}%
\else
\language=\csname l@#1\endcsname
\fi
#2}}
\providecommand{\BIBdecl}{\relax}
\BIBdecl

\bibitem{Parker:Diss:14}
J.~T. Parker, ``Approximate message passing algorithms for generalized bilinear
  inference,'' Ph.D. dissertation, The Ohio State University, Columbus, OH,
  Aug. 2014.

\bibitem{Candes:PROC:10}
E.~J. Cand{\`e}s and Y.~Plan, ``Matrix completion with noise,'' \emph{Proc.
  IEEE}, vol.~98, no.~6, pp. 925--936, Jun. 2010.

\bibitem{Candes:JACM:11}
E.~J. Cand{\`e}s, X.~Li, Y.~Ma, and J.~Wright, ``Robust principal component
  analysis?'' \emph{J. ACM}, vol.~58, no.~3, p.~11, May 2011.

\bibitem{Chandrasekaran:JO:11}
V.~Chandrasekaran, S.~Sanghavi, P.~A. Parrilo, and A.~S. Willsky,
  ``Rank-sparsity incoherence for matrix decomposition,'' \emph{SIAM J.
  Optim.}, vol.~21, pp. 572--596, 2011.

\bibitem{Zhou:ISIT:10}
Z.~Zhou, J.~Wright, X.~Li, E.~J. Cand{\`e}s, and Y.~Ma, ``Stable principal
  component pursuit,'' in \emph{Proc. IEEE Int. Symp. Inform. Thy.}, Austin,
  TX, Jun. 2010.

\bibitem{jtp_Rubinstein2010}
R.~Rubinstein, A.~Bruckstein, and M.~Elad, ``Dictionaries for sparse
  representation modeling,'' \emph{Proceedings of the IEEE}, vol.~98, no.~6,
  pp. 1045--1057, 2010.

\bibitem{Lee:NIPS:01}
D.~D. Lee and H.~S. Seung, ``Algorithms for non-negative matrix
  factorization,'' in \emph{Proc. NIPS}, 2001, pp. 556--562.

\bibitem{Ling:IP:15}
S.~Ling and T.~Strohmer, ``Self-calibration and biconvex compressive sensing,''
  \emph{Inverse Problems}, vol.~31, no.~11, p. 115002, 2015.

\bibitem{Zhu:TSP:11}
H.~Zhu, G.~Leus, and G.~B. Giannakis, ``Sparsity-cognizant total least-squares
  for perturbed compressive sampling,'' \emph{IEEE Trans. Signal Process.},
  vol.~59, no.~5, pp. 2002--2016, May 2011.

\bibitem{Waters:NIPS:11}
A.~E. Waters, A.~C. Sankaranarayanan, and R.~G. Baraniuk, ``{SpaRCS}:
  {R}ecovering low-rank and sparse matrices from compressive measurements,'' in
  \emph{Proc. NIPS}, 2011, pp. 1089--1097.

\bibitem{Candes:TIT:11}
E.~J. Cand{\`e}s and Y.~Plan, ``Tight oracle inequalities for low-rank matrix
  recovery from a minimal number of noisy random measurements,'' \emph{IEEE
  Trans. Inform. Theory}, vol.~57, no.~4, pp. 2342--2359, 2011.

\bibitem{Agarwal:AS:12}
A.~Agarwal, S.~Negahban, and M.~J. Wainwright, ``Matrix decomposition via
  convex relaxation: {O}ptimal rates in high dimensions,'' \emph{Ann.
  Statist.}, vol.~40, no.~2, pp. 1171--1197, 2012.

\bibitem{Wright:II:13}
J.~Wright, A.~Ganesh, K.~Min, and Y.~Ma, ``Compressive principal component
  pursuit,'' \emph{Inform. Inference}, vol.~2, no.~1, pp. 32--68, 2013.

\bibitem{Bioucas:GRSM:13}
J.~M. Bioucas-Dias, A.~Plaza, G.~Camps-Valls, P.~Scheunders, N.~M. Nasrabadi,
  and J.~Chanussot, ``Hyperspectral remote sensing data analysis and future
  challenges,'' \emph{IEEE Geoscience and Remote Sensing Magazine}, vol.~1,
  no.~2, pp. 6--36, 2013.

\bibitem{Kamilov:TSP:12}
U.~S. Kamilov, V.~K. Goyal, and S.~Rangan, ``Message-passing de-quantization
  with applications to compressed sensing,'' \emph{IEEE Trans. Signal
  Process.}, vol.~60, no.~12, pp. 6270--6281, Dec. 2012.

\bibitem{Fletcher:NIPS:11}
A.~K. Fletcher, S.~Rangan, L.~R. Varshney, and A.~Bhargava, ``Neural
  reconstruction with approximate message passing ({NeuRAMP}),'' in \emph{Proc.
  NIPS}, 2011.

\bibitem{Schniter:TSP:15}
P.~Schniter and S.~Rangan, ``Compressive phase retrieval via generalized
  approximate message passing,'' \emph{IEEE Trans. Signal Process.}, vol.~63,
  no.~4, pp. 1043--1055, Feb. 2015, (see also \emph{arXiv:1405.5618}).

\bibitem{Montanari:Chap:12}
A.~Montanari, ``Graphical models concepts in compressed sensing,'' in
  \emph{Compressed Sensing: Theory and Applications}, Y.~C. Eldar and
  G.~Kutyniok, Eds.\hskip 1em plus 0.5em minus 0.4em\relax Cambridge Univ.
  Press, 2012.

\bibitem{Rangan:ISIT:11}
S.~Rangan, ``Generalized approximate message passing for estimation with random
  linear mixing,'' in \emph{Proc. IEEE Int. Symp. Inform. Thy.}, Aug. 2011, pp.
  2168--2172, (full version at \emph{arXiv:1010.5141}).

\bibitem{Parker:TSP:14a}
J.~T. Parker, P.~Schniter, and V.~Cevher, ``Bilinear generalized approximate
  message passing---{Part I: D}erivation,'' \emph{IEEE Trans. Signal Process.},
  vol.~62, no.~22, pp. 5839--5853, Nov. 2014, (See also arXiv:1310:2632).

\bibitem{Parker:TSP:14b}
------, ``Bilinear generalized approximate message passing---{Part II:
  A}pplications,'' \emph{IEEE Trans. Signal Process.}, vol.~62, no.~22, pp.
  5854--5867, Nov. 2014, (See also arXiv:1310:2632).

\bibitem{Kabashima:14}
Y.~Kabashima, F.~Krzakala, M.~Mezard, A.~Sakata, and L.~Zdeborova, ``Phase
  transitions and sample complexity in {B}ayes-optimal matrix factorization,''
  \emph{arXiv:1402.1298}, 2014.

\bibitem{Dempster:JRSS:77}
A.~Dempster, N.~M. Laird, and D.~B. Rubin, ``Maximum-likelihood from incomplete
  data via the {E}{M} algorithm,'' \emph{J. Roy. Statist. Soc.}, vol.~39, pp.
  1--17, 1977.

\bibitem{Krzakala:JSM:12}
F.~Krzakala, M.~M\'ezard, F.~Sausset, Y.~Sun, and L.~Zdeborov\'a,
  ``Probabilistic reconstruction in compressed sensing: {A}lgorithms, phase
  diagrams, and threshold achieving matrices,'' \emph{J. Stat. Mech.}, vol.
  P08009, 2012.

\bibitem{Vila:TSP:13}
J.~P. Vila and P.~Schniter, ``Expectation-maximization {G}aussian-mixture
  approximate message passing,'' \emph{IEEE Trans. Signal Process.}, vol.~61,
  no.~19, pp. 4658--4672, Oct. 2013.

\bibitem{Kamilov:TIT:14}
U.~S. Kamilov, S.~Rangan, A.~K. Fletcher, and M.~Unser, ``Approximate message
  passing with consistent parameter estimation and applications to sparse
  learning,'' \emph{IEEE Trans. Inform. Theory}, vol.~60, no.~5, pp.
  2969--2985, May 2014.

\bibitem{Herman:JSTSP:10}
M.~A. Herman and T.~Strohmer, ``Generalized deviants: {A}n analysis of
  perturbations in compressed sensing,'' \emph{IEEE J. Sel. Topics Signal
  Process.}, vol.~4, no.~2, pp. 342--349, Apr. 2010.

\bibitem{Parker:ASIL:11}
J.~T. Parker, V.~Cevher, and P.~Schniter, ``Compressive sensing under matrix
  uncertainties: {A}n approximate message passing approach,'' in \emph{Proc.
  Asilomar Conf. Signals Syst. Comput.}, Pacific Grove, CA, Nov. 2011, pp.
  804--808.

\bibitem{Krzakala:ICASSP:13}
F.~Krzakala, M.~M\'ezard, and L.~Zdeborov\'a, ``Compressed sensing under matrix
  uncertainty: {O}ptimum thresholds and robust approximate message passing,''
  in \emph{Proc. IEEE Int. Conf. Acoust. Speech \& Signal Process.}, 2013, pp.
  5519--5523.

\bibitem{Gribonval:ICASSP:12}
R.~Gribonval, G.~Chardon, and L.~Daudet, ``Blind calibration for compressed
  sensing by convex optimization,'' in \emph{Proc. IEEE Int. Conf. Acoust.
  Speech \& Signal Process.}, 2012, pp. 2713--2716.

\bibitem{Bilen:TSP:14}
C.~Bilen, G.~Puy, and R.~Gribonval, ``Convex optimization approaches for blind
  sensor calibration using sparsity,'' \emph{IEEE Trans. Signal Process.},
  vol.~62, no.~18, pp. 4847--4856, 2014.

\bibitem{Schulke:NIPS:14}
C.~Sch{\"u}lke, F.~Caltagirone, F.~Krzakala, and L.~Zdeborov{\'a}, ``Blind
  calibration in compressed sensing using message passing algorithms,'' in
  \emph{Proc. NIPS}, 2014.

\bibitem{Schniter:JSTSP:11}
P.~Schniter, ``A message-passing receiver for {BICM-OFDM} over unknown
  clustered-sparse channels,'' \emph{IEEE J. Sel. Topics Signal Process.},
  vol.~5, no.~8, pp. 1462--1474, Dec. 2011.

\bibitem{Kamilov:ICASSP:13}
U.~S. Kamilov, A.~Bourquard, E.~Bostan, and M.~Unser, ``Autocalibrated signal
  reconstruction from linear measurements using adaptive {GAMP},'' in
  \emph{Proc. IEEE Int. Conf. Acoust. Speech \& Signal Process.}, 2013, pp.
  5925--5928.

\bibitem{Asif:ALL:09}
M.~S. Asif, W.~Mantzel, and J.~Romberg, ``Random channel coding and blind
  deconvolution,'' in \emph{Proc. Allerton Conf. Commun. Control Comput.},
  2009, pp. 1021--1025.

\bibitem{Ahmed:TIT:14}
A.~Ahmed, B.~Recht, and J.~Romberg, ``Blind deconvolution using convex
  programming,'' \emph{IEEE Trans. Inform. Theory}, vol.~60, no.~3, pp.
  1711--1732, 2012.

\bibitem{Hedge:TSP:11}
C.~Hegde and R.~G. Baraniuk, ``Sampling and recovery of pulse streams,''
  \emph{IEEE Trans. Signal Process.}, vol.~59, no.~14, pp. 1505--1517, 2011.

\bibitem{Choudhary:deconv1:14}
S.~Choudhary and U.~Mitra, ``Fundamental limits of blind deconvolution {Part I:
  A}mbiguity kernel,'' \emph{arXiv:1411.3810}, 2014.

\bibitem{Choudhary:deconv2:15}
------, ``Fundamental limits of blind deconvolution {Part II:
  S}parsity-ambiguity trade-offs,'' \emph{arXiv:1503.03184}, 2015.

\bibitem{Li:deconv1:15}
Y.~Li, K.~Lee, and Y.~Bresler, ``Identifiability in blind deconvolution with
  subspace or sparsity constraints,'' \emph{arXiv:1505.03399}, 2015.

\bibitem{Li:deconv2:15}
------, ``Identifiability in blind deconvolution under minimal assumptions,''
  \emph{arXiv:1507.01308}, 2015.

\bibitem{Choudhary:bilinear:14}
S.~Choudhary and U.~Mitra, ``Identifiability scaling laws in bilinear inverse
  problems,'' \emph{arXiv:1402.2637}, 2014.

\bibitem{Zhou:ICML:11}
T.~Zhou and D.~Tao, ``Godec: {R}andomized low-rank \& sparse matrix
  decomposition in noisy case,'' in \emph{Proc. Int. Conf. Mach. Learning},
  2011.

\bibitem{Kyrillidis:12b}
A.~Kyrillidis and V.~Cevher, ``Matrix {ALPs}: {A}ccelerated low rank and sparse
  matrix reconstruction,'' \emph{arXiv:1203.3864}, 2012.

\bibitem{Aravkin:UAI:14}
A.~Aravkin, S.~Becker, V.~Cevher, and P.~Olsen, ``A variational approach to
  stable principal component pursuit,'' in \emph{Proc. Conf. Uncertainty
  Artificial Intell.}, 2014.

\bibitem{Pearl:Book:88}
J.~Pearl, \emph{Probabilistic Reasoning in Intelligent Systems}.\hskip 1em plus
  0.5em minus 0.4em\relax San Mateo, CA: Morgan Kaufman, 1988.

\bibitem{Kschischang:TIT:01}
F.~R. Kschischang, B.~J. Frey, and H.-A. Loeliger, ``Factor graphs and the
  sum-product algorithm,'' \emph{IEEE Trans. Inform. Theory}, vol.~47, pp.
  498--519, Feb. 2001.

\bibitem{Cooper:AI:90}
G.~F. Cooper, ``The computational complexity of probabilistic inference using
  {B}ayesian belief networks,'' \emph{Artificial Intelligence}, vol.~42, pp.
  393--405, 1990.

\bibitem{Murphy:UAI:99}
K.~P. Murphy, Y.~Weiss, and M.~I. Jordan, ``Loopy belief propagation for
  approximate inference: {A}n empirical study,'' in \emph{Proc. Uncertainty
  Artif. Intell.}, 1999, pp. 467--475.

\bibitem{Javanmard:II:13}
A.~Javanmard and A.~Montanari, ``State evolution for general approximate
  message passing algorithms, with applications to spatial coupling,''
  \emph{Inform. Inference}, vol.~2, no.~2, pp. 115--144, 2013.

\bibitem{Rangan:ISIT:14}
S.~Rangan, P.~Schniter, and A.~Fletcher, ``On the convergence of generalized
  approximate message passing with arbitrary matrices,'' in \emph{Proc. IEEE
  Int. Symp. Inform. Thy.}, Jul. 2014, pp. 236--240, (full version at
  \emph{arXiv:1402.3210}).

\bibitem{Vila:ICASSP:15}
J.~Vila, P.~Schniter, S.~Rangan, F.~Krzakala, and L.~Zdeborov{\'a}, ``Adaptive
  damping and mean removal for the generalized approximate message passing
  algorithm,'' in \emph{Proc. IEEE Int. Conf. Acoust. Speech \& Signal
  Process.}, 2015.

\bibitem{GAMPmatlab}
S.~Rangan, P.~Schniter, J.~T. Parker, J.~Ziniel, J.~Vila, M.~Borgerding
  \emph{et~al.}, ``{GAMP}matlab,''
  {\footnotesize\verb+https://sourceforge.net/projects/gampmatlab/+}.

\bibitem{Neal:Jordan:98}
R.~Neal and G.~Hinton, ``A view of the {EM} algorithm that justifies
  incremental, sparse, and other variants,'' in \emph{Learning in Graphical
  Models}, M.~I. Jordan, Ed.\hskip 1em plus 0.5em minus 0.4em\relax MIT Press,
  1998, pp. 355--368.

\bibitem{Manton:SCL:03}
J.~H. Manton and W.~D. Neumann, ``Totally blind channel identi􏰻cation by
  exploiting guard intervals,'' \emph{Syst. Control Lett.}, vol.~48, pp.
  113--119, 2003.

\bibitem{Hua:TSP:96}
Y.~Hua, ``Fast maximum likelihood for blind identification of multiple {FIR}
  channels,'' \emph{IEEE Trans. Signal Process.}, vol.~44, pp. 661--672, 1996.

\bibitem{jtp_de-Jong1987}
P.~{de Jong}, ``A central limit theorem for generalized quadratic forms,''
  \emph{Probab. Th. Rel. Fields}, vol.~75, pp. 261--277, 1987.

\end{thebibliography}

\end{document}